\documentstyle[12pt,epsfig]{article}
\renewcommand{\theequation}{\mbox{\arabic{section}.\arabic{equation}}}
\begin{document}\def\p{\phi}\def\P{\Phi}\def\a{\alpha}\def\e{\epsilon}
\def\be{\begin{equation}}\def\ee{\end{equation}}\def\l{\label}
\def\0{\setcounter{equation}{0}}\def\b{\beta}\def\S{\Sigma}\def\C{\cite}
\def\r{\ref}\def\ba{\begin{eqnarray}}\def\ea{\end{eqnarray}}
\def\n{\nonumber}\def\R{\rho}\def\X{\Xi}\def\x{\xi}\def\la{\lambda}
\def\d{\delta}\def\s{\sigma}\def\f{\frac}\def\D{\Delta}\def\pa{\partial}
\def\Th{\Theta}\def\o{\omega}\def\O{\Omega}\def\th{\theta}\def\ga{\gamma}
\def\Ga{\Gamma}\def\t{\times}\def\h{\hat}\def\rar{\rightarrow}
\def\vp{\varphi}\def\inf{\infty}\def\le{\left}\def\ri{\right}
\def\foot{\footnote}\def\ve{\varepsilon}\def\N{\bar{n}(s)}\def\cS{{\cal S}}
\def\k{\kappa}\def\sq{\sqrt{s}}\def\bx{{\bf x}}\def\La{\Lambda}
\def\bb{{\bf b}}\def\bq{{\bf q}}\def\cp{{\cal P}}\def\tg{\tilde{g}}
\def\cf{{\cal F}}\def\bN{{\bf N}}\def\Re{{\rm Re}}\def\Im{{\rm Im}}
\def\K{{\bf K}}
\footnotesize
\begin{flushright}
{\it\large E2-2000-217}
\end{flushright}
\vskip 3cm
\begin{center}
{\huge\bf Very High Multiplicity Hadron Processes}\\
\vskip 0.5cm
{\large\it J.Manjavidze$^a$\foot{Permanent address: Institute of
Physics, Tbilisi, Georgia}} and {\large\it A.Sissakian$^b$}
\vskip 1cm
{\large (To appear in {\it Physics Reports})}
\vskip 3cm

$^a$Joint Institute for Nuclear Research, Djelepov Lab. of Nucl.
Problems, 141980 Dubna, Russia, E-mail: joseph@nusun.jinr.ru\\
Tel.: (09621) 6 35 17, Fax: (09621) 6 66 66
\vskip 0.5cm
$^b$Joint Institute for Nuclear Research, Bogolyubov Lab. of
Theor. Physics, 141980 Dubna, Russia, E-mail: sisakian@jinr.ru\\ Tel.
(09621) 6 22 68, Fax:  (09621) 6 58 91 \end{center}

\newpage
\begin{abstract}

The paper contains a description of a first attempt to understand the
extremely inelastic high energy hadron collisions, when the
multiplicity of produced hadrons considerably exceeds its mean value.
Problems with existing model predictions are discussed. The real-time
finite-temperature $S$-matrix theory is built to have a possibility
to find model-free predictions. This allows to include the
statistical effects into consideration and build the phenomenology.
The questions to experiment are formulated at the very end of the
paper.
\end{abstract}
\vskip 0.5cm

\vskip 0.5cm

\newpage
\tableofcontents
\newpage
\section{Introduction}\0

The intuitive feeling that hadron matter should be maximally
perturbed in the high energy extremely inelastic collisions was a
main reason of our effort to consider such processes. It was a
hope to observe new dynamical phenomena, or new degrees of freedom,
unattainable in other ordinary hadron reactions.  This paper presents
a first attempt to describe the particularity of considered processes,
to give a review of existing models prediction and, at the end, we
will offer the field-theoretical formalism for hadron inelastic
processes.

Thus, considering hadrons mean multiplicity $\N$ as a natural scale of
the produced hadrons multiplicity $n$ at given CM energy $\sqrt s$, we
would assume that
\be
n>>\N.
\l{0.1}\ee
At the same time we wish to have
\be
n<<n_{max}=\sqrt{s}/m,
\l{0.2}\ee
where $m\simeq0.1$ Gev is the characteristic hadron mass. The last
restriction is introduced to weaken the unphysical constraints from
the finite, for given $s$, phase space volume. We should assume
therefore that $s$ is high enough.

The multiple production is the process of colliding particles where
kinetic energy is dissipated into the mass of produced particles
\C{fer}. Then one may validate that the entropy $\cS$ accedes its
maximum in the domain (\r{0.1}) since the multiplicity $n$
characterizes the rate of stochastization, i.e. the level of incident
energy dissipation over existing (free) degrees of freedom.

There is also another quantitative definition of our reactions. Let
$\ve_{max}$ be the energy of the fastest particle in the given frame
and let $E$ be the total incident energy in the same frame. Then the
difference ($E-\ve_{max}$) is the energy spent on production of the
less energetic particles. It is useful to consider the inelasticity
coefficient
\be
\k=\f{E-\ve_{max}}{E}=1-\f{\ve_{max}}{E}\leq 1.
\l{0.2a}\ee
It defines the portion of spent energy. Therefore, we wish to
consider processes with
\be
1-\k<<1.
\l{0.2b}\ee
So, the produced particles have comparatively small energies.

This property may be used for experimental triggering of our
processes. Indeed, using the energy conservation law,
\be
n(1-\k)>1.
\l{0.2c}\ee
Following (\r{0.2}) we will assume that
\be
1-\k>>\f{m}{E}.
\l{0.2d}\ee
Therefore, the kinetic energy of produced particles in our processes
can not be arbitrarily small.

Using thermodynamical terminology, we wish investigate the production
and properties of comparatively `cold' multi-hadron (mostly of
$\pi$-mesons) state. We would like to note from the very beginning
that we have only a qualitative scenario of such states which may be
produced and the review of the corresponding why's may be considered
as the main purpose of this paper. Although at the end of this paper
(Appendix K) we will describe principal features of possible
field-theoretical solution of our problem.

It should be noted the absence of any authentic experimental
information concerning discussed processes. Moreover, actually the
hadron inelastic interactions with a set peculiar to them of unsolved
theoretical problems will be considered. Nevertheless we suggest to
work in this field in spite of these difficulties because the
system with extremal properties may be more transparent since the
asymptotics always simplify a picture.  We would demonstrate this
idea and will try to put it in the basis of developed theoretical
methods.

Absence of experimental information about so high inelastic hadron
processes is the consequence of the smallness of corresponding cross
sections. Besides this it was unclear for what purpose the
experimental efforts should be done. We would like to convince the
reader that the discussed problem is interesting and important. For
instance, we will discuss a possibility that asymptotics over $n$ may
replace in definite sense the asymptotics over $\sqrt s$. A short
address to experimentalists will be given in concluding Sec.4.1.2.

We hope that the paper would be useful both for theorists and
experimentalists. By this reason the main text of this paper will
contain only the qualitative discussion of the problems and results.
The quantitative proof, formulation of pure theoretical methods
etc. are added in the Appendices. Considered extremal problem is a
good theoretical laboratory and is described in the Appendices and
theoretical methods may be applied for other physical problems.

We would like to point out that the special technique was built for
the problem discussed, see Sec.2.

-- {\it Having the very high multiplicity (VHM) state it is natural
to use the thermodynamical methods.} We will offer for this purpose
the real-time $S$-matrix interpretation of thermodynamics. It can be
shown in what quantitative conditions it will coincide with simpler
canonical imaginary-time Matsubara formalism. We will give also the
generalization of the real-time finite-temperature perturbation
theory in the case of local temperature $T=1/\b$ distribution, when
$\b=\b(x,t)$. This will allow to use the thermodynamical description
if the system is far from equilibrium.

-- {\it The particle spectrum in the VHM region is soft. It is just a
situation when the collective phenomena should be important.} To
describe this phenomena, the decomposition on correlators will be
adopted.  The origin of this decomposition lie in the Mayer's `group
decomposition'. In multiple production physics this decomposition is
known also as the `multi-component description' \C{slep}. It based on
the idea that the multiple production process may include various
mechanisms.

In Sec.3 we will investigate model predictions for VHM region. We
would like to note two main conclusions:

-- Existing multiperipheral-type models are unable to describe
the VHM region.

-- The infrared region of the pQCD becomes important even if the
constraint (\r{0.2}) is taken into account.

\section{Qualitative inside of the problem}\0

In Sec.2.1 we will try to formulate the phenomenology of our problem,
i.e. the way the VHM processes may be described and what type of
phenomena one may expect. The importance of thermodynamical methods
will became evident and we will offer in Sec.2.2 a general
description of the corresponding formalism.

It is important to note that we may classify the possible asymptotics
over $n$. We will find that there exist only three classes of
asymptotics.  This will simplify consideration definitely restricting
the possibilities.

In Sec.2.3 we will use the thermodynamical language to give a physical
interpretation of these classes of asymptotics.  We will see in result
that in our choice of the VHM final state this should lead to
reorganization of multiple production dynamics: we will get out of
the habitual multiperipheral picture in the VHM domain.

Moreover, one may assume that the semiclassical approximation becomes
exact in the VHM domain.  This naturally leads to the idea to search
such a scheme of calculation which depends on the choice of final
state.  Quantitative description of this idea may be realized as is
described in Appendix K.

\subsection{Phenomenology of VHM processes}

The VHM production phenomena includes two sub-problems. First of all
it is the dynamical problem of incident energy degradation into the
secondary particle energies and the second one is the description of
the final state.

We will start discussion in Sec.2.1.1 from the second part of the
problem to explain that the statistical methods are essential for us.

In Sec.2.1.2 we will try to outline at least qualitatively the main
mechanisms of hadron production. The peculiarity of hadron production
phenomena consists in the presence of hidden constraints, the
consequence of local non-Abelian gauge symmetry.  The constraints may
prevent thermalization and the incident energy dissipation is
confused in this case.  Just the `confusing' effect is dominant
in the hadron multiple production processes if $n\sim\N$.

The expected change of dominant mechanism of hadron production is
discussed Sec.2.1.3. It is important that, in spite of hidden
constraints, the system may freely evolve to define the VHM state.
Such a VHM state should be in equilibrium. Formal definition of the
`equilibrium' notion will be given in Sec.2.2.2.

The problem described contains small parameters $(\N/n)<<1$ and
$(1-\k)<<1$. To have the possibility of estimation of contributions in
accordance with these parameters one should include them into the
formalism. This becomes possible if and only if the integral
quantities are calculated. So, the multiplicity $n$ is an index
only if the multiple production amplitudes $a_n(p_1,...,p_n)$ are
considered.  But the cross section $\s_n(s)$ is a nontrivial
$function$ of $n$.  We will calculated by this logic mostly
integrals of $|a_n|^2$, living out the consideration the amplitudes,
see also Sec.2.2, where the first realization of this idea is
offered (the $naive$ attempt to realize this idea one may find in
\C{VIR}).  This is a general methodological feature of our
consideration.

\subsubsection{Formulation of the problem}

The multiple production cross section $\s_n(s)$ falls down rapidly in
the discussed very high multiplicity (VHM) domain (\r{0.1}) and for
this reason the multiplicities $n\sim n_{max}$ are not accessible
experimentally. At the LHC energy $\N\simeq 100$ is valid and we will
assume that $n\sim\N^2\simeq10~000$ is just the discussed VHM region
($n_{max}\simeq 100~000$ at the LHC energy). We will explain later why
\be
n\sim\N^2
\l{0.1'}\ee
is chosen for the definition of VHM region.

Generally speaking, having the state of a large number of particles,
it is reasonable to depart from an exact definition of the number $n$
of created particles, their individual energies $\ve_i$, momenta
$q_i$, etc.  since they cannot be defined exactly by experiment.
Indeed, for instance, full reconstruction of kinematics is a
practically impossible task because of neutral particles, neutrinos,
the more so as $n\sim10~000$ is considered. We suppose that nothing
will happen if $n$ is measured with $\D n\neq0$ accuracy since $(\D
n/n)<<1$ is easily attainable in the VHM region. Besides it is
practically impossible to deal with theory which operates by the
$N=3n-4$ ($\sim 30~000$!) variables.

Artificial reduction of the set of the necessary variables may lead
to a temporary success only. Indeed, the last thirty years of
multiple production physics development was based on the inclusive
approach \C{0.1}, when the measured quantities (cross sections)
depend on a few dynamical parameters only. But later on the
experiment and its fractal analyses shows that the situation is not
so simple, also as, for instance, for the classical turbulence. So,
the event-by-event experimental data shows that the particle density
fluctuation is unexpectedly large \C{0.2} and the fractal dimension
$D_f$ is not equal to zero \C{0.3}.

We know that if the fractal dimension is non-trivial, then the system
is extremely `non-regular' \C{0.4}. So, $D_f\simeq$0.3 for the
perimeter of Great Britain and $D_f\simeq$0.5 for Norway.  The
discrepancy marks the fact that the shore of Norway is much more
broken then of Great Britain \C{flat}. It is noticeable that the
fractal dimension $D_f$ crucially depends on the type of reaction,
incident energy and so on.

It is evident that one may choose from $N=3n-4$ an arbitrary finite
set of variables to characterize the multiple production process. But
the fractal analyses shows that such approach would lead to the same
effect as if one may hear, for example, only the first violin of
Mahler's music.

So, it is important to understand when the restricted set of
dynamical variables will admit to describe the process (state)
$completely$.  The same problem was solved in statistical
physics, where the `rough' description by a restricted number of
(thermodynamical) parameters is a basis of its success, see the
discussion of rough variables description e.g. in the review of
Uhlenbeck \C{uhl}.  We will search the same solution desiring to
build a complete theory of the VHM hadron reactions.

We want to note that just VHM process may be in this sense `simple':
at all evidence, the system becomes `quiet' in the VHM region and by
this reason its `rough' thermodynamical description is available. It
seems natural, therefore, to start investigation of multiple
production phenomena from the (extremely rare) VHM processes.

\subsubsection{Soft channel of hadron production}

The dominant inelastic hadron processes at $n\sim\N$ are saturated
by production of low transverse momentum hadrons \C{1.5}. One
of the approaches explains this phenomena by the nonperturbative
effect of quarks created from vacuum.

Corresponding dynamics looks as follows. At the expanse of transverse
kinetic motion color charges may separated at large distances.
Nevertheless the transverse motion is suppressed since separation
leads to increasing polarization of vacuum, because of confinement
phenomenon.  Then, as in QED \C{psf}, the vacuum becomes unstable in
regard to the tunneling creation of real fermions.  Just on their
creation, the transverse kinetic energy is spent and, as a result,
particles cannot have, with exponential accuracy, high transverse
energies. This picture is attractive being simple and transparent,
but despite numerous efforts \C{string}, there is no quantitative
description of this phenomenon till now.  Briefly, the problem is
connected with the unknown mechanism of strong coloured electric
fields formation among distant colored charges, see also \C{t'hooft}.

One may use other terms. The soft channel of multiple production means
the long-range correlation among hadron coloured constituents. Under
this special correlation the non-Abelian gauge field theory
conservation laws constraint were implied.  They are important in
dynamics since each conservation law decreases the number of
independent degrees of freedom at least on one unity (this may
explain why hadrons $\bar{n}(s) << n_{max}(s)$), i.e. it has the
nonperturbative effect.  Moreover, in so-called integrable systems
each independent integral of motion (in involution) reduces the
number of degrees of freedom on two units.  As a result there is no
thermalization in such systems \C{zak} and the corresponding mean
multiplicity $\bar{n}(s)$ should be equal to zero, see Appendix K.

Existence of multiple production, $\bar{n}(s)>>1$, testifies the
statement that the thermalization phenomena exist in hadron
processes, i.e. the system of Yang-Mills fields is not completely
integrable. But the most probable process with $n\sim\N$ did not
lead to the final state with maximal entropy since $\bar{n}(s)
<<n_{max}$, i.e.  the definite restrictions on the dissipation
dynamics should be taken into account.  Such problems, being
intermediate, are mostly complicated ones.

The quantitative theory of this phenomena may lead to deep revision
of the main notions of existing quantum field theory \C{jmp, tmf}, see
Appendix K.  So, the dynamical display of hidden conservation laws of
the hadron system are probably unstable since we expect that the
system is not completely integrable, solitary field configurations
$u_c(x,t)$ \C{solit}.  Then the quantum theory should be
able to describe quantum excitations of these fields, i.e.  to count
the fluctuations of `curved' manifolds. The canonical perturbation
theory methods, formulated in terms of creation and annihilation of
particles in the external field $u_c(x,t)$, are too
complicated, see \C{fad.PR} and references cited therein. For this
reason, existing calculations usually do not exceed the semiclassical
approximation.  We hope that, as described in Appendix K, the
quantization scheme would be able to solve this problem (see also the
example described there).

Another approach assumes that the special `$t$-channel' ladder type
Feynman diagrams are able to describe the $n\sim\N$ region
\C{lipatov}. This approach did not take into account the confinement
nonperturbative effects introducing the hardly controllable
supposition that the free quarks and gluons may form a complete set
of states.  Formally this is right, but at all evidence, the
decomposition on this Fock basis is realized in the non-Abelian gauge
field theories on zero measure \C{tmf, hep} (see Appendix K).
Nevertheless one may reject this argument assuming that the process
is happening on a sufficiently small distances.

Corresponding contribution came from the so called `hard Pomeron'
\C{kuraev}.  But the intrinsic problems of the accuracy of chosen
logarithmic approximations \C{accuracy}, the understanding of the so
called non-logarithmic corrections \C{zakharov}, of the fate of the
infrared divergences stay unsolved till now.

\subsubsection{Multiple production as a process of dissipation}

So, the multiple production of soft hadron phenomena seems unsolvable
on the day-to-day level of understanding and we may with pure
conscience move it away.  This is why the hadron inelastic reactions
lost some popularity, migrating the last two decades to the class of
`non-interesting' problems.  Yet, in a number of modern fundamental
experiments, multiple production plays, at least, the role of
background to the investigated phenomena and for this reason we should
be ready for the quantitative estimation of it.

Our hope to describe such a complicated problem as the multiple
production phenomenon in the VHM conditions is based on the following
idea.  At the very beginning of this century, a couple P.  and
T.Euhrenfest, had offered a model to visualize Boltzmann's
interpretation of the irreversibility phenomena in statistics.  The
model is extremely simple and fruitful \C{kac}.  It considers the two
boxes with $2N_b$ numerated balls. Choosing the label of the balls
$randomly$ one must take the ball with the corresponding label from
one box and put it into another one. One may repeat this action an
arbitrary number of times $t$.

Starting from the highly `nonequilibrium' state with all balls in one
box, $N_b>>1$, it is seen stationary with $t$ tendency to
equalization of the number of balls in the boxes Fig.1. The
stationarity means that the number of balls in the other box rises
$\sim t$ at least on an early stage of the process. This signifies
presence of an $irreversible$\foot{One can say that the opposite flow
is never seen. `What never?  No never!  What never?  Well, hardly
ever.' This dialog was taken from \C{mayer}} flow (of balls) toward
the preferable (equilibrium) state.  One can hope \C{kac} that this
model reflects a physical reality of nonequilibrium processes with
initial state very far from equilibrium. A theory of such processes
with (irreversible) flow toward a state with maximal entropy should
be sufficiently simple being close to the stationary Markovian.

The VHM production process may be, at least on an early stage,
stationary Markovian. If this is so then one may neglect long-range
effects, non-perturbative as well, since they are not Markovian as
follows from the experience described in Sec.2.1.2.

This is possible if the VHM process is happening so fast (being
the short-range phenomenon) that the confinement forces became
`frozen'.  It can be shown that the quantitative reason for this
phenomena is a fast (exponential) reduction with $n$ of the soft
channel contribution into the hadron production process. So, we
expect a change of the multiple production dynamics in the VHM region.

Thus, the main input idea consists from two general propositions.
The first of them is following:

({\bf I}) {\it The hadron VHM production processes should be close
to the stationary Markovian.}

`Freezing' the confinement constraints, the entropy $\cS$ may exceed
for given energy $\sqrt s$ its available maximum in the VHM domain.
Then one can assume that the VHM final state is in `equilibrium', or
is close to it.  So,

({\bf II}) {\it The VHM final state should be close to equilibrium}\\
is our second basic proposition.

We would select and appreciate particle physics models in accordance
with these propositions.

\subsection{$S$-matrix interpretation of thermodynamics}

The field-theoretical description of statistical systems at a finite
temperature is based usually on the formal  analogy between imaginary
time and inverse  temperature $\b=1/T$ \C{bloch}. This
analogy is formulated by Schwinger \C{psf} as the `euclidean
postulate' and it assumes that (i) the system is in equilibrium, i.e.
it should allow the arbitrary rearrangement of states of temporal
sequence in the described process\foot{In other terms, one may have
the possibility to apply the ergodic hypothesis.}, and (ii) there is
not special space-time long-range correlations among states of the
process, i.e, for instance, the symmetry constraints should not play
a crucial role.  We do not know {\it ad hoc} whether or not to apply
the `euclidean postulate' for given $n$ and $s$, even if (\r{0.1}) is
satisfied. For this reason we are forced to formulate the theory in
the natural real-time terms.

The  first important quantitative attempt to build the real-time
finite-temperature field theory \C{jack} discovered the formal problem
of the so called `pinch-singularities'. Further investigation of the
theory has allowed to demonstrate the  cancellation mechanism of
these unphysical singularities \C{sem}. This is attained by doubling
the degrees of freedom \C {sch, kel}:  the Green functions of the
theory represent $2$ $\times$ $2$ matrix \C{mahan}.  It surely makes
the theory more complicated, but the operator formalism of the
thermo-field dynamics \C{umez} shows the unavoidable character  of
this complication.

Notice that the canonical real-time finite temperature
field-theoretical description \C{sch, kel} of the statistical systems
based on the Kubo-Martin-Schwinger (KMS) \C{sch, mar, kubo} boundary
condition for a field is:
\be
\Phi (t)=\Phi (t-i\beta).
\l{22.1}
\ee
It, without fail, leads to the $equilibrium$ fluctuation-dissipation
conditions \C{haag} (see also \C{chu}). By this reason it can not
be applied in our case, where the dissipation problem is solved. The
origin of this boundary condition is shown in Appendix A.

We will use a more natural microcanonical formalism for particle
physics\foot{ The statistical methods for particles physics was
discussed also in \C{shum}.}.  The thermodynamical `rough' variables
are introduced in these approach as the Lagrange multipliers of
corresponding conservation laws.  The physical meaning of these
`rough' variables are defined by the corresponding equations of
state.

We shall use the $S$-matrix approach which is natural for the
description of the time evolution. (The $S$-matrix  description was
used also in \C{dash, carr}.) For this purpose the amplitudes \be
<p_1,p_2,...,p_n|q_1,q_2,...,q_m>=a_{nm}(p,q)
\l{22.2}
\ee
of the $m$- into $n$-particles transition will be introduced. The
in- and out-states must be composed from mass-shell particles \C{pei}.
Using these amplitudes we will calculate
\ba
&R_{nm}(p,q)=|a_{nm}(p,q)|^2=
\n\\
&=<p_1,p_2,...,p_n|q_1,q_2,...,q_m>
<q_1,q_2,...,q_m|p_1,p_2,...,p_n>.
\l{22.3}\ea
This will lead to the doubling of the degrees of freedom.

The temperature description  will be introduced (see also
\C{kaj}) noting that, for instance,
\ba
&d\Ga_n=|a_{nm}(p,q)|^2d\O_n(p),
\n\\
&d\O_n(p)=\prod_{1}^{n}\frac{d^3
p_i}{(2\pi )^3 2\e (p_i)},~\e(p)= (p^2 +m^2)^{1/2},
\l{22.4} \ea
is the differential measure of the final state. It is a first example
where the usefulness of the probability-like quantity
$\sim|a_{nm}|^2$ is seen.

The measure (\r{22.4}) is defined on the energy-momentum shell:
\be
\sum_{i=1}^np_i=P.
\l{22.4a}\ee
It should be underlined that $a_{nm}(p,q)$ are the translationally
invariant amplitudes and four equalities:
\be
\sum_{i=1}^np_i=\sum_{i=1}^mq_i
\l{22.4b}\ee
are obeyed identically. So, (\r{22.4a}) are the constraints and to
take them into account one may multiply $d\Ga_n$ on
$$
\prod_{k=1}^ne^{i\a p_k},
$$
where $\a$ is the time-like 4-vector. It is evident that integration
over $\a$ with factor $e^{-i\a P}$ gives the constraints (\r{22.4a}).

One may simplify the calculation assuming that all calculations are
performed, for example, in the CM frame $P=(E,{\bf 0})$. Then one may
ignore the space components considering $\a=(\a_0,{\bf 0})$. This
is the equivalent of the assumption that only the energy conservation
law is important.

The last step is the substitution $\a_0=i\b$, where $\b$ is our
Lagrange multiplier. To define its physical meaning one should solve
the equation of state:
\be
E=-\f{\pa}{\pa\b}\ln\int d\Ga_n\prod_{k=1}^ne^{-\b\e(p_k)}\equiv
-\f{\pa}{\pa\b}\ln\int d\Ga_n(\b).
\l{3'}\ee
Such a definition of temperature as the Lagrange multiplier of the
energy conservation law is obvious for microcanonical description
\C{mar}.

The initial-state temperature will be introduced in the same way,
taking into account (\r{22.4b}).  So, we will construct the
two-temperature theory.  It is impossible to use the KMS boundary
condition in such a two temperatures description (the equation of
state can be applied at the very end of the calculations).

It should be notice that the `density matrix' $R_{n,m}(p,q)$, defined
in (\ref{22.3}), describes the `closed-path motion' in the functional
space. So, if
\be
<p_1,p_2,...,p_n|q_1,q_2,...,q_m>=<n,out|e^{iS(\P_+)}|m,in>
\l{o1}\ee
and
\ba
<p_1,p_2,...,p_n|q_1,q_2,...,q_m>^*=
<q_1,q_2,...,q_m|p_1,p_2,...,p_n>=
\n\\
=<m,in|e^{-iS(\P_-)}|n,out>
\l{o2}\ea
then, by definition,
\be
\P_+(\s_{\infty})=\P_-(\s_{\infty})=\P (\s_{\infty}),
\l{o3}\ee
with some `turning-point' fields $\P (\s_{\infty})$, where
$\s_{\infty}$ is the remote hypersurface. The value of $\P
(\s_{\infty})$ specifies the environment of the system. We will show
that (\r{o3}) coincides with the KMS boundary condition in some
special cases. Here consequences of the vacuum boundary condition:
\be
\P (\s_{\infty})=0
\l{03a}\ee
are analyzed.

One should admit also that below boundary conditions are not unique:
one can consider arbitrary organization of the environment of the
considered system. The $S$-matrix interpretation is able to show the
way as an arbitrary boundary condition may be adopted. This should
extend the potentialities of the real-time finite-temperature
field-theoretical methods.

\subsubsection{Example}

It seems useful to illustrate the above microcanonical approach
by the simplest example, see also \C{kaj}. By definition, the $n$
particles production cross section
\be
\s_n(s)=\int d\O_n(p)\d(q_1+q_2-\sum^{n}_{i=1}p_i)|a_{n}(p,q)|^2,
\l{18}\ee
where $a_n(p,q)\equiv a_{n2}(p,q)$ is the ordinary $n$ particle
production amplitude in accelerator experiments.

Considering the Fourier transform of energy-momentum conservation
$\d$-function one can introduce the generating function $\R_n$, see
\C{kaj} and references cited therein\foot{The generating functionals
method was developed in \C{bog}.}.  We may find in the result that
$\s_n$ is defined by the equality:
\be
\s_n(E)=\int^{+i\infty}_{-i\infty}\f{d\b}{2\pi}e^{\b E}\R_n(\b),~
E=\e(q_1)+\e(q_2),
\l{3a}\ee
where
\be
\R_n(\b)=
\int\le\{\prod^n_{i=1}\f{d^3p_ie^{-\b\e(p_i)}}{(2\pi)^32\e(p_i)}
\ri\}|a_n|^2=\int d\Ga_n(\b).
\l{19}\ee
Most probable value of $\b$ in (\r{3a}) is defined by the equation
of state (\r{3'}). Inserting (\r{19}) into (\r{3a}) we find the
expression (\r{18}) if the momentum conservation shell is neglected.
The last one is possible since the cross sections are always measured
in the definite frame.

Let us consider the simplest example of noninteracting particles
\C{kaj}:
$$
\R_n(\b)=\le\{2\pi mK_1(\b m)/\b\ri\}^n,
$$
where $K_1$ is the Bessel function. Inserting this expression into
(\r{3'}) we can find that in the nonrelativistic case ($n\sim
n_{max}$)
$$
\b_c=\f{3}{2}\f{(n-1)}{(\sqrt{s}-nm)},
$$
i.e., we find the well known equality:
\be
E_{kin}=\f{3}{2}T,
\l{o4}\ee
where $E_{kin}=(\sqrt{s}-nm)/(n-1)$ is the mean kinetic energy and
$T=1/\b_c$ is the temperature (Boltzmann constant was taken equal to
one).

It is important to note that the equation (\r{3'}) has a unique real
solution $\b_c(s,n)$ rising with $n$ and decreasing with $s$ \C{mar}.

The expansion of the integral (\r{3a}) near $\b_c(s,n)$ unavoidably
gives an asymptotic series with zero convergence radii since
$\R_n(\b)$ is the essentially nonlinear function of $\b$, see also
Sec.2.2.2.  This means that, generally speaking, fluctuations in the
vicinity of $\b_c(s,n)$ may be arbitrarily high and in this case
$\b_c(s,n)$ has no physical sense. But if the fluctuations are
Gaussian, then $\R_n(\b)$ coincides with the partition function of
the $n$ particle state and $\b_c(s,n)$ may be interpreted as the
inverse temperature. We will put the observation of this important
fact in the basis of our thermodynamical description of the VHM
region.

\subsubsection{Relaxation of correlations}

The notion of `equilibrium' over some parameter $X$ in our
understanding is a requirement that the fluctuations in the vicinity
of its mean value, $\bar{X}$, have a Gaussian character. Notice, in
this case, one can use this variable for a `rough' description of the
system.  We would like to show now that the corresponding
equilibrium condition would have the meaning of the correlations
relaxation condition of Bogolyubov \C{bog}\foot{The term `vanishing
of correlations' was used by N.N.Bogolyubov for this phenomena.}, see
also \C{baldin}.  Let us define the conditions when the fluctuations
in the vicinity of $\b_c$ are Gaussian \C{15}.  Firstly, to estimate
the integral (\r{3a}) in the vicinity of the extremum, $\b_c$, we
should expand $\ln\R_n(\b+\b_c)$ over $\b$:

\be
\ln\R_n(\b+\b_c)=\ln\R_n(\b_c)-\sqrt{s}\b+\f{1}{2!}\b^2\f{\pa^2}
{\pa\b^2_c}
\ln\R_n(\b_c)-\f{1}{3!}\b^3\f{\pa^3}{\pa\b^3_c}\ln\R_n(\b_c)+...
\l{21c}\ee
and, secondly, expand the exponent in the integral (\r{3a}) over, for
instance, $${\pa^3\ln\R_n(\b_c)}/{\pa\b^3_c},...,$$ etc. In the
result, if higher terms in (\r{21c}) are neglected, $k$-th term of
the perturbation series
\be
\R_{n,k}\sim\le\{\f{{\pa^3\ln\R_n(\b_c)}/{\pa\b^3_c}}
{({\pa^2\ln\R_n(\b_c)}/{\pa\b^2_c})^{3/2}}\ri\}^k
\Ga\le(\f{3k+1}{2}\ri).
\l{22c}\ee
Therefore, because of Euler's $\Ga((3k+1)/2)$ function, the
perturbation theory near $\b_c$ leads to the asymptotic series. The
supposition to define this series formally, for instance, in the Borel
sense is not interesting from physical point of view. Indeed, such
a formal solution assumes that the fluctuations near $\b_c$ may be
arbitrarily high. Then, by this reason, the value of $\b_c$ loses its
significance: arbitrary values of $(\b-\b_c)$ are important in this
case.

Nevertheless it is important to know that our asymptotic series
exists in some definite sense, i.e. we can calculate the integral
over $\b$ by expanding it over $(\b-\b_c)$. Therefore, if the
considered series is asymptotic, we may estimate it by first term if
\be
{\pa^3\ln\R_n(\b_c)}/{\pa\b^3_c}<<
({\pa^2\ln\R_n(\b_c)}/{\pa\b^2_c})^{3/2}.
\l{23c}\ee
One of the possible solutions of this condition is
\be
{\pa^3\ln\R_n(\b_c)}/{\pa\b^3_c}\approx 0.
\l{24c}\ee
If this condition is satisfied, then the fluctuations are
Gaussian with dispersion
$$\sim\{{\pa^2\ln\R_n(\b_c)}/{\pa\b^2_c}\}^{1/2},$$
see (\r{21c}).

Let us consider now (\r{24c}) carefully. We will find by computing
derivatives that this condition means the following approximate
equality:
\be
\f{\R^{(3)}_n}{\R_n}-3\f{\R^{(2)}_n\R^{(1)}_n}{\R^2_n}+2
\f{(\R^{(1)}_n)^3}{\R^3_n}\approx 0,
\l{25c}\ee
where $\R^{(k)}_n$ means the $k$-th derivative. For identical
particles,
\be
\R^{(k)}_n(\b_c)=n^k(-1)^k\int d\Ga_n(\b_c)\prod^k_{i=1}\e(q_i),
\l{26c}\ee
Therefore, the left hand side of (\r{25c}) is the 3-point correlator
$K_3$ since $d\Ga_n(\b_c)$ is a density of states for given $\b$:
\ba
&K_3\equiv\int d\O_3(q)\le(<\prod^3_{i=1}\e(q_i)>_{\b_c}-
3<\prod^2_{i=1}\e(q_i)>_{\b_c} <\e(q_3)>_{\b_c}+\ri.
\n\\
&+\le.2\prod^3_{i=1}<\e(q_i)>_{\b_c}\ri),
\l{27c}\ea
where the index $\b_c$ means that averaging is performed with the
Boltzmann factor $\exp\{-\b_c\e(q)\}$.

Notice, in distinction with Bogolyubov, $K_3$ is the $energy$
correlation function. So, in our interpretation, one can introduce
the notion of temperature $1/\b_c$ if and only if the $macroscopic$
energy flows, measured by the corresponding correlation functions, are
to die out.

As a result, to have all the fluctuations in the vicinity of $\b_c$
Gaussian, we should have $K_m\approx0$, $m\geq3$. Notice, as follows
from (\r{23c}), the set of minimal conditions actually looks as
follows:
\be
K_m<<K_2,~m\geq3.
\l{27d}\ee
If the experiment confirms this conditions then, independently from
the number of produced particles, the final state may be described
with high enough accuracy by one parameter $\b_c$ and the energy
spectrum of particles is Gaussian. In this conditions one may return
to the statistical \C{fer} and the hydrodynamical models \C{fein}.

Considering $\b_c$ as a physical (measurable) quantity, we are forced
to assume that both the total energy of the system, $\sqrt{s}=E$, and
the conjugate to it, variable $\b_c$, may be measured simultaneously
with high accuracy.

\subsubsection{Connection with Matsubara theory}

We would like to show now that the ordinary big partition
function of the statistical system coincide with
\be
\sum_{n,m}\int_{(\b_1,z_1;\b_2,z_2)}
R_{nm}(p,q)=\R(\b,z),
\l{densm}\ee
where $R_{nm}(p,q)$ is defined by (\r{22.3}). The summation and
integration are performed with constraints that the mean energy of
particles in the initial(final) state is $1/\b_1(1/\b_2)$. On may
interpret $1/\b$ in the first approximation as the temperature and
$z_1(z_2)$ as the activity for initial(final) state.

Direct calculation, see Appendix B, gives the following expression
for generating functional:
\be
\R(\b,z)=e^{-i{\bf N}(\phi_i^*\phi_j)}R_0(\p),
\l{22.22a}\ee
where the particle number operator ($\h{\p}(x)=\d/\d\p(x)$)
\ba
&{\bf N}(\p_i^*\p_j)=-
\int dxdx'(\h{\p}_+(x)D_{+-}(x-x',\b_2,z_2)\h{\p}_-(x')-
\n\\
&-\h{\p}_-(x)D_{-+}(x-x',\b_1,z_1)\h{\p}_+(x'))
\l{p22}\ea
and
\be
R_0(\p)=Z(\phi_+)Z^* (-\phi_-),
\l{p23}\ee
where $Z(\p)$ is defined in (\r{22.12}):
$$
Z(\phi)=\int D\Phi e^{iS(\Phi)-iV(\Phi+\phi)}
$$
and, for the vacuum boundary condition $\P (\s_{\infty})=0$,
\ba
D_{+-}(x-x',\b,z)=-i\int d\O_1(q)e^{iq(x-x')}e^{-\b\e(q)}z(q)
\l{22.23a}
\\
D_{-+}(x-x',\b,z)=i\int d\O_1(q)e^{-iq(x-x')}e^{-\b\e(q)}z(q)
\l{22.24a}\ea
are respectively the positive and negative frequency
correlation functions at $z=1$.

It is evident,
\be
R_{nm}(p,q)=\le.
\prod_{k=1}^n\le\{e^{\b_1\e(p_k)}\f{\d}{\d z_1(p_k)}\ri\}
\prod_{k=1}^m\le\{e^{\b_2\e(q_k)}\f{\d}{\d z_2(q_k)}\ri\}
\R(\b,z)\ri|_{z_i=0}.
\l{p2}\ee
Notice, defining $R_{nm}(p,q)$ through the generating
functional we extract the Boltzmann factors $e^{-\b\e}$ since the
energy-momentum conservation $\d$-functions were extracted from
amplitudes $a_{nm}(p,q)$.

We suppose that $Z(\p)$ may be computed perturbatively. As a result,
($\h{j}=\d/\d j$ is the variational derivative)
\be
R(\b,z)=
e^{-iV(-i\hat{j}_+)+iV(-i\hat{j}_-)}
e^{ \frac{i}{2} \int dx dx'j_i (x)D_{ik}(x-x';\b,z)j_k (x')},
\l{22.29a}\ee
where $D_{ik}(x-x')$ is the matrix Green function.
These Green functions are defined on the Mills \C{mil} time contours
$C_{\pm}$ in the complex time plane ($C_-=C_+^*$), see Fig.2. This
definition of the  time contours coincides with the Keldysh' time
contour \C{kel}.

The generating functional (\r{22.29a}) has the same structure
as the generating functional of Niemi and Semenoff \C{sem}. The
difference is only in the definition of Green functions $D_{ik}$.
This choice is a consequence of the boundary condition (\r{22.8}).
So, if (\r{22.37}) is used, then the Green function is defined by
eq.(\r{22.51}). Notice also that if $\b_1=\b_2=\b$ then a new Green
function obey KMS boundary condition, see (\r{22.54}).

Following Niemi and Semenoff \C{sem} one can write (\r{22.29a}) in
the form:
\be
\R(\b)=\int D_{NS}\P e^{iS_{NS}(\P)},
\l{nise}\ee
where the functional measure $D_{NS}\P$ and the action $S_{NS}(\P)$
are defined on the closed complex time contour $C_{NS}$, see Fig.3.
The choice of initial time $t_i$ and $t_f$ is arbitrary. Then one can
perform shifts: $t_i\to-\infty$ and $t_f\to+\infty$. In result, ({\bf
i}) if $\b_1=\b_2=\b$, ({\bf ii}) if contributions from imaginary
parts $C_{+-}$ and $C_{-\b}$ of the contour $C_{NS}$ have disappeared
in this limit, ({\bf iii}) if the integral (\r{nise}) may be
calculated perturbatively then this integral is a compact form of the
representation (\r{22.29a}).

Notice that the requirements ({\bf i}) - ({\bf iii}) are the
equivalent of the Euclidean postulate of Schwinger. In this frame one
can consider another limit $t_f\to t_i$. Then the $C_{NS}$ contour
reduces to the Matsubara imaginary time contour, Fig.4.

Later on we will use this $S$-matrix interpretation of
thermodynamics. But one should take in mind that corresponding
results will hide assumptions ({\bf i}) - ({\bf iii}).

We would like to mention the ambivalent role of external particles in
our $S$-matrix interpretation of thermodynamics. In the ordinary
Matsubara formalism the temperature is measured assuming that the
system under consideration is in equilibrium with the thermostat, i.e.
temperature is the energy characteristics of $interacting$ particles.
In our definition the temperature is the mean energy of produced,
i.e.  non-interacting, particles. It can be shown that both
definitions lead to the same result.

Explanation of this coincidence is the following. Let us consider the
point of particle production as the coordinate of fictitious
`particle'.  This `particle' interacts since the connected
contributions into the amplitudes $a_{nm}$ only are considered, and
has the equal to produced particles momentum and so on.  The set of
these `particles' form a system.  Interaction among these `particles'
may be described by corresponding correlation functions, see
Sec.2.3.3.

Let us consider now the limit $t_f\to t_i$. In this limit (\r{nise})
reduces to
\be
\R(\b)=\int D_{M}\P e^{-S_M(\P)},
\l{mats}\ee
where the imaginary time measure $D_{M}\P$ and action $S_M(\P)$ are
defined on the Matsubara time contour. The periodic boundary
condition (\r{22.1}) should be used calculating integral (\r{mats}).
The rules and corresponding problems in the integral (\r{mats}) can
be calculated as described in many textbooks, see also \C{land}.

In the limit considered the time was eliminated in the formalism and
the integral in (\r{mats}) performed over all states of the
`particles' system with the weight $e^{-S_M(\P)}$. Notice, the
doubling of degrees of freedom has disappeared and our fictitious
`particles' became real ones.

On the other hand, the produced particles may be considered as the
probes through which we measure the interacting fields. As was
mentioned above, their mean energy defines the temperature, if the
energy correlations are relaxed. If even one of the conditions ({\bf
i}) - ({\bf iii}) is not satisfied then one cannot reduce our
$S$-matrix formalism to the imaginary time Matsubara theory. Then one
can ask:  is there any possibility, staying in the frame of
$S$-matrix formalism, to conserve the statistics formalism. This
question is discussed in the Appendix C, where the Wigner functions
approach is applied. It may be shown that the formalism may be
generalized to describe the kinetic phase of the nonequilibrium
process, where the temperature should have the local meaning
\C{conf}. The comparison with the `local equilibrium hypothesis' is
discussed at the end of Appendix C.

\subsection{Classification of asymptotics over multiplicity}

Our further consideration will based on the model
independent (formal) classification of asymptotics \C{0.6}.

\subsubsection{`Thermodynamical' limit}

We will consider the generating function:
\be
T(s,z)=\sum_{n=1}^{n_{max}} z^n\s_n(s),~s=(p_1+p_2)^2>>m^2,~
n_{max}=\sqrt{s}/m.
\l{21.3}\ee
This step is natural since the number of particles is not conserved
in our problem. So, the total cross section and the averaged
multiplicity will be:
\be
\s_{tot}(s)=T(s,1)=\sum_n\s_n(s),~
\N=\sum_n n (\s_n(s)/\s_{tot})=
\le.\frac{d}{dz}\ln T(s,z)\ri|_{z=1}.
\l{21.4}\ee

At the same time, the inverse Mellin transform gives
\be
\s_n=\le.\f{1}{n!}\f{\pa^n}{\pa z^n}T(s,z)\ri|_{z=0}=
\f{1}{2\pi i}\oint\f{dz}{
z^{n+1}}T(s,z)= \f{1}{2\pi i}\oint\frac{dz}{z}e^{(-n\ln z+\ln
T(s,z))}.
\l{21.5}\ee
The essential values of $z$ in this integral are defined by the
equation (of state):
\be
n=z\f{\pa}{\pa z}\ln T(z,s).
\l{21.6}\ee
Taking into account the definition of the mean multiplicity $\N$,
given in (\r{21.4}), we can conclude that the solution of (\r{21.6})
$z_c$ is equal to one at $n=\N$.  Therefore, $z>1$ are essential in
the VHM domain.

The asymptotics over n ($n<<n_{max}$ is assumed) are governed by the
smallest solution $z_c$ of (\r{21.6}) because of the asymptotic
estimation of the integral (\r{21.5}):
\be
\s_n(s)\propto e^{-n\ln z_c(n,s)}.
\l{21.7}\ee

Let us assume that in the VHM region and at high energies,
$\sqrt s\to\infty$, there exist such a value of $z_c(n,s)$ that we can
neglect in (\r{21.3}) the dependence on the upper boundary $n_{max}$.
This formal trick with the thermodynamical limit allows to consider
$T(z,s)$ as the nontrivial function of $z$ for finite $s$.

Then, it follows from (\r{21.6}) that
\be
z_c(n,s)\to z_s~{\rm at}~n\in {\rm VHM},
\l{21.8}\ee
where $z_s$ is the leftmost singularity of $T(z,s)$ in the right half
plane of complex $z$.  One can say that the singularity of $T(z,s)$
attracts $z_c(n,s)$ if $n\in$VHM.  We will put this
observation in the basis of VHM processes phenomenology.

We would like to underline once more that actually $T(z,s)$ is
regular for arbitrary finite $z$ if $s$ is finite. But
$z_c(n,s)$ behaves in the VHM domain as if it is attracted by the
(imaginary) singularity $z_s$.   And just this $z_c(n,s)$ defines
$\s_n$ in the VHM domain. We want to note that actually the energy
$\sqrt s$ should be high enough to use such an estimation.

\subsubsection{Classes and their physical content}

One can notice from the estimation (\r{21.7}) that $\s_n$ weakly
depends on the character of the singularity. Therefore it is enough
to classify only the possible positions of $z_s$.  We may distinguish
following possibilities:
\ba
&({\bf A})~z_s=\infty:~\s_n<O(e^{-n})
\n\\
&({\bf B})~z_s=1:~\s_n>O(e^{-n})
\n\\
&({\bf C})~1<z_s<\infty:~\s_n=O(e^{-n}),
\l{21.9}\ea
i.e., following this classification, the cross section may
decrease faster ({\bf A}), slower ({\bf B}), or as ({\bf C})
an arbitrary power of $e^{-n}$. It is evident, if all these
possibilities may be realized in nature, then we should expect the
asymptotics ({\bf B}).

As was explained in Sec.2.2.1, $\s_n$ has the meaning of the $n$
particle partition function in the energy representation. Then
$T(z,s)$ should be the `big partition function'. Taking this
interpretation into account, as follows from Lee-Yang theorem
\C{lee}, $T(z,s)$ can not be singular at $|z|<1$.

At the same time, the direct calculations based on the physically
acceptable interaction potentials give the following restriction from
above:
\be
({\bf D})~\s_n<O({1/n})
\l{21.2}\ee
This means that $\s_n$ should decrease faster than any power of $1/n$.

It should be noted that our classification predicts rough
(asymptotic) behavior only and did not exclude local increase of
the cross section $\s_n$.

One may notice that
\be
-\f{1}{n}\ln\f{\s_n(s)}{\s_{tot}(s)}=\ln z_c(n,s)+O(1/n).
\l{21.10}\ee
Using thermodynamical terminology, the asymptotics of $\s_n$ is
governed by the physical value of the activity $z_c(n,s)$. One can
introduce also the chemical potential $\mu_c(n,s)$. It defines the
work needed for one particle creation, $\ln z_c(n,s)=\b_c(n,s)
\mu_c(n,s)$, where $\bar{\e}(n,s)=1/\b_c(n,s)$ is the produced
particles mean energy. So, one may introduce the chemical potential
if and only if $\b_c(n,s)$ and $z_c(n,s)$ may be used as the `rough'
variables.

Then the above formulated classification has a natural
explanation. So,  ({\bf A}) means that the system is stable with
reference to particle production and the activity $z_c(n,s)$ is the
increasing function of $n$, the asymptotics ({\bf B}) may realized if
and only if the system is unstable. In this case $z_c(n,s)$ is the
decreasing function. The asymptotics ({\bf C}) is not realized in
equilibrium thermodynamics \C{langer}.

We will show that the asymptotics ({\bf A}) reflects the
multiperipheral processes kinematics: created particles form jets
moving in the CM frame with different velocities along the incoming
particles directions, i.e.  with restricted transverse momentum, see
Sec.3.1.1.  The asymptotics ({\bf B}) assumes the
condensation-like phenomena, see Sec.3.3.  The third type asymptotics
({\bf C}) is predicted by stationary Markovian processes with the
pQCD jets kinematics, see Sec.3.2.2. The DIS kinematics may be
considered as the intermediate, see Sec.3.2.1.

This interpretation of classes (\r{21.9}) allows to conclude that we
should expect reorganization of production dynamics in the VHM
region: the soft channel ({\bf A}) of particle production should
yield a place to the hard dynamics ({\bf C}), if the ground state of
the investigated system is stable with reference to the particle
production.  Otherwise we will have asymptotics ({\bf B}).

\subsubsection{Group decomposition}

Let us consider the system with several correlation scales. For
example, in statistics one should distinguish correlation length
among particles (molecules) and correlation length among droplets if
the two-phase region is considered. In particle physics, one should
distinguish in this sense correlation among particles produced in
result of resonance decay and correlations among resonances. In pQCD
one may distinguish correlations of particles in jet and correlation
among jets.

There exist many model description of this physical picture. In
statistics Mayer's group decomposition \C{mayer} is well known. In
particle physics one should note also the many-component formalism
\C{slep}\foot{The example considered in \C{kuvsh} illustrate this
approach.}.  We will consider the generating functions (functionals)
formalism \C{bog} considering mostly jet correlations.  In many
respects it overlaps the above mentioned approaches.

The generating function $T(z,s)$ may be written in the form:
\be
\ln T(z,s)=\sum_{k=1}^\infty\f{(z-1)^k}{k!}C_k(s)=
\sum_{l=1}^\infty z^lb_l,
\l{g1}\ee
where the coefficients $C_k$ are the moments of the multiplicity
distribution
\be
P_n(s)=\s_n(s)/\s_{tot}(s).
\l{g2}\ee
So,
\be
C_1(s)=\sum_n nP_n(s)=\N,~C_2(s)=\sum_n n(n-1)P_n(s)-\N^2
\l{g3}\ee
and so on. Using the connection with the inclusive distribution
functions $f_k(q_1,q_2,...,q_k)$:
\be
T(z,s)=\sum_{k=1}^\infty\f{(z-1)^k}{k!}\int d\O_k(q)
f_k(q_1,q_2,...,q_k;s),
\l{g4}\ee
it is easy to find that
\ba
&C_1(s)=\int d\O_1(q) f_1(q;s)=\bar{f}_1(s),
\n\\
&C_2(s)=\int d\O_2(q) \{f_2(q_1,q_2;s)-f_1(q_1;s)f_1(q_2;s)\}=
\bar{f}_2(s)-\bar{f}_1^2(s),
\l{g5}\ea
etc. Generally,
\be
\f{1}{k!}C_k(s)=\sum_{l=1}^\infty\f{(-1)^l}{l}
\sum_{\{k\}_l=0}^\infty\d\le(\sum_{i=1}^lk_i-k\ri)
\prod_{i=1}^l\le\{\f{\bar{f}_{k_i}(s)}{k_i!}\ri\},
\l{g6}\ee
where $\{k\}_l=k_1,k_2,...,k_l$ and
$$
\bar{f}_{k}(s)=\int d\O_k(q)f_k(q_1,q_2,...,q_k;s).
$$
One may invert formulae (\r{g6}):
\be
\f{1}{l!}\bar{f}_{k}(s)=
\sum_{\{n_k\}=0}^\infty\d\le(\sum_{k=1}^\infty kn_k-l\ri)
\prod_{k=1}^\infty \f{1}{n_k!}\le(\f{C_k(s)}{k!}\ri)^{n_k}
\l{g7}\ee

The Mayer's group coefficients $b_l$ in (\r{g1}) have the following
connection with $C_k$:
\be
b_l(s)=\sum_{k=0}^\infty\f{(-1)^k}{l!k!}C_{k+l}(s).
\l{g23}\ee

It seems useful to illustrate the effectiveness of the generating
function method by the following example. We will consider the
transformation (multiplicity $n\to$ activity $z$) to show the origin
of the Koba-Nielsen-Olesen scaling (KNO-scaling)\foot{In private
discussion with one of the authors (A.S.) in the summer of 1973,
Z.Koba noted that the main reason of investigation leading to the
KNO-scaling was just the generating functional method of Bogolyubov
\C{conf}}.

If $C_m=0$, $m>1$, then $\s_n$ is described by the Poisson
formulae:
\be
\s_n(s)=\s_{tot}(s)e^{-\bar n}\f{\N^n}{n!}.
\l{k8}\ee
It corresponds to the case of absence of correlations.

Let us consider more weak assumption:
\be
C_m(s)=\ga_m\le(C_1(s)\ri)^m,
\l{k9}\ee
where $\ga_m$ is the energy independent constant, see also \C{matv},
where a generalization of KNO scaling on the semi-inclusive processes
was offered.  Then
\be
\ln T(z,s)=\sum_{m=1}\f{\ga_m}{m!}\{(z-1)\bar n(s)\}^m.
\l{k10}\ee
To find the consequences of this assumption, let us find the most
probable values of $z$. The equation of state
$$
n=z\f{\pa}{\pa z}\ln T(z,s)
$$
has solution $\bar z(n,s)$ increasing with $n$ since $T(z,s)$ is
an increasing function of $z$, if and only if, $T(z,s)$ is
nonsingular at finite $z$. As was mentioned above, the last
condition has deep physical meaning and practically assumes the
absence of the first order phase transition \C{lee}.

Let us introduce a new variable:
\be
\la=(z-1)\bar n(s).
\l{k12}\ee
The corresponding equation of state looks as follows:
\be
\f{n}{\bar n(s)}=\le(1+\f{\la}{\bar
n(s)}\ri)\f{\pa}{\pa\la}\ln T'(\la).
\l{k13}\ee
So, with $O(\la/\bar n(s))$ accuracy, one can assume that
\be
\la\simeq\la_c(n/\bar n(s)).
\l{k14}\ee
are essential. It follows from this estimation that such scaling
dependence is rightful at least in the neighborhood of $z=1$, i.e. in
vicinity of main contributions into $\s_{tot}$. This gives:
\be
\bar n(s)\s_n(s)=\s_{tot}(s)\psi(n/\bar n(s)),
\l{k15}\ee
where
\be
\psi(n/\bar n(s))\simeq T(\la_c(n/\bar n(s)))\exp
\{n/\bar n(s)\la_c(n/\bar n(s))\}\leq O(e^{-n})
\l{k16}\ee
is the unknown function. The asymptotic estimation follows from the
fact that $\la_c=\la_c(n/\bar n(s))$ should be a nondecreasing
function of $n$, as follows from nonsingularity of $T(z,s)$.

The estimation (\r{k14}) is right at least at $s\rar\infty$. The
range validity of $n$, where solution of (\r{k14}) is acceptable,
depends from exact form of $T(z,s)$.  Indeed, if $\ln T(z)\sim\exp
\{\ga\la(z)\}$, $\ga=const>0$, then (\r{k14}) is right at all
values of $n$ and it is enough to have the condition $s\rar\infty$.
But if $\ln T(z,s)\sim(1+ a\la(z))^\ga$, $\ga=const>0$, then (\r{k14})
is acceptable if and only if $n<<\bar n^2(s)$.

Representation (\r{k15}) shows that just $\bar{n}(s)$ is the natural
scale of multiplicity $n$ \C{kno}. This representation was offered
first as a reaction on the so called Feynman scaling for
inclusive cross section:
\be
f_k(q_1,q_2,...,q_k)\sim\prod^{k}_{i=1}\f{1}{\e(q_i)}.
\l{k17}\ee

As follows from estimation (\r{k16}), the limiting KNO prediction
assumes that $\s_n=O(e^{-n})$. In this regime $T(z,s)$ should be
singular at $z=z_c(s)>1$. The normalization condition
$$\f{\pa T(z,s)}{\pa z}|_{z=1}=\bar{n}(s)$$ gives:
$z_c(s)=1+\ga/\bar{n}(s)$, where $\ga>0$ is the constant.
Notice, such behavior of the big partition function $T(z,s)$ is
natural for stationary Markovian processes described by logistic
equations \C{vol}.  In the field theory such equation describes the
QCD jets \C{jet}.

We wish to generalize expansion (\r{g1}) to take into account the
possibility of many-component structure of the multiple production
processes \C{slep}.  Let us consider particle production through
the generation, for instance, of jets. In this case decay of a
particle of high virtuality $|q|>>m$ forms a jet of lower virtuality
particles.  It is evident that one should distinguish correlation
among particles in the jet, and correlation among jets.

Let $\o_{n_i}(m_i)$ be the $probability$ that $i$-th jet of mass
$m_i$ includes $n_i$ particles, $1\leq n_i\leq n$, where
\be
\sum_{i=1}^{N_j}n_i=n
\l{g8}\ee
The jets are the result of particles decay. Then let us assume that
$\bar{N}_1(m_i,p_i)$ defines the mean number of jets of mass $m_i$
and momentum $p_i$:
\ba
&\R^{(1)}_n(\b)=\sum_{N_j}\f{1}{N_j}\sum_{\{n\}_{N_j}}
\d\le(\sum_{i=1}^{N_j}n_i-n\ri)\t
\n\\
&\t\int\prod_{i=1}^{N_j}
\le\{\f{dm_i}{2m_i}\f{d^3p_i}{(2\pi)^3}
e^{-\b\e(p_i)}\bar{N}_1(m_i,p_i)\o_{n_i}(m_i)\ri\},
\l{g9}\ea
where $\{n\}_{N_j}=(n_1,n_2,...,n_{N_j})$. Notice the Boltzmann
factor $e^{-\b\e}$, where $\e(p)=m+p^2/2m$ is the jets energy, play
the same role as the corresponding factor in (\r{19}) and introduced
to take into account the energy conservation law.  We consider the
VHM domain and for this reason $(p^2/2m)<<1$ is assumed.

It is useful to avoid the particles number conservation law (\r{g8}).
For this purpose we will introduce
\be
\R^{(1)}(\b,z)=\sum_n\R^{(1)}_n(\b)=\exp\le\{\int\f{dm}{2m}
\f{d^3p}{(2\pi)^3}
e^{-\b\e(p)}\bar{N}_1(m,p)(t(z,m)-1)\ri\},
\l{g10}\ee
where
\be
t(z,m)=\sum_n z^n\o_n(m).
\l{g11}\ee
Comparing (\r{g10}) with (\r{g1}) we may conclude that $t(z,m)$ plays
the role of activity of jets. Then the generalization is evident:
\ba
&\R(\b,z)=\sum_k\R^{(k)}(\b,z)=
\n\\
&=\exp\le\{\sum_k\int\prod_{i=1}^k\le\{dm_id\O_1(p_i)
e^{-\b\e(p_i)}(t(z,m_i)-1)\ri\}\ri.
\t\n\\
&\le.\t\bar{N}_k(m_1,p_1,...,m_k,p_k)\ri\},
\l{g12}\ea
where $\bar{N}_k$ has the same meaning as $C_k$, i.e. $\bar{N}_k$ is
the correlation function of $k$ jets.

\subsubsection{Energy-multiplicity asymptotics equivalence}

Let us consider the following `bootstrap' regime when $\R(\b,z)$ is
defined by the equation:
\be
\R(\b,z)\propto\int\f{dm}{2m}\f{d^3p}{(2\pi)^3}
e^{-\b\e(p)}\bar{N}_1(m,p)t(z,m).
\l{g13}\ee
Inserting here the strict expression (\r{g12}) we find a nonlinear
equation for $t(z,m)$.

The solution of (\r{g13}) assumes that
\be
\int\f{dm}{2m}\f{d^3p}{(2\pi)^3}e^{-\b\e(p)}\bar{N}_1t>>
\le\{\int\f{dm}{2m}\f{d^3p}{(2\pi)^3}e^{-\b\e(p)}
\bar{N}_1t\ri\}^2
\l{g14}\ee
and
\be
\int\f{dm}{2m}\f{d^3p}{(2\pi)^3}e^{-\b\e}\bar{N}_1t>>
\int\le(\f{dm_1}{2m_1}\f{d^3p_1}{(2\pi)^3}e^{-\b\e}t\ri)
\le(\f{dm_2}{2m_2}\f{d^3p_2}{(2\pi)^3}e^{-\b\e}t\ri)
\bar{N}_2
\l{g15}\ee

To solve eq.(\r{g13}) in the VHM region, where the leftmost
singularity over z is important, let us consider the $anzats$:  \be
t(z,m)=\f{\vp(z,m)}{(1-(z-1)a(m))^{\k_0}},~\k_0>0,
\l{g16}\ee
where $\vp(z,m)$ is the polynomial function of $z$, $\vp(z=1,m)=1$.
Using the normalization condition:
\be
\bar{n}_j=\le.\f{\pa}{\pa z}t(z,m)\ri|_{z=1}
\l{g17}\ee
we can find:
\be
a(m)\k_0=\bar{n}_j-\vp'(1,m),~\vp'(1,m)\equiv\le.\f{\pa}{\pa
z}\vp(z,m)\ri|_{z=1}.
\l{g18}\ee

The partition function of the jet $t(z,m)$ defined by $anzats$
(\r{g16}) is singular at
\be
z_s(m)=1+\f{1}{a(m)}.
\l{g19}\ee
This singularity would be significant in the VHM region if $z_s(m)$ is
decreasing function of $m$. This means an assumption that
$$
\f{\vp'(1,m)}{\bar{n}_j(m)}\to0~{\rm at}~m\to\infty.
$$
So, in first approximation we will choose
\be
a(m)=\bar{n}_j(m).
\l{g20}\ee
This choice may be confirmed by concrete model calculations.

Taking into account the energy conservation law, conditions (\r{g14})
and (\r{g15}) are satisfied if
\be
\exp\le\{-n\f{\bar{n}_j(s)-\bar{n}_j(s/4)}{\bar{n}_j(s)\bar{n}_j(s/4)}
\ri\}<<1.
\l{g21}\ee
at $n\in$VHM. Therefore, (\r{g16}) obey eq.(\r{g13}) with
exponential accuracy in the VHM region, i.e. if
\be
n>>\f{\bar{n}_j(s)\bar{n}_j(s/4)}{\bar{n}_j(s)-\bar{n}_j(s/4)}
=\f{\k_0}{z_s(s/4)-z_s(s)}.
\l{g22}\ee
We assume here that one can find so large $n$ and $s$ that with
exponential accuracy the factors $\sim\bar{N}_k$ did not play an
important role. But at low energies the condition (\r{0.2}) is
important and the factors $\sim\bar{N}_k$ should be taken into
account.

Notice now important consequence of our `bootstrap' solution: it
means that we can leave production of the heavy jets only, if $n\in
VHM$.  On the other hand, let us choose $n=z_0\bar{n}_j(s)$, where
$z_0>1$ is the function of $s$, and consider $s\to\infty$. Then the
condition (\r{g22}) defines $z_0$: if
\be
z_0>>\f{\bar{n}_j(s/4)}{\bar{n}_j(s)-\bar{n}_j(s/4)}
\l{g22'}\ee
then we are able to obey the inequalities (\r{g14}) and (\r{g15}).

The jet mean multiplicity, see Sec.3.2.2,
\be
\ln\bar{n}_j(s)\sim\sqrt{\ln s}.
\l{g26}\ee
Then
\be
\f{\bar{n}_j(s/4)}{\bar{n}_j(s)-\bar{n}_j(s/4)}
=\le\{e^{\ga/\sqrt{\ln s}}-1\ri\}^{-1}\sim\sqrt{\ln s}<<\bar{n}_j(s).
\l{g50}\ee
at $s\to\infty$. Therefore, (\r{g22'}) may be satisfied outside the
VHM domain.

Let us compare now the solutions of the equation of state. Inserting
(\r{g16}) into (\r{21.6}) we can find for a jet of mass $\sqrt s$ that
\be
z^1_c=z^1_s-\f{\k_0}{n}=1+\f{1}{\bar{n}_j(s)}-\f{\k_0}{n}.
\l{g24}\ee
The two-jet contribution of the masses $\sim {\sqrt s}/2$
gives:
\be
z^2_c=z^2_s-\f{\k_0}{n}=1+\f{1}{\bar{n}_j(s/4)}-\f{\k_0}{n}.
\l{g25}\ee
At arbitrary finite energies $(z^2_c-z^1_c)>0$ and, as follows from
(\r{g50}), they decrease $\sim(1/{\bar{n}_j(s)}\sqrt{\ln s})$ with
energy.

Noting the normalization condition, $T(z=1,s)=\s_{tot}(s)$, and
assuming that the vacuum is stable, i.e. $\s_n\leq O(e^{-n})$, we can
conclude that

-- if $n\in VHM$ then $z_s$ attracts $z_c$, i.e. $z_s\to z_c$, and if
$z_c-1<<1$ then this contributions should be significant in
$\s_{tot}$;

-- if $s\to\infty$, then $z_c-1<<1$, and if $n$ satisfy the inequality
(\r{g22}), or if $z_0$ satisfy the inequality (\r{g22'}), then
considered contributions are significant in $\s_{tot}$.\\
It is the (energy-multiplicity) asymptotics equivalence principle. One
of the simplest consequences of this principle is the prediction that
the mean transverse momentum of created particles should increase
with multiplicity at sufficiently high energies.

This principle is the consequence of independence of contributions in
the VHM domain on the type of singularity in the complex $z$ plane
and of the energy conservation law. Just the last one shifts the
two-jet singularity to the right and $z_c(s)<z_c(s/4)$.

We would like to notice also that this effect, when the mostly
`energetic population' survives was described mathematically by
V.Volterra \C{vol}. Intuitively evident is that one may find the
`energetic population' searching the VHM one, or, it is the same,
giving it a rich supply, i.e. to give the population enough energy.
This is our (energy-multiplicity) equivalence (($\e-n$)-equivalence)
principle.

Notice, if the amount of supply is too high then few populations may
grow.  This is the case when the difference $(z^2_c-z^1_c)>0$  tends
to zero at high energies.

We would like to note that singular at finite $z$ partition functions
was predicted in the $(\la\p^3)_6$-theory \C{tay}, in QCD jets
\C{jet}, in the generalized Bose-Einstein distribution model
\C{shih}. In all of this models decay of the essentially
nonequilibrium initial state (highly virtual parton, heavy resonance,
etc) was described.

One may distinguish the phases of the media by a characteristic
correlation length. Then the phase transition may be considered as
the process of changing correlation length. Our `bootstrap' solution
predicts just such phenomena: at low multiplicities the
long-range correlations among light jets is dominant. The `bootstrap'
solution predicts that for VHM processes just the short-range
correlations among particles of the heavy jet become dominant. The
($\e-n$)-equivalence means that this transition is a pure dynamical
effect.

\section{Model predictions}\0

Multiple production phenomena was first observed more than seventy
years ago \C{1.1}. During this time a vast experimental information
was accumulated concerning hadron inelastic interactions, see the
review papers \C{1.5}.

Now we know that at high energy $\sqrt{s}$:\\ (i) The total cross
section $\s_{tot}(s)$ of hadron interactions is enhanced almost
completely by the inelastic channels;\\ (ii) The mean
multiplicity of produced hadrons $\N$ slowly (logarithmically) grows
with $\sqrt{s}$;\\ (iii) The interaction radii of
hadrons $\bar b$ slowly (logarithmically) increase with energy.;
\\ (iv) The multiplicity distribution $\s_n(s)$ is wider than the
Poisson distribution.;\\ (v) The mean value of transverse
momentum $k$ of produced hadrons is restricted and is
independent of the incident energy $\sqrt s$ and produced particle
multiplicity $n$; \\(vi) The one-particle energy
spectrum $d\s\sim d\ve/\ve$.

First of all, (i) means that the high energy hadron interaction may
be considered as the ordinary dissipation process. In this process
the kinetic energy of incident particles is spent in produced
particle mass formation.

The VHM process takes place in the vacuum and then it was assumed
on the early stages that the multiple production phenomena reflects a
natural tendency of the excited hadron system to get to equilibrium
with the environment \C{fer}.  In this way one can introduce as a
first approximation the model that the excited hadron system evolves
without any restrictions.  In this model we should have
$\N\sim\sqrt{s}$. The dissipation is maximal in this case and the
entropy $\cS$ exceeds its maximum.  This simple model has definite
popularity up to 70-th.  But the experimental data (ii) and (iii)
prohibit this model and it was forgotten.

Choosing the model we would like to hope that the considered model

--takes into account experimental conditions (i) - (vi) in the
$n\sim\N$ domain;

-- has natural asymptotics over multiplicity to the VHM
region.\\
It is necessary to remember also that

-- New channels of hadron production may arise in the VHM region.\\
It is impossible to understand all possibilities without those
offered in Sec.2.3.2 in the classification of asymptotics.

Thus, we will observe predictions of

-- {\it Multiperipheral models},\\
distinguishing the soft Pomeron models, see
\C{landshoff}.

-- {\it The dual-resonance model}\\
predictions for the VHM region are described also. It can be shown
that this models predict asymptotics (A) if $n>>\N^2$. Just this
result explains why VHM domain is defined by the condition
(\r{0.1'}).

We are forced since it allows to include pQCD, forbidden by the
multiperipheral models, considering

-- {\it Hard Pomeron model}\\
production of mini-jets. But we will find using Monte Carlo
simulations that the pQCD Pomeron is unable to adopt the hard
channels of hadron production. Then

-- {\it The deep inelastic processes}\\
for VHM region will be considered to generalize the DGLAP
kinematics in the case of heavy QCD jets production. The analysis
shows that transition to the VHM leads to the necessity to include
low-$x$ sub-processes. As a result we get out of the range of pQCD
validity.

-- {\it Multiple reduction of jets.}\\
We will see, that at very high energies in the VHM region the heavy
jets creation should be a dominant process if the vacuum is stable
with reference to the particle production.

We will consider also decay of the `false vacuum' to describe the
consequence of

-- {\it Phase transition}\\
in the VHM domain. This channel is hardly seen for $n\sim\N$ since
the confinement constraints may prevent cooling of the system up to
phase transitions condition.

\subsection{Peripheral interaction}

\subsubsection{Multiperipheral phenomenology}

Later on multiple production physics was developed on the
basis of experimental observation (iv). The Regge pole
model naturally explains this experimental data and, at the
same time, absorbs all experimental information, (i)-(vi) .  At the
very beginning, adopting the Regge poles notion without its
microscopical explanation, this description was self-consistent. The
efforts to extend the Regge pole model to the relativistic hadron
reactions was ended by the Reggeon diagram technique, see \C{levin}
and references cited therein, and it was used later to construct the
perturbation theory for $\s_n$ \C{1.12}. It was shown that the
multiplicity distribution is wider then the Poissonian because of the
multi-Pomeron exchanges.

The leading energy asymptotics Pomeron contribution reflects the
created particle kinematics described in Appendix D, where the
available kinematical scenario in the frame of pQCD are described.
So, the longitudinal momentum of produced particles is large and is
strictly ordered.  At the same time, particles transverse momentum is
restricted.

Let us consider the inelasticity coefficient introduced in (\r{0.2a})
$\k=1- \ve_{max}/E<1$, where $\ve_{max}$ is the energy of the
fastest particle in the laboratory frame. Then the strict ordering of
particles in the Pomeron kinematics means that $\k$ is independent of
the index of the particle.  So, if the fastest particle has the
energy $\ve_{max}\simeq(1-\k)\sq$, then the following particle should
have the energy $\ve_1\simeq(1-\k) \ve_{max}\simeq(1-\k)^2\sq$, and
so on.  Following to this law, $(n-1)$-th particle would have the
energy $\ve_n\simeq(1-\k)^n\sq$.  In the laboratory frame the energy
should degrade to $\ve_n\simeq m$. Inserting here the above
formulated estimation of $\ve_n$ we can find that if the number of
produced particles is
\be
\N\simeq n_0\xi,~n_0=-\ln(1-\k)^2>0,~\xi=\ln(s/m^2),
\l{1.1a}\ee
then we may
expect the total degradation of energy. This degradation is the
necessary condition noting that the total cross section of slowly
moving particles may depend only slightly on energy and is seems
necessary for natural explanation of the weak dependence of the hadron
cross sections on the energy. This consideration would be
Lorentz-covariant if one can find the slowly moving particle in an
arbitrary frame.  Resulting estimation of mean multiplicity have good
qualitative experimental confirmation.

Notice that it was assumed deriving (\r{1.1a}) that the energy
degrades step by step. In other words, if we introduce a time
of degradation, then the time $\sim\xi$ is needed for complete
degradation of energy. Assuming the random walk in the normal to
incident particle plane, we can conclude that the points of particle
production are located on a disk (in the moving frame) of radii
$\bar{b}\sim\xi^{1/2}$. This means that the interaction
radii should grow with the energy of the colliding particles.

If $f(a+b\to c +...)$ is the cross section to observe particle $c$
inclusively in the $a$ and $b$ particles collision, then it was found
experimentally that the ratio
\be
\frac{f(\pi^+p\to\pi^- +...)}{\s_{tot}(\pi^+p)} =
\frac{f(K^+p\to\pi^- +...)}{\s_{tot}(K^+p)}=
\frac{f(pp\to\pi^- +...)}{\s_{tot}(pp)}
\l{1.2}\ee
is universal. This may be interpreted as the direct evidence
of fact that the hadron interactions have a large-distance character,
i.e. that the interaction radii should be large.

This picture assumes that the probability to have total
degradation of energy is
\be
\sim e^{-b^2/4\a'\xi},
\l{1.3}\ee
where $\a'$ is some dimensional constant (the slope of Regge
trajectory) and $\bb$ is the 2-dimensional impact parameter. This
formulae has also the explanation connected to the vacuum instability
with reference to the real particle production in the strong colour
electric field.

The above picture has natural restrictions. We can assume
that each of the produced particles may be the source of above
described $t$-channel cascade of the energy degradation. This means
that in the frame of the Pomeron phenomenology, we are able to
describe the production of
\be
n< \N^2
\l{1.4}\ee
particles only. If $n>\N^2$ then the density of particles in the
diffraction disk becomes large and (a) one should introduce
short-distance interactions, or (b) rise interaction radii.  It will
be shown that just (a) is preferable.

We will build the perturbation theory in the phenomenological frames
(i) - (vi). Considering the system with variable
number of particles the generating function
\be
T(z,s)=\sum_n z^n \s_n(s)
\l{i1}\ee
would be useful. One can use also the decomposition:
\be
T(z,s)=\s_{tot}(s)\exp\le\{(z-1)C_1(s)+\f{1}{2}(z-1)^2C_2(s)+...
\ri\},
\l{i2}\ee
where, by definition,
\be
C_1(s)=\N=\le.\f{\pa}{\pa z}\ln T(z,s)\ri|_{z=1}
\l{i3}\ee
is the mean multiplicity,
\be
C_2(s)=\le.\f{\pa^2}{\pa z^2}\ln T(z,s)\ri|_{z=1}
\l{i4}\ee
is the second binomial momentum, and so on.

Our idea is to assume that all $C_m$, $m>1$ may be calculated
perturbatively choosing
\be
P(0,s)=e^{(z-1)\N}
\l{i5}\ee
as the Born approximation `superpropagator'. It is evident that
(\r{i5}) leads to Poisson distribution. Then, having in mind (ii),
(iii) and (v) we will use following $anzats$:
\be
P(q,s)=e^{\a(o)-\a'q^2\ln s}e^{(z-1)\xi},
\l{i6}\ee
where the transverse momentum ${\bq}$ is conjugate to the
impact parameter $\bb$. So, the Born term (\r{i6}) is a Fourier
transform of the simple product of (\r{i5}) and (\r{1.3}). It contains
only one free parameter, the Pomeron intercept $\a(0)$. On the
phenomenological level it is not important to know the
dynamical (microscopical) origin of (\r{i6}).

For our purpose the Laplace transform of $P$ would be useful. If
$\N=n_0\ln s$, then
\be
\cp(\o,q^2)=\int_0^\infty d\xi e^{-\o\xi}P(q^2,s)=
\f{1}{\o+\a'q^2+\psi_0(z)}.
\l{i8}\ee
It is the propagator of two-dimensional field theory with mass
squared
$$
\psi_0(z)=(1-\a)+(1-z)n_0,~n_0>0.
$$
Knowing the Gribov's Reggeon calculus completed by the
Abramovski-Gribov-Kancheli (AGK) cutting rules \C{abarb} one can
investigate the consequences of this approach.

The LLA approximation of the pQCD \C{lipatov} gives
\be
\D=\a(0)-1=\f{12\ln2}{\pi}\a_s\approx 0.55,~\a_s=0.2.
\l{i10}\ee
But radiative corrections give $\D\approx0.2$ \C{fadin}. We will call
this solution as the BFKL model.

The quantitative origin of the restriction (\r{1.4}) is following.
The contribution of the diagram with $\nu$ Pomeron exchange gives,
since the diffraction radii increase with $s$, see (\r{i3}),
mean value of the impact parameter decreasing with $\nu$:
$$
\bar{\bb^2}\simeq 4\a'\ln (s/m^2)/\nu=a\a'\f{\N}{\nu},
$$
where $a={4}/{n_0}$. On the other hand, the number of necessary
Pomeron exchanges $\nu\sim n/\N$ since one Pomeron gives maximal
contribution (with factorial accuracy) at $n\simeq\N$. In result,
\be
\bar{\bb^2}\sim a\a'\f{\N^2}{n}.
\l{i11}\ee
Therefore, if the transverse momentum of created particles is a
restricted quantity, i.e. $\mu_0^2\bar{\bb^2}\sim 1$, where
$\mu_0$ is a constant, then the mechanism of particle
production is valid up to
\be
n\sim\N^2.
\l{o8}\ee

Following our general idea, it will be enough for us find the
position of singularity over $z$. Analysis shows that (\r{i10})
predict the singularity at infinity.

In Appendix E the Gribov's Reggeon diagram technique with cut
Pomerons is described and (\r{o8}) is derived.  It can be shown
using this technique that the model with the critical Pomeron,
$\D=0$, is inconsistent from the physical point of view \C{1.12}.

As was mentioned above, the model with $\D=\a(0)-1>0$ is natural for
the pQCD.  The concrete value of $\D$ will not be important for us.
We will assume only that
\be
0<\D<<1.
\l{l2}\ee
It is evident that the Born approximation (\r{i6}) with $\D>0$
violate the Froissart boundary condition. But it can shown that the
sum of `eikonal' diagrams\foot{The eikonal approximation in a
quantum field theory was developed in \C{eikon}.} solves this
problem, see \C{karen} and references cited therein.

The interaction radii may increase with increasing number of
produced particles if $\D>0$ and then the restriction (\r{o8}) is not
important.  In the used eikonal approximation, see Appendix F, \be
z\simeq z_c=1+\f{1}{\N}\ln\f{n}{\N}
\l{l11}\ee
are essential. Then the interaction radii $\bar{\bb^2}\sim B^2\simeq
4\a'\xi(\D\xi+\ln({n}/{\N}))$ for this values of $z$. Notice that
\be
B^2\sim\xi^2~ >>\xi
\l{l12}\ee
even for $n\sim n_{max}\sim\sqrt s$.

Nevertheless, using (\r{l11}) one can find that the cross section
decrease faster than any power of $e^{-n}$:
\be
-\ln\le(\f{\s_n(s)}{\s_{tot}(s)}\ri)=\f{n}{\N}\ln\f{n}{\N}(1+O(\N/n))
\l{l13}\ee
Generally speaking, although this estimation is right in the VHM
region, there may be large corrections because of Pomeron
self-interactions. But careful analyses shows \C{1.12} that these
contributions can not change drastically the estimation (\r{l13}).

\subsubsection{Dual resonance model}

The search of dynamical source of the Regge description shows
the different dynamical nature of the Regge and Pomeron poles. The
established resonance-Regge pole duality, e.g. \C{logun} led to the
Veneziano representation of the Regge amplitudes \C{venez}. The
Reggeon pole gives the decreasing $\sim s^{-1/2}$ contribution, but
careful investigation shows that the mass spectrum of dual to Regge
pole resonances increase exponentially. This prediction was confirmed
by experiment, see the discussion of this question in \C{haged}.

The field theory development is marked by considerable efforts to
avoid the problem of colour charge confinement. Notice that the
classical string has the same excitation spectrum. The remarkable
attempt in this direction based on the string model, in its various
realizations, see e.g. \C{bran}.  But, in spite of remarkable success
(in formalism especially) there is not an experimentally measurable
prediction of this approach till now, e.g.  \C{str.-exp.}.

We would like to describe in this section production of `stable'
hadrons through decay of resonances \C{tor, dual-}.

Our consideration will use the following assumptions.

{\bf A}. The string interpretation of the dual-resonance model bring
to the observation that the mass spectrum of resonances, i.e.  the
total number $\R(m)$ of mass $m$ resonance excitations, grows
exponentially:
\be
\R(m)=(m/m_0)^{\ga} e^{\b_0m},~\b_0={\rm const},~m>m_0.
\l{rms}\ee
Note also that the same hadron mass spectrum (\r{rms}) was predicted
in the `bootstrap' approach \C{haged, hag}.  Moreover, it predicts
that
\be
\ga=-5/2.
\l{crex}\ee

{\bf B}. The mass $m$ resonance creation cross section $\s^R(m)$ has
the Regge pole asymptotics:
\be
\s^R(m)=g^R\f{m_0}{m},~g^R=const.
\l{rcs}\ee
It was assumed here that the intercept of the Regge pole trajectory
$\a^R=1/2$. So, only the meson resonances would be taken into
account.

{\bf C}. If $\s^R_n(m)$ describes the decay of a mass $m$ resonance
into the $n$ hadrons, then the mean multiplicity of hadrons
\be
\bar{n}^R(m)=\f{\sum_n n\s^R_n(m)}{\s^R(m)}.
\l{rmm}\ee
Following the Regge model,
\be
\bar{n}^R(m)=\bar{n}^R_0\ln\f{m^2}{m_0^2}.
\l{rmm2}\ee

{\bf D}. We will assume that there is a definite vicinity of
$\bar{n}^R(m)$ where $\s^R_n(m)$ is defined by $\bar{n}^R(m)$
only. So, in this vicinity
\be
\s^R_n(m)=\s^R(m)e^{-\bar{n}^R(m)}({{\bar{n}^R(m)})^n}/{n!}.
\l{pois}\ee
This is the direct consequence of the Regge pole model, if $m/m_0$ is
high enough.

Following our idea, we will distinguish the `short-range'
correlations among hadrons and the `long-range' correlations among
resonances. The `connected groups' would be described by resonances
and the interactions among them should be described introducing for
this purpose the correlation functions among strings.  So, we will
consider the `two-level' model of hadron creation: the first level
describes the short-range correlation among hadrons and the second
level is connected to the correlations among resonances.

The exact calculations are given in Appendix G.

Comparing A and B solutions we can see the change of attraction
points with rising $n$: at $n\simeq\bar{n}^2(s)= \bar{n}_0^R
\ln(\sqrt{s}/m_0)$ the transition from ({\bf A}) asymptotics to
({\bf C}) in (\r{21.9}) should be seen. At the same time one should
see the strong KNO scaling violation at the tail of the multiplicity
distribution.

We have neglected the resonance interactions deriving these results.
This assumption seems natural since at $\bar{n}(s)<<n<\bar{n}^2(s)$
the inequality (\r{res1}) should be satisfied, see discussion of
inequalities (\r{g14}) and (\r{g15}) in Sec.2.3.4.

\subsection{Hard processes}

\subsubsection{Deep inelastic processes}

The role of soft colour partons in the high energy hadron interactions
is the most intriguing modern problem of particle physics. So, the
collective phenomena and symmetry breaking in the non-Abelian gauge
theories, confinement of coloured charges and the infrared divergences
of the pQCD are the phenomena just of the soft colour particles
domain.

It seems natural that the very high multiplicity (VHM) hadron
interaction, where the energy of the created particles is small,
should be sensitive to the soft colour particle densities. Indeed,
the aim of this Section is to show that even in the hard by definition
deep inelastic scattering (DIS), see also \C{kuraev} , the soft color
particles role becomes important in the VHM region \C{dis}.

To describe the hadron production in pQCD terms the parton-hadron
duality is assumed. This is natural just for the VHM process
kinematics:  because of the energy-momentum conservation law,
produced (final-state) partons cannot have high relative momentum
and, if they were created at small distances, production of $q\bar q$
pairs from the vacuum will be negligible (or did not play an
important role).  Therefore, if the `vacuum' channel is negligible,
only the pQCD contributions should be considered \C{0.6, siss}. All
this means that the multiplicity, momentum etc. distributions of
hadron and colored partons are the same. (This reduces the problem
practically to the level of QED.)  xx

Let us consider now $n$ particles (gluons) creation in the DIS
\C{murad}.  We would like to calculate $D_{ab}(x,q^2;n)$, where

\be
\sum_n D_{ab}(x,q^2;n)= D_{ab}(x,q^2).
\l{1d}\ee
As usual, let $D_{ab}(x,q^2)$ be the probability to find parton $b$
with virtuality $q^2<0$ in the parton $a$ of $\sim\la$ virtuality,
$\la>>\La$ and $\a_s(\la)<<1$.  We may always choose $q^2$ and $x$ so
that the leading logarithm approximation (LLA) will be acceptable.
One should assume also that $(1/x)>>1$ to have the phase space, into
which the particles are produced, sufficiently large.

Then $D_{ab}(x,q^2)$ is described by ladder diagrams. From a
qualitative point of view this means the approximation of random walk
over coordinate $\ln(1/x)$ and the time is $\ln\ln|q^2|$. LLA means
that the `mobility' $\sim \ln(1/x)/\ln\ln\le|q^2\ri|$ should be large
\be
\ln(1/x)>>\ln\ln \le|q^2/\la^2\ri|.
\l{2}\ee
But, on other hand \C{lla},
\be
\ln(1/x)<<\ln\le|q^2/\la^2\ri|.
\l{3x}\ee
See also Appendix D.

The leading contributions, able to compensate the smallness of
$$\a_s(\la)<<1,$$ give integration over a wide range
$\la^2<<k_i^2<<-q^2$, where $k_i^2>0$ is the `mass' of a real, i.e.
time-like, gluon.  If the time needed to capture the parton into the
hadron is $\sim(1/\La)$ then the gluon should decay if
$k_i^2>>\la^2$.  This leads to the creation of (mini)jets. The mean
multiplicity $\bar{n}_j$ in the QCD jets is high if the gluon `mass'
$|k|$ is high: $\ln\bar{n}_j \simeq \sqrt{\ln(k^2/\la^2)}$.

Raising the multiplicity may (i) raise the number of (mini)jets $\nu$
and/or (ii) raise the mean value mass of (mini)jets $\bar{|k_i|}$. We
will see that the mechanism (ii) would be favorable.

But raising the mean value of gluon masses, $|k_i|$, decreases the
range of integrability over $k_i$, i.e. violates the condition
(\r{2}) for fixed $x$. One can remain the LLA taking $x\to0$. But
this may contradict to (\r{3x}), i.e. in any case the LLA becomes
invalid in the VHM domain and the next to leading order corrections
should be taken into account.

Noting that the LLA gives the main contribution, that the rising
multiplicity leads to the infrared domain, where the soft gluon
creation becomes dominant.

First of all, neglecting the vacuum effects, we introduce definite
uncertainty to the formalism. It is reasonable to define the level of
strictness of our computations. Let us introduce for this purpose the
generating function $T_{ab}(x,q^2;z)$:
\be
D_{ab}(x,q^2;n)=\f{1}{2\pi i}\oint\f{dz}{z^{n+1}}T_{ab}(x,q^2;z).
\l{4}\ee
At large $n$, the integral may be calculated by the saddle point
method. The smallest solution $z_c$ of the equation
\be
n=z\f{\pa}{\pa z}\ln T_{ab}(x,q^2;z)
\l{5}\ee
defines the asymptotic over $n$ behavior:
\be
D_{ab}(x,q^2;n)\propto \exp\{-n\ln z_c(x,q^2;n)\}.
\l{6}\ee
Using the statistical interpretation of $z_c$ as the fugacity it is
natural to write:
\be
\ln z_c(x,q^2;n)=\f{C_{ab}(x,q^2;n)}{\bar{n}_{ab}(x,q^2)}.
\l{7}\ee
Notice that the solution of eq.(\r{5}) $z_c(x,q^2;n)$ should be an
increasing function of $n$. At first glance this follows from the
positivity of all $D_{ab}(x,q^2;n)$. But actually this assumes that
$T_{ab}(x,q^2;z)$ is a regular function of z at $z=1$. This is a
natural assumption considering just the pQCD predictions.

Therefore,
\be
D_{ab}(x,q^2;n)\propto
\exp\{-\f{n}{\bar{n}_{ab}(x,q^2)}C_{ab}(x,q^2;n)\}.
\l{8a}\ee
This form of $D_{ab}(x,q^2;n)$ is useful since usually
$C_{ab}(x,q^2;n)$ is a slowly varying  function of $n$. So, for a
Poisson distribution $C_{ab}(x,q^2;n)\sim\ln n$. For KNO scaling
we have $C_{ab}(x,q^2;n)=const.$ over $n$.

We would like to note that, neglecting effects of vacuum
polarization,  we introduce into the exponent so high uncertainty
assuming $n\simeq n_p$ that it is reasonable to perform the
calculations with exponential accuracy. So, we would calculate
\be
-\bar\mu_{ab}(x,q^2;n)=\ln\f{D_{ab}(x,q^2;n)}{D_{ab}(x,q^2)}=
\f{n}{\bar{n}_{ab}(x,q^2)}C_{ab}(x,q^2;n)(1+O(1/n))
\l{9a}\ee

The $n$ dependence of $C_{ab}(x,q^2;n)$ defines the asymptotic
behavior of $\bar\mu_{ab}(x,q^2;n)$ and calculation of its
explicit form would be our aim.

We can conclude, see Appendix H, that our LLA is applicable in the
VHM domain till
\be
\o(\tau,z)<<\ln(1/x)<<\tau=\ln(-q^2/\la),
\l{37}\ee
where
\be
\o(\tau,z)=\int^\tau_{\tau_0}\f{d\tau'}{\tau'}w^{g}(\tau',z).
\l{32''}\ee
and
\be
w^{g}(\tau,z)=\sum_n z^nw^{g}_n(\tau)
\l{32'''}\ee
is the generating function of the multiplicity distribution in a gluon
jet.  In the frame of constraints (\r{37}),
\be
F^{ab}(q^2,x;w)\propto \exp\le\{4\sqrt{N\o(\tau,z)\ln(1/x)}\ri\}.
\l{36''}\ee

The mean multiplicity of gluons created in the DIS kinematics
\be
\bar{n}_g(\tau,x)=\f{\pa}{\pa z}\ln F^{ab}(q^2,x;w)\le.\ri|_{z=1}
=\o_1(\tau)\sqrt{4N\ln(1/x)/\ln\tau}>>\o_1(\tau),
\l{38}\ee
where
\be
\o_1(\tau)=\int^\tau_{\tau_0}\f{d\tau_1}{\tau_1}\bar{n}_j(\tau)
\l{39}\ee
and the mean gluon multiplicity in the jet $\bar{n}_j(\tau)$
has the following estimation \C{12}:
\be
\ln\bar{n}_j(\tau)\simeq \sqrt{\tau}
\l{40}\ee
Inserting (\r{40}) into (\r{39}),
$$
\o_1(\tau)=\bar{n}_j(\tau)/\sqrt{\tau}.
$$
Therefore, noting (\r{3x}),
\be
\bar{n}_g(\tau,x)\simeq\bar{n}_j(\tau)
\sqrt{4N\ln(1/x)/\tau\ln\tau}<<\bar{n}_j(\tau).
\l{41}\ee
This means that the considered `t-channel' ladder is important in the
narrow domain of multiplicities
\be
n\sim \bar{n}_g<<\bar{n}_j.
\l{42}\ee

So, in the VHM domain $n>>\bar{n}_g$ one should consider\\
(i) The ladder diagrams with a small number of rungs;\\
(ii) To take into account the multi-jet correlations assuming that
increasing multiplicity leads to the increasing number of rungs in
the ladder diagram.\\
To choose one of these possibilities one should consider the structure
of $\o(\tau,z)$ much more carefully. This will be done in the
Sec.3.5.

We can conclude that in the VHM domain, multiplicity production
unavoidably destroys the ladder LLA. To conserve this leading
approximation one should choose $x\to0$ and, in result, to get to the
multi-ladder diagrams, since in this case $\a_s\ln (-q^2/\la^2)\sim
1$ and $\a_s\ln(1/x)\sim 1$. Such theory was considered in \C{grib}.

\subsubsection{QCD jets}

As was mentioned above, the pQCD description is right if the colour
particles virtuality is bounded from below, $|q^2|\geq la^2$, where
$\la$ is chosen so that $\a_s(\la^2)<<1$. This kinematical
restriction leads to the infrared cutoff \C{140, 143} and may
essentially influence the particle production in the VHM region.
It is a special property of pQCD. Indeed, for example, careful
investigation of this question in the asymptotically free
$(\vp^3)_6$-theory \C{chang} shows that this restriction is
`unobservable' since their inclusion takes us beyond the LLA
\C{144}. At the same time, the condition $|q^2|\geq la^2$
essentially shrinks the phase space where particles are produced.

Particle (gluons) distribution in pQCD jets was investigated firstly
in \C{140, 141} and it was shown that the generating function is
singular at $z_s-1\sim (1/\bar{n})$. Let us consider this solution
stable with reference to the discussed cutoff.

The explicit formulae for one jet production may be written in the
form, see Appendix I:
\be
\s^{(1j)}_n(M)=a^{(1j)}(M,n)e^{-c_jn/\bar{n}_j(M)},~n\geq\bar{n}_j(M),
\l{1j}\ee
where $a^{(1j)}(n,M)$ is the polynomial function of $n$, $\bar{n}_j(M)$ is
the mean multiplicity in the mass $M$ jet and $c_j$ is a positive
constant.

The linear behavior of the exponent in (\r{j4}) over $n/\bar{n}$
has important consequences. So, let us assume that the total energy
$M$ is divided into two jets of masses $M_1$ and $M_2$ equally:
$M_1=M_2=M/2$. If, for instance,  $M_2<<M_1\simeq M$ then the
distribution will coincide with (\r{1j}), but the second jet
distribution would renormalize the coefficient $a^{(1j)}$.

Then the multiplicity distribution in the two-jet event would be
\be
\s^{(2j)}_n(M)=a^{(2j)}(M,n)e^{-c_jn/\bar{n}_j(M/2)},
\l{2j}\ee
where $n_1+n_2=n$ is the total multiplicity.

Comparing (\r{1j}) with (\r{2j}) we can see that with exponential
accuracy:
$$
\sim\exp\{-c_j\f{\bar{n}_j(M)-\bar{n}_j(M/2)}
{\bar{n}_j(M)\bar{n}_j(M/2)}n\}
$$
the (\r{1j}) would dominate in the VHM domain since the mean
multiplicity $\bar{n}_j(M)$ increases with $M$.

The experimental observation of this phenomena crucially depends on
the value of $a^{1j}$, $a^{2j}$,... but if (\r{1j}) is satisfied then
one can expect that the events in the VHM domain   would be enhanced
by QCD jets and the mass of jets would have a  tendency to be high
with growing multiplicity.

The singular at finite $z$ solutions arise in the field theory, when
the $s$-channel cascades (jets) are described \C{jet}. By definition
$T(z,s)$ coincides with the total cross section at $z=1$. Therefore,
the nearness of $z_c$ to one defines the significance of
the corresponding processes. It is evident that both $s$ and $n$
should be high enough to expect the jets creation.

Summarizing the above estimations, we may conclude that
\be
O(e^{-n})\leq\s_n< O(1/n),
\l{creg}\ee
i.e. the soft Regge-like channel of hadron creation is suppressed in
the VHM region in the high energy events with exponential accuracy.

\subsection{Phase transition - condensation}

The aim of this section is to find the experimentally observable
consequences of collective phenomena in the high energy hadron
inelastic collision \C{ph.tr.}. We will pay the main attention to the
phase transitions, leaving out other possible interesting collective
phenomena.

The statistics experience dictates that we should prepare the system
for the phase transition. The temperature in a critical domain and
the equilibrium media are just these conditions. It is evident that
they are not a trivial requirement considering the hadrons inelastic
collision at high energies.

The collective phenomena by definition suppose that the kinetic
energy of particles of media are comparable, or even smaller, than
the potential energy of their interaction. It is a quite natural
condition noting that, for instance, the kinetic motion may destroy
even completely at a given temperature $T$, necessary for the phase
transition long-range order. This gives, more or less definitely, the
critical domain.

The same idea as in statistics seems natural in the multiple
production physics.  We will assume ({\bf A}) that the collective
phenomena should be seen just in the very high multiplicity (VHM)
events, where, because of the energy-momentum conservation laws, the
kinetic energy of the created particles can not be high.

We will lean at this point on the $S$-matrix interpretation of
statistics \C{elpat}, see Sec.2.2.  It based on the $S$-matrix
generalization of the Wigner function formalism of Carruthers and
Zachariazen \C{carr} and the real-time finite temperature field
theory of Schwinger and Keldysh \C{sch, kel}, see Appendices A,B and
C.

It was mentioned that the $n$-particle partition function in this
approach coincides with the $n$ particle production cross section
$\s_n(s)$ (in the appropriate normalization condition). Then, the
cross section $\s_n(s)$ can be calculated applying the $n$-point
Wigner function $W_n(X_1,X_2,...,X_n)$. In the relativistic case
$X_k=(u,q)_k$ are the 4-vectors. So, the external particles are
considered as the `probes' to measure the state of the interacting
fields, i.e.  the low mean energy of probes means that the system is
`cold'.

The multiple production phenomena may be considered also as the
thermalization process of incident particle kinetic energy
dissipation into the created particle mass.  From this point of
view the VHM processes are highly nonequilibrium since the final
state of this case is very far from the initial one. It is known in
statistics \C{kac} that such processes aspire to be the stationary
Markovian with a high level of entropy production. In the case of
complete thermalization, the final state is in equilibrium.

The equilibrium we will classify as the condition in the frame of
which the fluctuations of corresponding parameter are Gaussian. So,
in the case of complete thermalization, the probes should have the
Gaussian energy spectra.  In other terms, the necessary and
sufficient condition of the equilibrium is the smallness of the mean
value of energy correlators \C{bog, elpat}. From the physical point
of view, the absence of these correlators means depression of the
macroscopic energy flows in the system.

The multiple production experiment shows that the created particle
energy spectrum is far from a Gaussian law, i.e. the final states
are far from equilibrium. The natural explanation of this phenomena
consist in the presence of (hidden) conservation laws in the
interacting Yang-Mills fields: it is known that the presence of
sufficient number of first integrals in involution prevents
thermalization completely.

But nevertheless the VHM final state may be equilibrium ({\bf B}) in
the above formulated sense. This means that the forces created by the
non-Abelian symmetry conservation laws may be frozen during the
thermalization process (remembering its stationary Markovian
character in the VHM domain). We would like to take into account that
the entropy $\cal S$ of a system is proportional to number of created
particles and, therefore, $\cal S$ should tend to its maximum in the
VHM region \C{fer}.

One may consider following the small parameter $(\bar{n}(s)/n)<<1$,
where $\bar{n}(s)$ is the mean value of the multiplicity $n$ at
a given CM energy $\sqrt s$.  Another small parameter is the energy
of the fastest hadron $\e_{max}$. One should assume that in the VHM
region $(\e_{max}/\sqrt s)\to0$.  So, the conditions:
\be
\f{\bar{n}(s)}{n}<<1,~\f{\e_{max}}{\sqrt s}\to0
\l{A}\ee
would be considered as the mark of the processes under consideration.
We can hope to organize the perturbation theory over them
having there small parameters. In this sense VHM processes may be
`simple', i.e. one can use for their description semiclassical
methods.

So, considering VHM events one may assume that the conditions ({\bf
A}) and ({\bf B}) are satisfied and one may expect the phase
transition phenomena.

The $S$-matrix interpretation of statistics is based on the following
definitions.  First of all, let us  introduce the generating function
\C{conf}:
\be
T(z,s)=\sum_n z^n \s_n(s).
\l{1a}\ee
Summation is performed over all $n$ up to $n_{max}=\sqrt{s}/m$ and,
at finite CM energy $\sqrt{s}$, $T(z,s)$ is a polynomial function of
$z$. Following our idea, see Sec.2.3, let us assume now that $z>1$
is sufficiently small and for this reason $T(z,s)$ depends on the
upper boundary $n_{max}$ only weakly.  In this case one may formally
extend summation up to infinity and in this case $T(z,s)$ may be
considered as a whole function.  This possibility is important being
the equivalent of the thermodynamical limit and it allows to classify
the asymptotics over $n$ in accordance with the position of
singularities over $z$.

Let us consider $T(z)$ as the big partition function, where $z$ is
`activity'. It is known \C{lee} that $T(z)$ should be regular inside
the circle of unit radius. The leftist singularity lies at $z=1$. This
singularity is manifestation of the first order phase transition
\C{lee, kac, langer}.

The origin of this singularity was investigated carefully in the
paper \C{kac}. It was shown that the position of singularities over
$z$ depends on the number of particles $n$ in the system: the two
complex conjugated singularities move to the real $z$ axis with
rising $n$ and in the thermodynamical limit, $n=\infty$, they pinch
the point $z=1$ in the first order phase transition case.  More
general analysis \C{langer} shows that if the system is in
equilibrium, then $T(z)$ may be singular only at $z=1$ and
$z=\infty$.

The position of the singularity over $z$ and the asymptotic behavior
of $\s_n$ are related closely. Indeed, for instance, inserting into
(\r{1a}) $\s_n\propto \exp\{-cn^\ga\}$ we find that $T(z)$ is singular
at $z=1$ if $\ga<1$. Generally, using the Mellin transformation
(\r{21.5}) one can find an asymptotic estimation (\r{21.7}):
\be
\s_n\propto e^{-n\ln z_c(n)},~z_c>1,
\l{1.3a}\ee
where $z_c$ is smallest solution of the equation of state
\be
n=z\f{\pa}{\pa z}\ln T(z).
\l{1.2a}\ee

Therefore, to have the singularity at $z=1$, we should consider
$z_c(n)$ as a decreasing function of $n$. On the other hand, at
constant temperature, $\ln z_c(n)\sim\mu_c(n)$ is the chemical
potential, i.e. is the work necessary for creation of one particle.
So, the singularity at $z=1$ means that the system is unstable: the
less work is necessary for creation of one more particle if $\mu(n)$
is a decreasing function of $n$.

The physical explanation of this phenomena is the following, see also
\C{andr}. The generating function $T(z)$ has following expansion:
\be
T(z)=\exp\{\sum_lz^lb_l\},
\l{1.4a}\ee
where $b_l$ are known as the Mayer's group coefficients \C{mayer}.
They can be expressed through the inclusive correlation functions and
may be used to describe formation of droplets of correlated
particles, see Sec.2.3.3.  So, if droplet consist of $l$ particles,
then
\be
b_l\sim e^{-\b\xi l^{(d-1)/d}}
\l{1.5}\ee
is the mean number of such droplets. Here $\xi l^{(d-1)/d}$ is the
surface energy of $d$-dimensional droplet.

Inserting this estimation into (\r{1.4a}),
\be
\ln T(z)\sim\sum_l e^{\b(l\mu -\xi l^{(d-1)/d})},~\b\mu=\ln z.
\l{1.6}\ee
The first term in the exponent $\b l\mu$ is the volume energy of
the droplet and being positive it tries to enlarge the droplet. The
second surface term $-\b\xi l^{(d-1)/d}$ tries to shrink it.
Therefore, the singularity at $z=1$ is the consequence of
instability:  at $z>1$ the volume energy abundance leads to unlimited
growth of the droplet.

In conclusion we wish to formulate once more the main assumptions.

(I). It was assumed first of all that the system under consideration
is in equilibrium. This condition may be naturally reached in the
statistics, where one can wait the arbitrary time till the system
becomes in equilibrium. Note, in the critical domain, the time of
relaxation $t_r\sim (T_c/(T-T_c))^\nu\to\infty$, $(T-T_c)\to+0$,
$\nu>0$, $T_c$ is the critical temperature.

We cannot give the guarantee that in the high energy hadron
collisions the final state system is in equilibrium. The reason of
this uncertainty is the finite time the inelastic processes and
presence of hidden (confinement) constraints on the dynamics. But if
the confinement forces are frozen in the VHM domain, i.e. the
production process is `fast', then the equilibrium may be reached.

We may formulate the quantitative conditions, when the
equilibrium is satisfied \C{bog}. One should have the Gaussian energy
spectra of created particles. If this condition is hardly
investigated in the experiment, then one should consider the
relaxation of `long-range' correlations. This excludes the usage of
relaxation condition for the `short-range' (i.e. resonance)
correlation

(II). The second condition consists in the requirement that the system
should be in the critical domain, where the (equilibrium)
fluctuations of the system become high.  Having no theory of hadron
interaction at high energies we can not define where lies the
`critical domain' and even whether it exists or not.

But, having the VHM `cold' final state, we can hope that the
critical domain is achieved.

The quantitative realization of this picture is given in Appendix J.
It is important to note that used there the semiclassical
approximation is rightfully in the VHM domain.

\section{Conclusion}\0

\subsection{Discussion of physical problems}

It seems useful to start the discussion of models outlined the main
problems, from authors point of view.

{\bf A.} {\it Soft colour parton problems}

The infrared region of soft colour parton interactions is a very
important problem of high energy hadron dynamics. Such
fundamental questions as the infrared divergences of pQCD, collective
phenomena in the coloured particles system and symmetry breaking are
the phenomena of the infrared domain.

The standard (most popular) hadron theory considers pQCD at small
distances (in the scale of $\La\simeq 0.2$Gev) as the exact theory.
This statement is confirmed by a number of experiments, namely
deep-inelastic scattering data, hard jets observation. But the pQCD
predictions have a finite range of validity since  the
non-perturbative effects should be taken into account at distances
larger then $1/\La$.

It is natural to assume, building the complete theory, that at large
distances the non-perturbative effects $lay~on$\foot{The corresponding
formalism was described e.g. in \C{fad.PR}.} the perturbative ones.
As a result pQCD loses its predictability $screened$ by the
non-perturbative effects.

Notice, the pQCD running coupling constant
$\a_s(q^2)=1/b\ln(q^2/\La^2)$ becomes infinite at $q^2=\La^2$ and we
do not know what happens with pQCD if $q^2<\La^2$. There are few
possibilities.  For instance, there is a suspicion \C{t'hooft} that
at $q^2\sim \La^2$ the properties of theory changed so drastically
(being defined on a new vacuum) that even the notions of pQCD
$disappeared$. This means that pQCD should be truncated from below on
the `fundamental'  scale $\La$. It seems natural that this infrared
cut-off would influence the soft hadrons emission.

The new possibility is described in Appendix K. This strict
formalism allows to conclude that pure pQCD contributions are
realized on zero measure, i.e. it is the phenomenological theory
only. The successive approach shows that the Yang-mills theory should
be described in terms of (action,angle)-like variables. The last one
means that the self-consistent description excludes such notions as
the `gluon'. As a result of this substitution new perturbation theory
would be free from infrared divergences, i.e. there is not necessity
to introduce the infrared cutoff parameter $\La$. (Moreover, in the
sector of vector fields (without quarks) the theory is ultraviolet
stable.) It seems important for this reason to investigate
experimentally just VHM events, where the soft colour partons
production is dominant.

It is important to try to raise the role of pQCD in the `forbidden'
area of large distances.  The VHM processes are at highly
unusual condition, where the non-perturbative effects must be
negligible.

{\bf B}. {\it Dissipation problems}

The highly nonequilibrium states decay (thermalization) which means
in the pQCD terms that the process of VHM formation should be
enhanced, at least in asymptotics over multiplicity and energy, by
jets.  It is the general conclusion of nonequilibrium thermodynamics
and it means that the very nonequilibrium initial state tends to
equilibrium (thermalized) as fast as possible.

The entropy $\cal S$ of a system is proportional to the number of
created particles and, therefore, $\cal S$ should tend to its maximum
in the VHM region \C{fer}. But the maximum of entropy testifies also
the equilibrium of the system.

{\bf C}. {\it Collective phenomena}

We should underline that the collective phenomena may take place if
and only if the particles interaction energy is  comparable to  the
kinetic one. The VHM  system considered may be `cold' and
`equilibrium'.  For this reason the VHM state is mostly adopted for
investigation of collective phenomena. One of possible states in
which the collective effects, see a.g. \C{qgp}, may be important is
the `could coloured plasma' \C{chelk}.

The fundamental interest presents the problem of vacuum structure of
Yang-Mills theory. For instance, if the process of cooling is `fast',
since the dissipation process of VHM final state formation should be
as fast as possible, then one may consider formation of vacuum domains
with various properties.  Then decay of these domains may lead to
large fluctuations, for instance, of the isotopic spin.

Another important question is the collective phenomena in the
VHM final state. The last one may be created `perturbatively', for
instance, by the formation of heavy jets. Then the colour charges
should be confined.  There are various predictions about this
process.  One of them predicts that there should be a first order
phase transition.

\subsection{Model predictions}

Now we can ask: what can models say concerning the above problems?

{\bf A}. {\it Soft coloured partons production}

The multiperipheral models predict fast decreasing of topological
cross section in the VHM domain $\N^2<<n<<n_{max}$, $\s_n<O(e^{-n})$.
At the same time the mean transverse momentum should $decrease$ in
this domain since the interaction radii should `increase' with $n$.

The BFKL Pomeron predicts the same asymptotics, $\s_n<O(e^{-n})$, but
the pQCD jet predicts  $\s_n=O(e^{-n})$. The $naive$ attempt to
insert into the BFKL Pomeron the production of particles via (mini)jets
seems impossible.

This `insertion' can be done into the DIS ladder but investigation of
the LLA kinematics in the VHM domain allows the conclusion that the
`low-x' contributions should be taken into account.

All this experience allows the assumption that in the VHM domain no
`$t$-channel ladder' diagrams play sufficient role. This means
the existence of a transition to the processes with jet dominance.
The pQCD is unable to predict the transition mechanism.

{\bf B}. {\it Transition into 'equilibrium'}

If the $t$-channel ladders are `destroyed' in the VHM region, then
jets, despite the small factor $O(1/s)$ in the cross sections, are
the only mechanism of particle production in the VHM domain.
Dominance of heavy jets in the VHM domain may naturally explain the
tendency towards equilibrium.

But the description of thermalization in terms of jets of massless
gluons production destroy this hope: the jet contribution
$\s_n=O(e^{-n})$ assumes `bremsstrahlung' of $soft$ gluons \C{khose}.
This prevents the equilibrium since ordering without fail
introduce the non-relaxing correlations.

{\bf C}. {\it Collective phenomena}

Considering the collective phenomena, we proposer to distinguish (a)
the collective phenomena connected with the vacuum and (b) the
collective phenomena produced in the VHM system. Following
the experience of Sec.3.3 we can conclude that the signal of vacuum
instability is inequality: $\s_n>O(e^{-n})$.

The case (b) will not effect the cross section $\s_n$. But if the
system became equilibrium in the VHM domain then the collective
phenomena may be investigated using ordinary thermodynamical methods.
For instance, noting that $-\{\ln(T(\b_c,z)/T(\b_c,z))\}/\b_c=
{\cal F}(\b_c,z)$ is the free energy one can measure the thermal
capacity
\be
\f{\pa}{\pa\b_c}{\cal F}(\b_c,z)=C(\b_c,z).
\l{o23}\ee
Then, comparing capacities of hadron and $\ga$-quanta systems we can
say whether or not the phase transition happened.

The connection  of the equilibrium  and relaxation of  correlations is
well known  \C{bog}.  Continuing  this idea, if  the VHM system  is in
equilibrium   one  may   assume  that   the  colour   charges   in
the pre-confinement phase  of VHM event  form the plasma. One  should
note here that the expected  plasma is `cold' and `dense'. For  this
reason no long-range confinement forces would act  among colour
charges.  Then, being `cold', in such a system various, collective
phenomena may be important.

\subsection{Experimental perspectives}

The experimental possibilities in the VHM domain are not clear up to
now. Nevertheless first steps toward formulation of trigger system
was done, see \C{tapp}.  By this reason we would like to restrict
ourselves by following two general questions.  It seems that these
questions are mostly important being in the very beginning of VHM
theory.

\vskip 0.3cm
{\bf I}. {\it For what values of multiplicity at a given energy the
VHM processes become hard}?\\
The answer to this question depends on the value of the incident
energy.  If we know the answer then it will appear possible to
estimate

-- the role of multiperipheral contributions,

-- the jet production rate,

-- the role of vacuum instabilities.

It seems that the experimental answer to this question is absent since
produced particles are soft, theirs mean transverse and longitudinal
momenta have the same value.  In our understanding this question
means: the total transverse energy may be extremely high.

It is interesting also to search the heavy jets, i.e. to observe the
fluctuations of particle density in the event-by-event experiment,
but this program seems vague since, for all evidence, fractal
dimensions tend to zero with increasing multiplicities.

\vskip 0.3cm
{\bf II}. {\it For what values of multiplicity does the VHM final
state become equilibrium}?\\ We hope that having answer on this
question we would be able

-- to investigate the status of pQCD,

-- to observe the phase transition phenomena directly,

-- to estimate the role of confinement constraints.

The equilibrium means that the energy correlation functions mean
values are small. It is interesting also, for example, the charge
equilibrium, when the mean value of charge correlation functions are
small.

Notice, the effect of the phase space boundary may lead to
`equilibrium'.  Indeed, if $\e(p)\simeq m+p^2/2m$ and $p^2<<m^2$ then
one may neglect the momentum dependence of the amplitudes $a_n$. In
this case the momentum dependence is defined by the Boltzmann
exponent $e^{-\b p^2/2m}$, $\b\to\infty$, only and we get naturally
to the Gaussian law for momentum distribution. Correlators should be
small in this case since there is no interactions among particles
($a_n$ are constants). But our question assumes that we investigate
the possibility of equilibrium when $n<<n_{max}$ and $p^2>>m^2$.

\renewcommand{\theequation}{A.\arabic{equation}}
\appendix\section{Appendix. Matsubara formalism and the KMS boundary
condition}\0

There are various approaches to build the real-time
finite-temperature field theories of Schwinger-Keldysh type (e.g.
\C{land}). All of them use various tricks for analytical
continuation of imaginary-time Matsubara formalism to real time
\C{kad}. The basis of the approaches is the introduction of
the Matsubara field operator
\be
\P_M ({\bf x}, \b)=e^{\b H}\P_S
({\bf x}) e^{-\b H},
\l{32}\ee
where
$\P_S ({\bf x})$ is the interaction-picture operator introduced
instead of the habitual Heisenberg operator
$$
\P ({\bf x}, t)=e^{itH}\P_S ({\bf x}) e^{-itH}.
$$
Eq.(\r{32})
introduces the averaging over the Gibbs ensemble instead of averaging
over zero-temperature vacuum states.

If the interaction switched on at the moment $t_i$ adiabatically
and switched off at $t_f$ then there is the unitary transformation:
\be
\P (x,t)=U(t_i ,t_f )U(t_i ,t)\P_S (x)U(t ,t_i).
\l{a32}\ee
Introducing the complex Mills time contours \C{mil} to connect $t_i$ to
$t$, $t$ to $t_f$ and $t_f$ to $t_i$ we form a `closed-time' contour
$C$ (the end-points of the contours $joined$ together). This allows
to write the last equality (\r{a32}) in the compact form:
$$
\P (x)
=T_C \{ \P (x) e^{i\int_C d^4 x' L_{int} (x')} \}_S,
$$
where $T_C$ is the time-ordering on the contour $C$ operator.

The generating functional $Z(j)$ of correlation (Green) functions has
the form:
$$
Z(j)=R(0)< T_C  e^{i\int_C d^4 x \{L_{int} (x) + j(x)\P (x)\}_S}>,
$$
where $<>$ means averaging over the initial state.

If the initial correlations have a little effect, we can perform
averaging over the Gibbs ensemble. This is the main assumption of the
formalism: the generating functional of the Green functions $Z(j)$
has the form in this case:
$$
Z(j)=\int D\P' <\P';t_i| e^{-\b H} T_C e^{i\int_C d^4 x
j(x)\P (x)}|\P'; t_i>
$$
with $\P' =\P' ({\bf x})$. In accordance with (\r{32}) we have:
$$
<\P';t_i| e^{-\b H} = <\P';t_i -i\b|
$$
and, as a result,
\be
Z(j)=\int D\P' e^{i\int_{C_{\b}} d^4 x \{ L(x) + j(x)\P (x)\}}
\l{33}\ee
where the path integration is performed with KMS periodic boundary
condition:
$$
\P (t_i) = \P (t_i -i\b).
$$
In (\r{33}) the contour $C_{\b}$ connects $t_i$ to $t_f$, $t_f$ to
$t_i$ and $t_i$ to $t_i -i\b$. Therefore it contains an
imaginary-time Matsubara part $t_i$ to $t_i -i\b$. A more symmetrical
formulation uses the following realization:  $t_i$ to $t_f$, $t_f$ to
$t_f - i\b/2$, $t_f - i\b/2$ to $t_i - i\b/2$ and $t_i -i\b/2$ to
$t_i -i\b$ (e.g. \C{sem}). This case also contains the imaginary-time
parts of the time contour.  Therefore, eq.  (\r{33}) presents the
analytical continuation of the Matsubara generating functional to
real times.

One can note that if this analytical continuation is possible for
$Z(j)$ then representation (\r{33}) gives good recipe of
regularization of frequency integrals in the Matsubara perturbation
theory, e.g. \C{land}. But it gives nothing new for our problem since
the Matsubara formalism is a formalism for equilibrium states only.

\renewcommand{\theequation}{B.\arabic{equation}}
\section{Appendix. Constant temperature formalism}\0

The starting point of our calculations is the $n$- into $m$-particles
transition amplitude $a_{nm}$, the derivation of which is well
known procedure  in the Lehmann-Symanzik-Zimmermann (LSZ) reduction
formalism \C{shir} framework, see also \C{slavnov}.  The
$(n+m)$-point Green function $G_{nm}$ are introduced for this purpose
through the generating functional $Z_{j}$ \C {vas}:

\be
G_{nm}(x,y)=(-i)^{n+m}\prod_{k=1}^{n}\hat{j}(x_k)\prod_{k=1}^{m}
\hat{j}(y_k)Z_j,
\l{22.5}\ee
where
\be
\hat{j}(x)=\frac{\delta}{\delta j(x)},
\l{*}
\ee
and
\be
Z_j=\int D\Phi e^{iS_j(\Phi)}.
\l{22.6}
\ee
The action
\be
S_j (\Phi)=S_0(\Phi)-V(\Phi)+\int dxj(x)\Phi (x),
\l{22.7}
\ee
where $S_0(\P)$ is the free part and $V(\P)$ describes the
interactions.  At the end of the calculations one can put $j=0$.

To provide the convergence of the integral (\ref {22.6}) over the
scalar field $\P$ the action $S_{j}(\P)$ must contain a positive
imaginary part.  Usually for this purpose Feynman's
$i\e$-prescription is used.  But it is better for us to use the
integral on the Mills complex time contour $C_+$ \C{mil, land}. For
example,
\be
C_\pm :~t\to t+i\ve, \ve\to+0,~ -\infty\leq t\leq+\infty
\l{o5}\ee
and after all the calculations return the time contour on the real
axis putting $\ve=0$.

In eq. (\ref{22.6}) the integration is performed over all field
configurations with standard vacuum boundary condition:
\be
\int d^4 x \partial_{\mu}(\Phi \partial^{\mu}\Phi)=
\int_{\sigma_{\infty}}d\sigma_{\mu}\Phi\partial^{\mu}\Phi=0,
\l{22.8}
\ee
which leads to zero contribution from the surface term.

Let us introduce now field $\p$ through the equation:
\be
-\frac{\d S_0(\p)}{\d\p(x)} =j(x)
\l{22.9}
\ee
and perform the shift $\P\rightarrow\P+\p$ in integral (\ref{22.6}),
conserving boundary condition (\ref{22.8}). Considering $\p$ as the
probe field created by the  source:
\ba
\phi(x)=\int dy G_0
(x-y)j(y),
\n \\
(\partial^2 +m^2)_x G_0 (x-y)=\delta (x-y),
\l{22.10}
\ea
only the connected Green function $G^{c}_{nm}$ will be
interesting for us. Therefore,
\be
G_{nm}^{c}(x,y)=(-i)^{n+m}\prod_{k=1}^{n}\hat{j}(x_k)
\prod_{k=1}^{m}\hat{j}(y_k)Z(\phi),
\l{22.11}
\ee
where
\be
Z(\phi)=\int D\Phi e^{iS(\Phi)-iV(\Phi+\phi)}
\l{22.12}
\ee
is the new generating functional.

To calculate the nontrivial elements of the $S$-matrix we
must put the external particles on the mass shell. Formally this
procedure means amputation of the external legs of $G^{c}_{nm}$
and further multiplication on the free particle wave functions.
As a result the amplitude of $n$- into $m$-particles transition
$a_{nm}$ in the momentum representation has the form:
\be
a_{nm}(q,p)=(-i)^{n+m}\prod_{k=1}^{n}\hat{\phi}(q_k)
\prod_{k=1}^{m}\hat{\phi}^* (p_k) Z(\phi).
\l{22.13}
\ee
Here we introduce the particle distraction operator
\be
\h{\p}(q)=\int dxe^{-iqx}\h{\p}(x),~\h{\p}(x)=\frac{\d}{\d\p(x)}.
\l{22.14}
\ee

Supposing that the momentum of  particles are insufficient for us the
probability of $n$- into $m$-particles transition is defined by the
integral:
\be
r_{nm}=\frac{1}{n!m!}\int d\O_n (q) d\O_m (p)
\delta^{(4)}(\sum_{k=1}^{n}q_k - \sum_{k=1}^{m}p_k) |a_{nm}|^2,
\l{22.15}
\ee
where
\be
d\O_n(q)=\prod_{k=1}^{n}d\O(q_k)=
\prod_{k=1}^{n}\f{d^3q_k}{(2\pi)^3 2\e(q_k)},
\l{22.16}\ee
is  the Lorentz-invariant phase space element. We assume that the
energy-momentum conservation $\d$-function was extracted from the
amplitude.

Note that $r_{nm}$ is the divergent quantity. To avoid this problem
with trivial divergence, connected integration over reference frame,
let us divide the energy-momentum fixing $\d$-function into two
parts:
\be
\d^{(4)}(\sum q_k -\sum p_k)=\int d^4 P
\d^{(4)}(P-\sum q_k)\d^{(4)}(P-\sum p_k)
\l{22.17}
\ee
and consider a new quantity:
\be
R(P)=\sum_{n,m}\frac{1}{n!m!}\int d\O_n (q) d\O_m (p)
\delta^{(4)}(P-\sum_{k=1}^{n} q_k)
\delta^{(4)}(P-\sum_{k=1}^{n} p_k) |a_{nm}|^2
\l{22.18}\ee
defined on the energy momentum shell (\r{22.4a}). Here we suppose
that the numbers of particles are not fixed. It is not too hard to
see that, up to a phase space volume,
\be
R=\int d^4P~R(P)
\l{22.19}
\ee
is the imaginary part of the amplitude $<vac|vac>$. Therefore,
computing $r(P)$ the  standard renormalization procedure  can be
applied and the new divergences will not arise in our formalism.

The Fourier  transformation of $\d$-functions in (\ref{22.18}) allows
one to write $R(P)$ in the form:
\be
R(P)=\int \frac{d^4
\a_1}{(2\pi)^4}\frac{d^4\a_2}{(2\pi)^4}
e^{iP(\a_1+\a_2)}
\R(\a),
\l{22.20}\ee
where
\be
\R(\a)=
\sum_{n,m} \frac{1}{n!m!}\int
\prod_{k=1}^{m}\{d\O(q_k)e^{-i\a_1 q_k}\}
\prod_{k=1}^{n}\{d\O(p_k)e^{-i\a_2 p_k}\} |a_{nm}|^2.
\l{22.21}
\ee

Introducing the Fourier-transformed probability $\R(\a)$ we assume
that the  phase-space volume is not fixed exactly, i.e. it is
proposed that the 4-vector $P$ is fixed with some accuracy if $\a_i$
are fixed.  The energy and momentum in our approach are still locally
conserved quantities since an amplitude $a_{nm}$ is translationally
invariant.  So, we can perform the transformation:  \be \a_1 \sum q_k
=(\a_1 -\s_1 )\sum q_k +\s_1 \sum q_k \to (\a_1 -\s_1 )\sum q_k +\s_1
P \l{} \ee since 4-momenta are conserved. The choice of $\s_1$ fixes
the reference  frame. This  degree of freedom of the theory was
considered in \C{psf}.

Inserting (\ref{22.13}) into (\ref{22.21}) we find  that
\ba
&\R(\a)=\exp \{-i\int dx dx'(
\h{\p}_+(x)D_{+-}(x-x',\a_2)\h{\p}_-(x')-
\n \\
&-\h{\p}_-(x)D_{-+}(x-x',\a_1)\h{\p}_+(x'))\}
Z(\p_+)Z^* (-\p_-),
\l{22.22}\ea
where $D_{+-}$ and $D_{-+}$ are  the positive and negative frequency
correlation functions correspondingly:
\be
D_{+-}(x-x',\a)=-i\int d\O_1(q)e^{iq(x-x'-\a)}
\l{22.23}
\ee
describes the process of particles creation at the  time $x_0$ and
its absorption at $x'_0$, $x_0>x'_0$, and $\a$ is the CM
4-coordinate. Function
\be
D_{-+}(x-x',\a)=i\int d\O_1(q)e^{-iq(x-x'+\a)}
\l{22.24}
\ee
describes the opposite process, $x_0<x'_0$. These functions obey the
homogeneous equations:
\be
(\pa^2 +m^2)_x G_{+-}=(\pa^2 +m^2)_x G_{-+}=0
\l{22.25}\ee
since the propagation of on mass-shell particles is described.

We suppose that $Z(\p)$ may be computed perturbatively. For this
purpose the following transformations will be used ($\h{X}\equiv \d/
\d X$ at $X=0$):
\ba
&e^{-iV(\phi)}=
e^{-i\int dx \hat{j}(x)\hat{\phi}'(x)}
e^{i\int dx j(x)\phi (x)}
e^{-iV(\phi ')}=
\n\\
&=e^{\int dx \phi(x)\hat{\phi}'(x)}
e^{-iV(\phi ')}=
\n\\
&=e^{-iV(-i\hat{j})}
e^{i\int dx j(x)\phi (x)},
\l{22.26}
\ea
where $\hat j$ was defined in (\ref{*}) and $\hat{\phi}$ in
(\ref{22.14}). At the end of the calculations, the auxiliary variables
$j$, $\p'$ can be taken equal to zero. Using the first equality in
(\ref{22.26}) we find that
\be
Z(\phi)=
e^{-i\int dx \hat{j}(x)\hat{\Phi}(x)}
e^{-iV(\Phi+\phi)}
e^{-\frac{i}{2}\int dx dx'
 j(x)D_{++}(x-x')j(x')},
\l{22.27}\ee
where $D_{++}$ is the causal Green function:
\be
(\partial^2 +m^2)_x G_{++} (x-y)=\delta (x-y).
\l{22.28}
\ee
Inserting (\ref{22.27}) into (\ref{22.22}) after simple manipulations
with differential operators, see (\ref{22.26}), we may find the
expression:
\ba
&\R(\a)=
e^{-iV(-i\hat{j}_+)+iV(-i\hat{j}_-)}\exp\{ \f{i}{2} \int dx dx'
\t\n\\
&\t(
 j_+ (x)D_{+-}(x-x',\a_1)j_- (x')-
 j_- (x)D_{-+}(x-x',\a_2)j_+ (x')-
\n\\
&- j_+ (x)D_{++}(x-x')j_+ (x')+
 j_- (x)D_{--}(x-x')j_- (x'))\},
\l{22.29}\ea
where
\be
D_{--}=(D_{++})^*
\l{22.30}
\ee
is the anticausal Green function.

Considering the system with a large number of particles, we
can simplify the calculations choosing the CM frame $P=(P_0 =E,\vec
0)$.  It is useful also \C{kaj, mar} to rotate the contours of
integration over $$ \a_{0,k}:~\a_{0,k}=-i\b_k, \Im\b_k =0, k=1,2.  $$
For the result, omitting the unnecessary constant, we will consider
$\R=\R(\b)$.

External particles play a double role in the $S$-matrix approach:
their interactions create and annihilate the system under consideration
and, on the other hand, they are probes through which the
measurement of a system is  performed. Since $\b_k$ are the conjugate
to the particles  energies quantities we will interpret them
as  the inverse temperatures in the initial ($\b_1$) and  final ($\b_2$)
states of interacting fields. They are the `good' parameters if and
only if the energy correlations are relaxed.

\vskip 0.2cm
\begin{center}
{\bf Kubo-Martin-Schwinger boundary condition}
\end{center}
\vskip 0.2cm

The  simplest (minimal) choice of $\P (\s_{\infty})\neq 0$ assumes
that the  system under consideration is surrounded by black-body
radiation. This interpretation restores Niemi-Semenoff's formulation
of the real-time finite temperature field theory \C{sem}.

Indeed, as follows from (\r{22.22}), the generating functional
$\R(\a)$ is defined by corresponding generating functional
\ba
&\R_0 (\p_{\pm})=Z(\phi_+)Z^* (-\phi_-)
=\int D\Phi_+ D\Phi_-
e^{iS_0(\Phi_+)-iS_0(\Phi_-)}
\t\n\\
&\t e^{-iV(\Phi_+ +\phi_+) + iV(\Phi_--\phi_-)},
\l{22.35} \ea
see (\ref{22.22}). The fields $(\p_+,\P_+)$ and $(\p_-,\P_-)$ were
defined on the time contours $C_+$ and $C_-$.

As was mentioned above, see (\r{o3}), the path integral (\ref{22.35})
describes the closed path motion in the space of fields $\P$. We want
to use this fact and introduce a more general boundary condition
which also guarantees the cancelation of the surface terms in the
perturbation  framework. We will introduce the equality:
\be
\int_{\s_{\infty}} d\s_{\mu}\P_+\pa^{\mu}\P_+ =
\int_{\s_{\infty}} d\s_{\mu}\P_- \pa^{\mu}\Phi_-.
\l{22.36} \ee The
solution of eq.(\ref{22.36}) requires that the fields $\P_+$ and
$\P_-$ (and their first derivatives $\partial_{\mu}\P_{\pm}$)
coincide on the boundary hypersurface $\s_{\infty}$:
\be
\P_{\pm}(\s_{\infty})=\P(\s_{\infty})\neq0,
\l{22.37}
\ee
where, by definition, $\Phi(\sigma_{\infty})$ is the arbitrary
`turning-point" field.

In the absence of the surface terms, the existence of a nontrivial
field $\P(\s_{\infty})$ has  the influence only on the  structure of
the Green functions
\ba
G_{++}=<T\Phi_+\Phi_+>,~G_{+-}=<\Phi_+\Phi_->,
\n \\
G_{-+}=<\Phi_-\Phi_+>, ~G_{--}=<\tilde{T}\Phi_-\Phi_->,
\l{22.38}
\ea
where $\tilde{T}$ is the antitemporal time ordering operator. These
Green functions must obey the  equations:
\ba
&(\partial^2 +m^2)_x
G_{+-} (x-y)= (\partial^2 +m^2)_x G_{-+} (x-y)=0, \n \\ &(\partial^2
+m^2)_x G_{++} (x-y)= (\partial^2 +m^2)_x^* G_{--} (x-y)=\delta
(x-y),
\l{22.39} \ea
and the general solution of these equations:
\ba
G_{ii}=D_{ii}+g_{ii},
\n \\
G_{ij}=g_{ij},~ i\neq j
\l{22.40}
\ea
contain the arbitrary terms $g_{ij}$ which are the solutions of
homogenous equations:
\be
(\partial^2 +m^2)_x g_{ij} (x-y)=0,~i,j=+,-.
\l{22.41}
\ee
The general solution of these equations (they are distinguished by the
choice of the time contours $C_{\pm}$)
\be
g_{ij}(x-x')=\int d\O_1(q) e^{iq(x-x')} n_{ij} (q)
\l{22.42}
\ee
are defined through the functions $n_{ij}$ which are the
functionals of the `turning-point' field $\Phi(\sigma_{\infty})$: if
$\Phi(\sigma_{\infty})=0$ we must have $n_{ij}=0$.

Our aim is to define $n_{ij}$. We can suppose that
$$n_{ij}\sim <\Phi(\sigma_{\infty})\cdots\Phi(\sigma_{\infty})>.$$
The simplest supposition gives:
\be
n_{ij}\sim <\Phi_{i}\Phi_{j}>\sim <\Phi^2(\sigma_{\infty})>.
\l{22.43}
\ee
We will find the exact definition of
$n_{ij}$ starting from the $S$-matrix interpretation of the  theory.

It was noted previously that the turning-point field
$\P(\s_{\infty})$ may be arbitrary. We will suppose that on the
remote $\s_{\infty}$ there are only free, on the mass-shell,
particles. Formally it follows from (\ref{22.40} - \ref{22.42}).
This assumption is natural also in the $S$-matrix framework \C{pei}.
In other respects the choice of boundary condition is arbitrary.

Therefore, we wish to describe the evolution of the system in a
background field of mass-shell particles. The restrictions connected
with energy-momentum conservation laws will be taken into account
and in other respects background particles are free.  Then our
derivation is the same as in \C {psf}.  Here we restrict ourselves
mentioning only the main quantitative points.

Calculating the  product $a_{nm}a^*_{nm}$ we describe a time ordered
processes of particle creation and absorption described by $D_{+-}$
and $D_{-+}$.  In presence  of the background particles, this
time-ordered picture is slurred over because of the possibility to
absorb particles before their creation appears.

The processes of creation and absorption are described in vacuum by
the product of operators $\hat\p_+\hat\p_-$ and $\hat\p_-\hat\p_+$.
We can derive (see also \C{psf}) the generalizations of
(\ref{22.22}). The presence of the background particles  will lead to
the following generating functional:
\be
R_{cp}=e^{-i{\bf
N}(\phi_i^*\phi_j)}R_0(\phi_{\pm}),
\l{22.44} \ee
where $R_0 (\p_{\pm})$ is the generating functional for the vacuum
case, see (\ref{22.35}). The operator $$ {\bf N}(\phi_i^*\phi_j) $$
describes the external particles environment.

The operator $\hat\p^*_i(q)$ can be considered as the creation and
$\hat\p_i(q)$ as the annihilation operator  and the product
$\hat\p^*_i(q)\hat\p_j(q)$ acts as the activity operator. So, in the
expansion of $N(\hat\p^*_i\hat\p_j)$ we can leave only the first
nontrivial term:
\be
{\bf N}(\phi_i^*\phi_j)= \int d\O(q)
\h{\p}^*_i (q) n_{ij} \h{\p}_j (q),
\l{22.45}\ee
since  no special correlation among background particles should
be expected.  If the external (nondynamical) correlations are present
then the higher powers of $\hat\p^*_i\hat\p_j$ will appear in
expansion (\ref{22.45}).  Following the interpretation of
$\hat\p^*_i\hat\p_j$, we conclude that $n_{ij}$ is the mean
multiplicity of background  particles.

Computing $\R_{cp}$ we   must  conserve the translation invariance
of amplitudes in the background field. Then, to take into account
the energy-momentum conservation laws one should adjust to each
vertex of in-going $a_{nm}$ particles the factor
$e^{-i\a_1q/2}$ and for each out-going particle we have
correspondingly $e^{-i\a_2q/2}$.

So, the product $e^{-i\a_kq/2}e^{-i\a_jq/2}$ can be interpreted as
the probability factor of the one-particle (creation+annihilation)
process. The $n$-particles $(creation+annihilation)$ process
probability is the simple product of  these factors if  there is no
special correlations among background  particles. This
interpretation is evident in the CM frame $\a_k=(-i\b_k,\vec0)$.

After these preliminaries, it is not too hard to find that in the
CM frame we have:
\ba
&n_{++}(q_0)=n_{--}(q_0)=
\frac{\sum_{n=0}^{\infty}ne^{-\frac{\beta_1+\beta_2}{2}|q_0|n}}
{\sum_{n=0}^{\infty}e^{-\frac{\beta_1+\beta_2}{2}|q_0|n}}=
\n \\
&=\frac{1}{e^{\frac{\beta_1 +\beta_2}{2}|q_0|}-1}=
\tilde {n}(|q_0|\frac{\beta_1 +\beta_2}{2}).
\l{22.47}
\ea
Computing $n_{ij}$ for $i\neq j$ we must take into account the
presence of one more particle:
\ba
n_{+-}(q_0)= \theta (q_0)
\frac{\sum_{n=1}^{\infty}ne^{-\frac{\beta_1+\beta_1}{2}q_0 n}}
{\sum_{n=1}^{\infty}e^{-\frac{\beta_1+\beta_1}{2}q_0 n}}+
 \Theta (-q_0)
\frac{\sum_{n=0}^{\infty}ne^{\frac{\beta_1+\beta_1}{2}q_0 n}}
{\sum_{n=0}^{\infty}e^{\frac{\beta_1+\beta_1}{2}q_0 n}}=
\n \\
= \Theta (q_0)(1+\tilde {n}(q_0 \beta_1))+
 \Theta (-q_0)\tilde {n}(-q_0 \beta_1)
\l{22.48}
\ea
and
\be
n_{-+}(q_0)=
 \Theta (q_0)\tilde {n}(q_0 \beta_2)+
 \Theta (-q_0)(1+ \tilde {n}(-q_0 \beta_2)).
\l{22.49}
\ee
Using (\ref{22.47}), (\ref{22.48}) and (\ref{22.49}), and
the definition (\ref{22.40}) we find the Green functions:
\be
G_{i,j}(x-x',(\beta))=\int \frac{d^4 q}{(2\pi)^4} e^{iq(x-x')}
\tilde{G}_{ij} (q, (\beta))
\l{22.50}
\ee
where
\ba
i\tilde{G}_ij (q, (\beta))=
\left( \matrix{
\frac{i}{q^2 -m^2 +i\epsilon} & 0 \cr
0 & -\frac{i}{q^2 -m^2 -i\epsilon} \cr
}\right)
+\n \\ \n \\+
2\pi \delta (q^2 -m^2 )
\left( \matrix{
\tilde{n}(\frac{\beta_1 +\beta_2}{2}|q_0 |) &
\tilde{n}(\beta_2 |q_0 |)a_+ (\beta_2) \cr
\tilde{n}(\beta_1 |q_0 |)a_- (\beta_1) &
\tilde{n}(\frac{\beta_1 +\beta_2}{2}|q_0 |) \cr
}\right)
\l{22.51}
\ea
and
\be
a_{\pm}(\beta)=-e^{\frac{\beta}{2}(|q_0|\pm q_0)}.
\l{22.52}
\ee
The corresponding
generating functional has the standard form:
\be
R_{p}(j_{\pm})=e^{-iV(-i\hat{j}_+)+iV(-i\hat{j}_-)}
e^{\frac{i}{2}\int dx dx' j_i (x)G_{ij}(x-x',(\beta))j_j(x')}
\l{22.53}\ee
where the summation over repeated indexes  is assumed.

Inserting (\ref {22.53}) in the equation of state (\ref{3'}) we
can find that $\beta_1 =\beta_2 =\beta (E)$. If $\beta (E)$ is a
`good' parameter then $G_{ij}(x-x';\beta )$ coincides with the Green
functions of the real-time finite-temperature field theory and the
KMS boundary condition:
\be
G_{+-}(t-t')=G_{-+}(t-t'-i\beta),\;\;\;
G_{-+}(t-t')=G_{+-}(t-t'+i\beta),
\l{22.54}
\ee
is restored. The eq.(\ref{22.54}) can be deduced from (\ref{22.51}) by
direct calculations.

\renewcommand{\theequation}{C.\arabic{equation}}
\section{Appendix. Local temperatures}\0

We  start this consideration from the assumption that the temperature
fluctuations are large scale. We can assume that the temperature is a
`good' parameter in a cell the dimension of which is much smaller
then the fluctuation scale of temperature.  (The `good' parameter
means that the corresponding fluctuations are Gaussian.)

Let us divide the remote hypersurface $\s_\infty$ on a $N_c$ and let
us propose that we can measure the energy and momentum of groups of
in- and out-going particles in each cell. The 4-dimension of cells
can not be arbitrary small because of the quantum uncertainty
principle.

To describe this situation we decompose the $\d$-function of the
initial state constraint (\r{22.4a}) on the product of $(N_c+1)$
$\d$-functions:
$$
\delta^{(4)}(P-\sum^{m}_{k=1}q_k)= \int
\prod^{N_c}_{\nu =1}\{dQ_{\nu}\delta
(Q_{\nu}-\sum^{m_{\nu}}_{k=1}q_{k,\nu})\}
\delta^{(4)}(P-\sum^{N_c}_{\nu =1}Q_{\nu}),
$$
where $q_{k,\nu}$ are the momentum of $k$-th in-going particle in the
$\nu$-th cell and $Q_{\nu}$ is the total 4-momenta of $n_{\nu}$
in-going particles in this cell, $\nu =1,2,...,N_c$. Therefore,
$$
\sum^{N}_{\nu =1}\sum^{m_{\nu}}_{k=1}q_{k,\nu}=P.
$$

The same
decomposition will be used for the second $\d$-function of outgoing
particle constraints.  We must take into account the multinomial
character of particle decomposition on $N$ groups. This will give
the coefficient:
$$
\frac{n!}{n_{1}!\cdots
n_{N}!}\delta_{K}(n-\sum^{N}_{\nu =1}n_{\nu}) \frac{m!}{m_{1}!\cdots
m_{N}!}\delta_{K}(m-\sum^{N}_{\nu =1}m_{\nu}), $$ where $\d_{K}$ is
the Kronecker's symbol. The summation over
$$
\{n_1,n_2,...,n_{N_c}\}=\{n\}_{N_c},~
\{m_1,m_2,...,m_{N_c}\}=\{m\}_{N_c}
$$
is assumed.

In result, the quantity
\ba
&R_{N_c}(P,Q)=
\sum_{\{n,m\}_{N_c}} \int |a_{nm}|^2\t
\n\\
&\t\prod^{N_c}_{\nu =1}\le\{\frac{d\O_{m_\nu}(q_k)}{m_{\nu}!}
\delta^{(4)}(Q_{\nu}-\sum^{m_{\nu}}_{k=1}q_{k,\nu})
\frac{d\O_{n_\nu}(p_k)}{n_\nu!}
\delta^{(4)}(P_{\nu}-\sum^{n_\nu}_{k=1}p_{k,\nu})\ri\}
\l{34'}\ea
defines the probability to find in the $\nu$-th cell the fluxes of
in-going particles with total 4-momentum  $Q_{\nu}$ and of out-going
particles with the total 4-momentum $P_{\nu}$. The sequence of these
two measurements is not fixed.

The Fourier transformation of $\d$-functions in (\ref{34'}) gives:
$$
R_{N_c}(P,Q)=\int \prod^{N}_{k=1}
\frac{d^4 \a_{1,\nu}}{(2\pi)^4}\frac{d^4 \a_{2,\nu}}{(2\pi)^4}
e^{i\sum^{N}_{\nu =1}(Q_{\nu}\a_{1,\nu} +P_{\nu}\a_{2,\nu})}
\R_{N_c}(\a),
$$
where
$$
\R_{N_c}(\a)=\R_{N_c}(\a_{1,1},\a_{1,2}...,\a_{1,N_c};
\a_{2,1},\a_{2,2},...,\a_{2,N})
$$
has the form:
\be
\R_{N_c}(\a)=\int
\prod_{\nu =1}^{N_c}\le\{\prod^{m_{\nu}}_{k=1}
\frac{d\O_{m_\nu}(q)}{m_{\nu}!}
e^{-i\a_{1,\nu}q_{k,\nu}}
\prod^{n_{\nu}}_{k=1} \frac{d\O_{n_\nu}(p)}{n_{\nu}!}
e^{-i\a_{2,\nu}p_{k,\nu}}\ri\}
|a_{nm}|^2.
\l{35'}\ee
Inserting
$$
a_{nm}(p,q)=(-i)^{n+m}\prod_{k=1}^{m}\hat{\phi}(q_{k,\nu})
\prod_{k=1}^{n}\hat{\phi}^* (p_{k,\nu}) Z(-\phi).
$$
into (\ref{35'}) we find:
\ba
&\R_{N_c}(\a)=
\exp\le\{ i\sum_{\nu =1}^{N_c} \int dx dx' [
\h{\p}_+(x)D_{+-}(x-x';\a_{2,\nu})\h{\p}_-(x')-\ri.
\n\\
&\le.-\h{\p}_-(x)D_{-+}(x-x';\a_{1,\nu})\h{\p}_+(x')]\ri\}
\R_0(\p),
\l{36'} \ea
where $D_{+-}(x-x';\a)$, and $D_{-+}(x-x';\a)$ are the
positive and negative frequency correlation functions.

We must integrate over sets $\{Q\}_{N_c}$ and $\{P\}_{N_c}$ if the
distribution of momenta over cells is not fixed. In the result,
\be
R(P)=\int D^{4}\a_1(P) D^{4}\a_2 (P)\R_{N_c}(\a),
\l{37'}\ee
where the differential measure
$$
D^{4}\a(P)=\prod^{N_c}_{\nu =1}\f{d^4\a_\nu}{(2\pi)^4}
K(P,\{\a\}_{N_c})
$$
takes into account the energy-momentum conservation laws:
$$
K(P,\{\a\}_{N_c})=
\int \prod^{N}_{\nu =1} d^4 Q_{\nu}
e^{i\sum^{N_c}_{\nu =1}\a_{\nu}Q_{\nu}}
\d^{(4)}(P-\sum^{N_c}_{\nu =1}Q_{\nu}).
$$
The explicit integration gives that
$$
K(P,\{\a\}_{N_c})\sim\prod^{N_c}_{\nu =1} \d^{(3)}(\a
-\a_{\nu}),
$$
where $\vec{\a}$ is 3-vector of the CM frame. Choosing CM frame,
$\a=(-i\b,\vec{0})$,
$$
K(E,\{\b\}_{N_c})=\int^{\infty}_{0} \prod^{N_c}_{\nu =1} dE_{\nu}
e^{\sum^{N_c}_{\nu =1}\b_{\nu}E_{\nu}}
\delta(E-\sum^{N_c}_{\nu =1} E_{\nu}).
$$

In this frame
$$
\R_{N_c}(P)=\int D\b_1(E)D\b_2(E)\R_{N_c}(\b),
$$
where
$$
D \beta (E)=\prod^{N_c}_{\nu=1}\frac{d \beta_{\nu}}{2\pi i}
K(E,\{\b\}_{N_c})
$$
and
$\R_{N_c}(\b)$ was defined in (\ref{36'}) with $\a_{k,\nu}=
(-i\b_{k,\nu}, \vec{0}), ~\Re\b_{k,\nu} >0, ~k=1,2$.

We will calculate integrals  over $\b_k$ using the stationary phase
method. The equations for the most probable values of $\b_k$:
\be
-\f{\pa}{\pa\b_{k,\nu}}\ln K(E,\{\b\}_{N_c})=
\f{\pa}{\pa\b_{k,\nu}}\ln\R_{N_c}(\b),~k=1,2,
\l{38'}\ee
always have unique positive solutions ${\b}^c_{k,\nu}(E)$. We
propose that the fluctuations of $\b_{k}$ near $\b^c_{k,\nu}$ are
small, i.e. are Gaussian. This is the basis of the local-equilibrium
hypothesis \C{zub}. In this case $1/\b^c_{1,\nu}$ is the temperature
in the initial state in the measurement cell $\nu$ and
$1/\b^c_{2,\nu}$ is the temperature of the final state  in the
$\nu$-th measurement cell.

The last formulation (\ref{37'}) implies that the 4-momenta
$\{Q\}_{N_c}$ and $\{P\}_{N_c}$ cannot be measured. It is possible
to consider another formulation also. For instance, we can suppose
that the initial set $\{Q\}_{N_c}$ is fixed (measured) but
$\{P\}_{N_c}$ is not.  In this case we will have a mixed experiment:
$\b^c_{1,\nu}$ is defined by the equation:
$$
E_{\nu}=-\f{\pa}{\pa\b_{1,\nu}}\ln\R_{N_c}
$$
and
$\b^c_{2,\nu}$ is defined by the second equation in (\ref{38'}).

Considering the continuum limit, $N_c\to\infty$, the dimension of the
cells tends to zero. In this case we are forced by quantum
uncertainty principle to assume that the 4-momenta sets $\{Q\}$ and
$\{P\}$ are not fixed. This formulation becomes pure thermodynamical:
we must assume that just $\{\b_1\}$ and $\{\b_2\}$ are measurable
quantities.  For instance, we can fix $\{\b_1\}$ and try to find
$\{\b_2\}$ as a function of the total energy $E$ and the functional of
$\{\b_1\}$.  In this case eqs.(\ref{38'}) become the functional
equations.

In the  considered microcanonical description, the finiteness of
temperature does not touch the quantization mechanism. Indeed, one
can see from (\ref{36'}) that all thermodynamical information is
confined in the operator exponent
$$
e^{{\bf N}(\phi_i^* \phi_j)}=
\prod_{\nu}\prod_{i\neq j}e^{i\int \hat{\phi}_{i} D_{ij}\hat{\phi}_{j}}
$$
the expansion of which describes the environment, and the `mechanical'
perturbations are  described by the functional $\R_0(\phi)$. This
factorization was achieved by the introduction of the auxiliary field
$\p$ and is independent from the choice of boundary conditions, i.e.
from the choice of the systems environment.

\vskip 0.2cm
\begin{center}
{\bf Wigner functions}
\end{center}
\vskip 0.2cm

We will adopt the Wigner functions formalism in the
Carruthers-Zachariazen formulation \C{carr}. For the sake of
generality, the $m$ into $n$ particles transition will be considered.
This will allow the inclusion of the heavy ion-ion collisions.

In the previous  section, the generating functional $\R_{N_c}(\b)$
was calculated by means of dividing the `measuring device' on the
remote hypersurface $\s_\infty$ into $N_c$ cells
\be
\R_{N_c}(\a)=e^{-i{\bf N}(\p;\b,z)}\R_0(\p),
\l{39'}\ee
where
\ba
&{\bf N}(\p;\b,z)=
\le\{\sum_{\nu =1}^{N_c} \int dx dx'\ri.\t
\n\\
&\t\le.(\h{\p}_+(x)D_{+-}(x-x';\b_{2,\nu},z_2)\h{\p}_-(x')\ri.-
\n\\
&\le.-\h{\p}_-(x)D_{-+}(x-x';\b_{1,\nu},z_1)\h{\p}_+(x'))\ri\}
\l{40'}\ea
is the particle number operator. The frequency correlation functions
$D_{+-}$ and $D_{+-}$ are defined by equalities:
\ba
D_{+-}(x-x';\b_{2,\nu},z_2)=-i\int
d\O_1(q)e^{iq_{2,\nu}(x-x')} e^{-\b_{2,\nu}\e(q_{2,\nu})}
z_2(q_{2,\nu})
\l{a3.5}
\\
D_{-+}(x-x',\b_{1,\nu},z_1)=i\int d\O_1(q)e^{-iq_{1,\nu}(x-x')}
e^{-\b_{1,\nu}\e(q_{1,\nu})}z_1(q_{1,\nu})
\l{b3.5}\ea
It was assumed that the dimension of the device cells tends to zero
($N_c\to\infty$).  Now we wish to specify the cells coordinates. In
the result we will get to the Wigner function formalism.

Let us introduce Wigner variables \C{wig}:
\be
x-x'=r,~x+x'=2y:~~x=y+r/2,~x'=y-r/2.
\l{3.6}\ee
Then
\ba
&{\bf N}(\p;\b,z)=-i\sum_{\nu =1}^{N_c} \int d\O(q)dr\t
\n\\
&\t\le(\h{\p}_+(y+r/2)\h{\p}_-(y-r/2)z_2(q_{2,\nu})e^{iq_{2,\nu}r}
e^{-\b_{2,\nu}\e(q_{2,\nu})}\ri.+
\n\\
&+\le.\h{\p}_-(y+r/2)\h{\p}_+(y-r/2)z_1(q_{1,\nu})
e^{-iq_{1,\nu}r}e^{-\b_{1,\nu}\e(q_{1,\nu})}\ri)dy
\l{3.10}\ea
The Boltzmann factor, $e^{-\beta_{i,\nu}\e(q_{i,\nu})}$, can be
interpreted as the probability to find a particle with the energy
$\e(q_{i,\nu})$ in the final ($i=2$) or initial ($i=1$) state. The
total probability, i.e. the process of creation and further
absorption of $n$ particles, is defined by multiplication of these
factors. Besides $e^{iq_{2,\nu}r}$ is the out-going particle
momentum measured in the $\nu$-th cell.

Generally it is impossible to adjust the 4-index of cell $\nu$ with
coordinate $y$. For this reason the summation over $\nu$ and the
integration over $r$ are performed in (\r{3.10}) independently. But
let us assume that the 4-dimension of the cell $L$ is higher then the
scale of the characteristic quantum fluctuations $L_q$,
\be
L>>L_q.
\l{o6}\ee
One can divide the 4-dimensional $y$ space into the $L$ dimensional
cells. Then, because of (\r{o6}), the quantum fluctuations can not
take away particles from this cell. Then we can adjust the index of
the measurements cell with the index of the $y$ space cell.

As a result,
\ba
&{\bf N}(\p;\b,z)=-i\int dy\int d\O_i(q)dr\t
\n\\
&\t\le(
\h{\p}_+(y+r/2)\h{\p}_-(y-r/2)z_2(q_2,y)e^{iq_2r}
e^{-\b_2(y)\e(q_2)}+\ri.
\n\\
&\le.+\h{\p}_-(y+r/2)\h{\p}_+(y-r/2)z_1(q_1,y)\ri)
e^{-iq_1r}e^{-\b_1(y)\e(q_1)},
\l{3.10'}\ea
where
\be
\int dy=\sum_\nu \int_{C(\nu)}dy,
\l{o7}\ee
and $C(\nu)$ is the dimension $L$ of the $y$ space cell with index
$\nu$. Notice that the momentum $q$ did not carry the index $\nu$ (or
the index $y$ of the space cell).

Our formalism allows the introduction of more general `closed-path'
boundary conditions. The presence of external black-body radiation
will only reorganize the differential operator
$\exp\{\hat{N}(\phi_i^* \phi_j)\}$ and a new generating functional
$\R_{cp}$ has the same form:

$$
\R_{cp}(\b,z)=e^{-i{\bf N}(\p;\b,z)}\R_0 (\phi).
$$
The calculation of operator $\hat{N}(\phi_i^* \phi_j)$
is strictly the same as in Appendix B. Introducing the
cells we will find that
$$
\hat{N}(\p_i^* \p_j)=
\int dr dy \h{\p}_i(r+y/2) \tilde{n}_{ij}(y)\h{\p}_j(r-y/2),
$$
where the occupation number $\tilde{n}_{ij}$ carries the cell index
$y$:
$$
\tilde{n}_{ij}(r,y)=\int d\O_1(q) e^{iqr}n_{ij}(y,q)
$$
and ($q_0 =\epsilon (q)$)
$$
n_{++} (y,q_0)=n_{--}(y,q_0)=\tilde{n}(y,(\beta_1 +\beta_2)|q_0|/2)=
\frac{1}{e^{(\beta_1 +\beta_2)(y)|q_0|/2}-1},
$$
$$
n_{+-}(y,q_0)=\Theta (q_0)(1+\tilde{n}(y,\beta_2 q_0))+
\Theta (-q_0)\tilde{n}(y,-\beta_1 q_0),
$$
$$
n_{-+}(y,q_0)=n_{+-}(y,-q_0).
$$
For simplicity the CM system was used. Other calculations are the
same as the constant temperature case.

\renewcommand{\theequation}{D.\arabic{equation}}
\section{Appendix. Multiperipheral kinematics}\0

First of all \C{kuraev}, two light-like 4-momenta
$$
p_{1,2}=P_{1,2}-P_{2,1}m^2/s
$$
are introduced. Here $P_{1,2}$ are momenta of colliding particles. The
final state particles momenta have the following representation:
\ba
&p'_1=\a_1'p_2+\b_1'p_1+p'_{1\bot},~
p'_2=\a_2'p_2+\b_2'p_1+p'_{2\bot},~
\n\\
&k_i=\a_ip_2+\b_ip_1+k_{i\bot}.
\l{m1}\ea
Sudakov's parameters, $\a,~\b$, are not independent. The mass shell
conditions and the energy-momentum conservation laws give:
\ba
&s\a'_1\b'_1=m^2+(p'_{1\bot})^2=E_{1\bot}^2,~
s\a'_2\b'_2=E_{1\bot}^2,~s\a_i\b_i=E_{i\bot}^2,
\n\\
&\a'_1+\a'_2+\sum\a_i=1,~
\b'_1+\b'_2+\sum\b_i=1,
\l{m2}\ea
where $E_{i\bot}$ is the transverse energy.

We have for the multiperipheral kinematics:
\ba
1\approx\b_1'>>\b_1>>...>>\b_n>>\b'_2\sim\f{m^2}{s},
\n\\
\f{m^2}{s}<<\a'_1<<\a_1<<...<<\a_n<<\a'_2\sim1
\l{m3}\ea
and the transverse momenta are restricted:
\be
|p'_{i\bot}|\sim|k_{i\bot}|\sim m.
\l{m4}\ee
It corresponds to small production angles in the considered CM frame:
\be
\th_i=\f{|k_{i\bot}|}{\sqrt{s}\b_i},~|\b_i|>>|\a_i|,
\l{m5}\ee
if the particle moves along ${\bf P}_1$, and a similar expression
exists for particles moving in the opposite direction, where
$|\b_i|<<|\a_i|$. In the `central region' of the CM frame
$|\b_i|\sim|\a_i|\sim (E_{i\bot}/E)<<1$ the angles of produced
particles are large and energies are small. It should be underlined
that all this excludes the (mini)jets formation.

The final-state particles phase space volume element is
\ba
&d\s_{2\to 2+n}=\f{(2\a_s)^{2+n}}{16\pi^{2n}}C^n_V
\f{d^2q_1}{q_1^2+m^2}\f{d^2q_2}{(q_1-q_2)^2+\la^2}\cdots
\f{d^2q_{n+1}}{(q_n-q_{n+1})^2+\la^2}\f{1}{q_{n+1}^2+\la^2}
\t\n\\
&\t\f{d\a_1}{\a_1}\Th(\a_2-\a_1)\cdots\f{d\a_n}{\a_n}
\prod^{n+1}_{i=1}\le(\f{s_i}{s_0}\ri)^{2\a(q_i^2)}=
\f{1}{q_{n+1}^2+\la^2}dZ_n,
\ea
where $C_V=3$ and the 4-momentum of produced particle
$$
k_i=(\a_i-\a_{i+1})p_2+(\b_i-\b_{i+1})p_1+(q_i-q_{i+1})_\bot=
-\a_{i+1}p_2+\b_ip_1+(q_i-q_{i+1})_\bot.
$$
The square of pairs invariant mass:
$$
s_1=(p_1'+k_1)^2=s|\a_2|,~s_{n+1}=(k_n+p_2')^2=\f{E_{n\bot}^2}{\a_n},~
s_i=E_{(i-1)\bot}^2\f{\a_{i+1}}{\a_{i-1}}.
$$
The energy conservation law takes the form:
$$
s_1s_2\cdots s_n=sE^2_{1\bot}\cdots E^2_{n\bot}.
$$
The trajectory of reggeized gluon is
$$
\a(q^2)=\f{q^2\a_s}{2\pi^2}\int \f{d^2k}{(k^2+\la^2)((q-k)^2+\la^2)},
$$
where $\la$ is the gluon `mass'. If this virtuality is large,
$\la>>m$, then the gluon decay creating a pQCD jet. But the
constraint on the multiperipheral kinematics prevent this
possibility.

\vskip 0.2cm
\begin{center}
{\bf Deep inelastic reactions}
\end{center}
\vskip 0.2cm

For the pure deep inelastic case, when one of the initial hadrons is
scattered at the angle $\theta$ have the energy $E'$ in the cms of
beams whereas the another is scattered at small angle and the large
transfer momentum $Q=4EE'\sin^2(\theta/2)>>m^2,$ is distributed to
the some number of the emitted particles due to evolution mechanism
we have \C{khose} ($\theta$ is small):
\ba
&d\s_n^{DIS}=\frac{4\alpha^2E^{{'}2}}{Q^4 M} dD_n dE'd\cos\th,
\n\\
&dD_n=(\frac{\alpha_s}{4\pi})^n\int_{m^2}^{Q^2}
\frac{d k_n^2}{k_n^2}\int_{m^2}^{k_n^2}\frac{d
k_{n-1}^2}{k_{n-1}^2}\cdots \int_{m^2}^{k_2}\frac{d
k_1^2}{k_1^2}\int_x^1 d\b_n\Th^{(1)}_n\int_{\b_n}^1
d\b_{n-1}\Th^{(1)}_{n-1}...
\n\\
&\t\int_{\beta_2}^1d \beta_1\Th^{(1)}_1
P(\frac{\beta_n}{\beta_{n-1}})...P(\beta_1), \quad
P(z)=2\frac{1+z^2}{1-z},
\ea
where $\Th^{(i)}=\Th(\th_{i+1}-\th_i)$ and the emission angle
$\th_i=|k_i|/(E\max (\a_i.\b_i))$.

\vskip 0.2cm
\begin{center}
{\bf Large angle production}
\end{center}
\vskip 0.2cm

For the large-angle particle production process the differential
cross section (as well as the total one) fall down with CM energy
$\sqrt{s}$. Let us consider for definiteness the process of
annihilation of electron-positron pairs to $n$ photons \C{gorsh}:
\ba
&d\s_n^{DL}=\frac{2\pi\alpha^2}{s}dF_n,
\n\\
&dF_n=\le(\frac{\alpha}{2\pi}\ri)^n
\prod^{n}_{i=1}d x_idy_i\Th(x_i-y_i)\Th(y_i-y_{i-1})
\Th(y_i)\Th(x_i)
\t\n\\
&\t\Th(\R-x_n)\Th(\R-y_n),\quad y_i=\ln\frac{1}{\beta_i},\quad
\rho=\ln\frac{s}{m^2},
\ea
The similar formulae takes place for subprocess of quark-antiquark
annihilation into $n$ large-angle moving gluons.

At the end, one can consider the following possibilities:\\
a) Pomeron regime (P);\\
b) Evolution regime (DIS);\\
c) Double logarithmic regime (DL);\\
d) DIS+P regime;\\
e) P+DL+P regime.\\
The description of every regime may be performed in terms of
effective ladder-type Feynman diagrams. This can be done using the
blocks $dZ_n$, $dD_n$ and $dF_n$.

\renewcommand{\theequation}{E.\arabic{equation}}
\section{Appendix. Reggeon diagram technique for generating function}\0

We will consider, see (\r{i8})
\be
\cp(q,\o;z)=\int_0^\infty d\xi e^{-\o\xi}P(q,\xi;z)=
\f{1}{\o+\a'_0q^2+\psi_0(z)},~\xi=\ln(s/m^2).
\l{s9}\ee
as the `propagator of the cut Pomeron'. It will be assumed also
that
$$
\psi_0(z)=-\D+(1-z)n_0,~n_0>0.
$$
So, the resonance short range correlations will be ignored in this
definition or propagator.  It was assumed also that the `bare' slope
$\a'$ is $z$ independent.

It should be underlined that the `propagator' (\r{s9}) is written
phenomenologically. It absorbs the assumptions that (i) the
diffraction cone shrinks with energy and (ii) the inclusive cross
sections are universal, see (\r{1.2})

The set of principal rules concerning multiperipheral kinematics of
Feynman diagrams is given in Appendix D. The reggeon calculus
supposes that the virtuality of each line of the Feynman diagram is
restricted. This ignores `hard jets', later known as the pQCD jets.

Then the $\nu$ Pomeron exchange eikonal diagram has only
$(\nu+1)$ ways of being cut.  If the cut line goes through $\mu$
Pomerons, then the corresponding contributions are:
\be
\P^\mu_\nu(\o,q)=
\int d\O_\nu \le(M^\mu_\nu(q_1,...,q_\nu)\ri)^2 {\cal Y}^\mu_\nu
\prod_{l=1}^\mu\cp(q_i,\o_i;z)\prod_{i=\mu+1}^\nu
\cp(q_I,\o_i;z=1),
\l{s1}\ee
where $M^\mu_\nu(q_1,...,q_\nu)$ is the `vertex function', the
combinatorial coefficient is
$$
{\cal Y}^\mu_\nu=\f{(-1)^{(\nu-\mu)}2^\nu\nu!}{\mu!(\nu-\mu)!}
$$
and the phase space element is
\be
d\O_\nu=\prod_{i=1}^\nu\f{d\o_ld^2q_l}{(2\pi)^3i}
\d(\o-\sum_{l=1}^\nu\o_l)\d^2(q-\sum_{i=1}^\nu q_l).
\l{s2}\ee

As usual, the contribution (\r{s1}) leads to the following mean
multiplicity of produced particles:
\be
\N^\mu_\nu=\le.\f{\pa}{\pa z}\ln\int d\O_\nu
(s/m^2)^\o\P^\mu_\nu(\o,q=0)\ri|_{z=1} \sim \mu \N.
\l{s3}\ee
Therefore,
\be
\mu\sim\N
\l{s4}\ee
is essential in the VHM region, where $n\sim \N^2$ is assumed.

The impact parameter representation:
\be
\vp^\mu_\nu(s,q)=\int d\O\le(\f{s}{m^2}\ri)^\o\f{d^2q}{(2\pi)^2}
e^{i\bq\bb}\P^\mu_\nu(\o,q)
\l{s5}\ee
would be useful also. The contribution (\r{s1}) describes
interactions with impact parameter
\be
<\bb^2>\simeq 4\a'\ln(s/m^2)/\nu.
\l{s6}\ee
Notice $<\bb^2>$ is the number of cut pomerons independent of
 $\mu$. But, remembering that $\mu\geq\nu$ and that the Regge model
is only able to describe large distance interactions,
$m^2<\bb^2>\geq1$, one can conclude that the Regge pole description is
valid only for
\be
n\leq \N^2.
\l{s7}\ee
Thus is why the VHM region is defined by $\N^2$.

\renewcommand{\theequation}{F.\arabic{equation}}
\section{Appendix. Pomeron with $\D>0$}\0

Then the cut Pomeron propagator in the impact parameter
representation
\be
\tg(\bb,\xi;z)=g(\bb,\xi)e^{(z-1)\N},
\l{l3}\ee
where
\be
g(\bb,\xi)=\f{1}{2\a'\xi}e^{\D\xi}e^{-\bb^2/4\a'\xi}
\l{l4}\ee
is the uncut Pomeron profile function. Using this definition one
can find that the contribution of the eikonal diagrams gives a
contribution:
\be
\cf_0(\bb,\xi; z)=\le( 1-e^{-\la^2g(\bb,\xi)}\ri)-
\f{1}{2}\le(1-e^{2\la^2g(\bb,\xi)(e^{(z-1)\N}-1)}\ri),
\l{l5}\ee
where $\la$ is a constant.

First bracket is essential for $\bb^2\leq4\a'\D\xi^2$.
So, with exponential accuracy, the first term is equal to
$$
\Th(4\a'\D\xi^2-\bb^2).
$$
Let us now consider the second bracket. For $z<1$
\be
\bb^2\leq4\a'\D\xi^2\le\{1+\f{1}{\xi}\ln\le(1-e^{(z-1)\N}\ri)\ri\}=
4\a'\D\xi^2\ga(\xi,z)
\l{l6}\ee
are essential. It is not hard to see that $\ga(\xi,z)$ decreases if
$z\to1$ and $\ga(\xi,z)$ is equal to zero for $z=1$. In this case, by
the definition of the generating function, the integral over $\bb$
of $\cf_0(\bb,\xi; z=1)$ defines the contribution to the total cross
section. So, the model predicts the production of particles in the
ring:
\be
4\a'\x^2\ga(\xi,z)\leq\bb^2\leq4\a'\x^2\D.
\l{l7}\ee
if $z<1$, i.e. if $n<\N$. Notice also that
$\ga(\xi,z=0)=\D+O(e^{-\xi})$. Then,
\be
\cf_0(\bb,\xi; z=0)=\f{1}{2}\cf_0(\bb,\xi; z=1).
\l{l8}\ee
So, the elastic part of the total cross section is half of the total
cross section. This means, using the optical analogy, that the
scattering on the absolutely black disk is well described.

The last conclusion means that the interaction radii should increase
with $n$ in  the VHM region. Indeed, as follows from (\r{l5}), \be
\cf_0(\bb,\xi; z)\simeq
\f{1}{2}\le(e^{2\la^2g(\bb,\xi)(e^{(z-1)\N}-1)}-1\ri),
\l{l9}\ee
at $z>1$ and, therefore,
\be
0\leq\bb^2\leq B^2=4\a'\xi(\D\xi+(z-1)\N)
\l{l10}\ee
are essential.

\renewcommand{\theequation}{G.\arabic{equation}}
\section{Appendix. Dual resonance model of VHM events}\0

Our purpose is to investigate the role of the exponential spectrum
(\r{rms}) in the asymptotic region over multiplicity $n$. In this
case one can validate heavy resonance creation and such formulation
of the problem have definite advantages.

(i) If creation of heavy resonances at $n\to\infty$ is expected, then
one can neglect the dependence on the resonance momentum ${\bf q_i}$ .
So, the `low-temperature' expansion is valid in the VHM region.

(ii) Having the big parameter $n$, one can construct the perturbations
expanding over $1/n$.

(iii) We will be able to show at the end the range of applicability of
these assumptions. \\
For this purpose, the following formal phenomena will be
used. The grand partition function
\be
T(z,s)=\sum_n z^n \s_n(s),~T(1,s)=\s_{tot}(s),~
n\leq\sqrt{s}/m_0\equiv n_{max}(s),
\l{gpf}\ee
will be introduced, see (\r{21.3}). Then the inverse
Mellin transformation, see (\r{21.5})
\be
\s_n(s)=\f{1}{2\pi i}\int\f{dz}{z^{n+1}}T(z,s).
\l{metr}\ee
will be performed expanding it in the vicinity of the solution
$z_c>0$ of the equation of state, see (\r{21.6}):
\be
n=z\f{\pa}{\pa z}\ln T(z,s).
\l{eqst1}\ee
It is assumed, and this should be confirmed at the end, that the
fluctuations in the vicinity of $z_c$ are Gaussian.

It is natural at first glance to consider $z_c=z_c(n,s)$ as an
increasing function of $n$. Indeed, this immediately follows from the
positivity of $\s_n(s)$ and the finiteness of $n_{max}(s)$ at finite
$s$.  But one can consider the `thermodynamical limit', see
Sec.2.3.1, or the limit $m_0\to0$.  Theoretically, the last one is
right because of the PCAC hypotheses and nothing should happen if
the pion mass $m_0\to0$.  In this sense $T(z,s)$ may be considered
as the whole function of $z$.  Then, $z_c=z_c(n,s)$ would be an
increasing function of $n$ if and only if $T(z,s)$ is a regular
function at $z=1$.

The proof of this statement is as follows. We should conclude, as
follows from eq.(\r{eqst1}), that
\be
z_c(n,s)\to z_s~{\rm at}~n\to\infty,~{\rm and~at}~s=const,
\l{asy}\ee
i.e. the singularity point $z_s$ $attracts$ $z_c$ in asymptotics
over $n$. If $z_s=1$, then $(z_c-z_s)\to +0$, when $n$ tends to
infinity \C{0.6}. The concrete realization of this possibility
is shown in Sec.3.3. But if $z_s>1$, then $(z_c-z_s)\to-0 $ in VHM
region, see Secs.3.1, 3.5.

On may use the estimation, see also (\r{21.7}):
\be
-\f{1}{n}\ln\f{\s_n(s)}{\s_{tot}(s)}= \ln z_c(n,s)+O(1/n),
\l{est0}\ee
where $z_c$ is the $smallest$ solution of (\r{eqst1}). It should be
underlined that this estimation is $independent$ of the character of
singularity, i.e. the position $z_s$ is only important with $O(1/n)$
accuracy.

\vskip 0.2cm
\begin{center}
{\bf Partition function}
\end{center}
\vskip 0.2cm

Introducing the `grand partition function' (\r{gpf}) the `two-level'
description means that
\ba
&\ln\f{T(z,\b)}{\s_{tot}(s)}=
\sum_k\f{1}{k!}\int
\prod^k_{i=1}\le\{d\O_k(q)dm_i\x(q_i,z)e^{-\b\e_i}\ri\}\t
\n\\
&\t N_k(q_1,q_2,...,q_k;\b)\equiv-\b{\cf}(z,s),
\l{vd}\ea
where $\e(q_i)=(q_i^2+m_i^2)^{1/2}$. This is our group decomposition.
The quantity $\x(q,z)$ may be considered as the local activity. So,
\be
\le.\f{\d T}{\d\x(q,z)}\ri|_{\x=1}\sim\s_{tot}B_1(q)
\l{r1cf}\ee
If the resonance decay forms a group of particles with total
4-momentum $q$, then $B_1(q)$ is the mean number of such groups. The
second derivative gives:
\be
\le.\f{\d^2T}{\d\x(q_1,z)\d\x(q_2,z)}\ri|_{\x=1}\sim\s_{tot}
\{B_2(q_1,q_2)-B_1(q_1)B_1(q_2)\}\equiv\s_{tot}K_2(q_1,q_2)
\l{r2cf}\ee
where $K_2(q_1,q_2)$ is the two groups correlation function, and so
on.  One can consider $B_k$ as Mayer's group coefficients, see
Sec.2.3.3.

The Lagrange multiplier $\b$ was introduced in (\r{vd}) to each
resonance: the Boltzmann exponent $\exp\{-\b\e\}$ takes into
account the energy conservation law $ \sum_i \e_i=E,$ where $E$ is
the total energy of colliding particles, $2E=\sqrt{s}$ in the CM
frame. This conservation law means that $\b$ is defined by equation:
\be
\sqrt{s}=\f{\pa}{\pa\b}\ln T(z,\b).
\l{eqst2}\ee
So, to define the state one should solve two equations of state
(\r{eqst1}) and (\r{eqst2}).

The solution  $\b_c$ of the eq.(\r{eqst2}) has the meaning of inverse
temperature of the gas of resonances if and only if the fluctuations
in vicinity of $\b_c$ are Gaussian, see Sec.2.2.2.

On the second level, we should describe the resonance decay into
hadrons. Using (\r{pois}) we can write in the vicinity
of $z=1$:
\be
\x(q,z)=\sum_n z^n \s^R_N(q)=g^R(\f{m_0}{m})e^{(z-1)\bar{n}(m)}.
\l{rgpf}\ee
The assumptions {\bf B} and {\bf D}, see (\r{rcs}), were used here

So,
\be
-\b{\cf}(z,s)=\sum_k\int
\prod^k_{i=1}\{dm_i\x(m_i,z)\}\tilde{B}_k(m;\b),
\l{vd1}\ee
where $m=(m_1,m_2,...,m_k)$ $\x$ was defined in (\r{rgpf}) and
\be
\tilde{B}_k(m;\b)=\int\prod_{i=1}^k\le\{
d\O_k(q)e^{-\b\e_i(q_i)}\ri\}
B_k(m;q).
\l{1}\ee
Assuming now that $|{\bf q_i}|<<m$ are essential,
\be
\tilde{B}_k(m;\b)\simeq B_k(m)
\prod_{i=1}^k\le\{\sqrt{\f{2m_i}{\b^3}}e^{-\b m_i}\ri\}
\l{est}\ee

Following the duality assumption, one may write:
\be
B_k(m)=\bar{B}_k(m)
\prod_{i=1}^k \le\{m_i^\ga e^{\b_0 m_i}\ri\}
\l{string}\ee
and $\bar{B}_k(m)$ is a slowly varying function of
$m=(m_1,m_2,...,m_k)$:
$$
\bar{N}_k(m)\simeq b_k
$$
In the result, the low-temperature expansion looks as follows:
\be
-\b{\cf}(z,s)=\sum_k \f{2^{k/2}m_0^k(g^R)^kb_k}{\b^{3k/2}}
\le\{\int_{m_0}^\infty dm m^{\ga+3/2}
e^{(z-1)\bar{n}^R(m)-(\b-\b_0)m}\ri\}^k.
\l{fe1}\ee
We should assume that $(\b-\b_0)\geq0$. In this sense one may
consider $1/\b_0$ as the limiting temperature and the above mentioned
constraint means that the resonance energies should be high enough.

\vskip 0.2cm
\begin{center}
{\bf Thermodynamical parameters}
\end{center}
\vskip 0.2cm

Remembering that the position of the singularity over $z$  is
essential, let us assume that the resonance interactions can not
renormalize it, i.e. that the sum (\r{fe1}) is convergent. Then,
leaving the first term in the sum (\r{fe1}),
\be
-\b{\cf}(z,s)=\f{m_0g^RC_1}{\b^{3/2}}
\int_{m_0}^\infty dm (m/m_0)^{\ga+3/2}
e^{(z-1)\bar{n}^R(m)-(\b-\b_0)m}.
\l{fe2}\ee
We expect that this assumption is satisfied if
\ba
&\int_{m_0}^\infty dm m^{\ga+3/2}
e^{(z-1)\bar{n}^R(m)-(\b-\b_0)m}>>
\f{2^{1/2}m_0(g^R)b_2}{b_1\b^{3/2}}\t
\n\\
&\t\le\{\int_{m_0}^\infty dm m^{\ga+3/2}
e^{(z-1)\bar{n}^R(m)-(\b-\b_0)m}\ri\}^2
\l{res1}\ea
for
\be
n\to\infty,~s\to\infty,~\f{nm_0}{\sqrt{s}}\equiv\f{n}{n_{max}} <<1
\l{restr}\ee

So, we would solve our equations of state with the following free
energy:
\be
-\b{\cal F}(z,s)=\f{\a}{\b^{3/2}}
\int_{m_0}^\infty d(\f{m}{m_0}) (\f{m}{m_0})^{\ga'-1}
e^{-\D(m/m_0)},
\l{fe3}\ee
where, using (\r{crex}),
\be
\ga'=\ga+2(z-1)\bar{n}^R_0+5/2=2(z-1)\bar{n}^R_0,~
\D=m_0(\b-\b_0)\geq0,~\a=const.
\l{nw}\ee

We have in terms of these new variables the following equation for
$z$,
\be
n=z\f{2\a\bar{n}^R_0}{\b^{3/2}}\f{\pa}{\pa\ga'}
\f{\Ga(\ga',\D)}{\D^{\ga'}}.
\l{eq1c}\ee
The equation for $\b$ takes the form:
\be
n_{max}=\f{\a m_0}{\b^{3/2}}\f{\Ga(\ga'+1,\D)}{\D^{\ga'+1}},
\l{eq2c}\ee
where $n_{max}=(\sqrt{s}/m_0)$ and $\Ga(\D,\ga')$ is the
incomplete $\Ga$-function:
$$
\Ga(\ga',\D)=\int^\infty_{\D} dx x^{\ga'-1}e^{-x}.
$$

\vskip 0.2cm
\begin{center}
{\bf Asymptotic solutions}
\end{center}
\vskip 0.2cm

Following physical intuition, one should expect the cooling of the
system when $n\to\infty$, for fixed $\sqrt{s}$, and heating when
$n_{max}\to\infty$, for fixed $n$.  But, as was mentioned above,
since the solution of eq.(\r{eq2c}) $\b_c$ is defined by the value of
the total energy, one should expect that $\b_c$ decreases in both
cases.  So, the solution
\be
\D_c\geq0,~ \f{\pa\D_c}{\pa n}<0~{\rm at}~n\to\infty,~
\f{\pa\D_c}{\pa s}<0~{\rm at}~s\to\infty
\l{sol2}\ee
is natural for our consideration.

The physical meaning of $z$ is activity. It defines at $\b=const$
the work needed for one particle creation. Then, if the system is
stable and $T(z,s)$ may be singular at $z>1$ only,
\be
\f{\pa z_c}{\pa n}>0~{\rm at}~n\to\infty,~
\f{\pa z_c}{\pa s}<0~{\rm at}~s\to\infty.
\l{sol3}\ee

One should assume solving equations (\r{eq1c}) and (\r{eq2c}) that
\be
z_c\D^{\ga'_c+1}\f{\pa}{\pa\ga'_c}
\f{\Ga(\ga'_c,\D_c)}{\D_c^{\ga'_c}}<<\Ga(\ga'_c+1,\D_c).
\l{ineq1}\ee
This condition contains the physical requirement that $n<<n_{max}$.
In the opposite case, the finiteness of the phase space for
$m_0\neq0$ should be taken into account.

As was mentioned above, the singularity $z_s$ attracts $z_c$ at
$n\to\infty$. For this reason one may consider the following
solutions.

A. $z_s=\infty$: $z_c>>\D$, $\D<<1$.

In this case
\be
\D^{-\ga'}\Ga(\ga',\D)\sim e^{\ga'\ln(\ga'/\D)}.
\l{a1}\ee
This estimation gives the following equations:
\be
n=C_1\ga'\ln(\ga'/\D)e^{\ga'\ln(\ga'/\D)},
\f{n}{n_{max}}=C_2 \D\ga'\ln(\f{\ga'}{\D})<<1,
\l{a2}\ee
where $C_i=O(1)$ are the unimportant constants. The inequality
is a consequence of (\r{ineq1}).

These equations have the following solutions:
\be
\D_c\simeq\f{n}{n_{max}\ln n}<<1,~ \ga'_c\sim\ln n>>1.
\l{a3}\ee
Using them one can see from (\r{est0}) that it gives
\be
\s_n<O(e^{-n}).
\l{a4}\ee

B. $z_s=+1$: $z_c\to1$, $\D_c<<1$.

One should estimate $\Ga(\ga',\D)$ near the singularity at $z=1$ and
in the vicinity of $\D=0$ to consider the consequence of this
solution.  Expanding $\Ga(\ga',\D)$ over $\D$ at $\ga'\to0$,
\be
\Ga(\ga',\D)=\Ga(\ga')-\D^{\ga'}e^{-\D}+O(\D^{\ga'+1})
\simeq
\f{1}{\ga'}+O(1).
\l{b1}\ee
This gives the following equations for $\ga'$:
\be
n=C_1'\f{\ga'\ln(1/\D)-1}{\ga'}e^{\ga'\ln(1/\D)}.
\l{b2}\ee
The equation for $\D$ has the form:
\be
n_{max}=C_2'e^{(\ga'+1)\ln(1/\D)}.
\l{b3}\ee
Where $C_i'=O(1)$ are unimportant constants.

At
\be
0<\ga'\ln(1/\D)-1<<1,~{\rm i.e.~at}~\ln(1/\D)<<n<<\ln^2(1/\D),
\l{b4}\ee
we find:
\be
\ga'_c\sim \f{1}{\ln(1/\D_c)}.
\l{b5}\ee
Inserting this solution into (\r{b3}):
\be
\D_c\sim\f{1}{n_{max}}.
\l{b6}\ee
It is remarkable that $\D_c$ in the leading approximation is $n$
independent. By this reason $\ga'_c$ becomes $n$ independent also:
\be
\ga'_c\sim \f{1}{\ln(n_{max})}:~
z_c=1+\f{1}{\bar{n}_0^R\ln(n_{max})}.
\l{b7}\ee
This means that
\be
\s_n=O(e^{-n})
\l{b8}\ee
and obeys the KNO scaling with mean multiplicity
$\bar{n}=\bar{n}_0^R\ln(n_{max})$.

\renewcommand{\theequation}{H.\arabic{equation}}
\section{Appendix. Correlation functions in DIS kinematics}\0

Considering particle creation in the DIS processes, one should
distinguish correlation of particles in the (mini)jets and the
correlations between (mini)jets. We will start from the description of
the jet correlations. One should introduce the inclusive cross
section for the $\nu$ jets creation
$$\P_\nu^{(r)_\nu}(k_1,k_2,....,k_\nu;q^2,x),$$
where $k_i$, $i=1,2,...,n$ are the jets 4-momentum in the DIS
kinetics, $-q^2>>\la^2$.  Having $\P_\nu$ we can find the
correlation functions $$N_\nu^{(r)}(k_1,k_2,....,k_\nu;q^2,x),$$
where $(r)=r_1,...,r_\nu$ and $r_i=(q,g)$ defines the sort of created
colour particle. It is useful to introduce the generating functional
\be
F^{ab}(q^2,x;w)=\sum_n\int d\O_n(k)\prod_{i=1}^n w^{r_i}(k_i)
\le|a^{ab}_n(k_1,k_2,....,k_n;q^2,x)\ri|^2,
\l{10a}\ee
where $a^{ab}_n$ is the amplitude, $d\O_n(k)$ is the phase space
volume and $w^{r_i}(k_i)$ are the arbitrary functions. It is evident
that
\be
F^{ab}(q^2,x;w)\le.\ri|_{w=1}=D^{ab}(q^2,x).
\l{11a}\ee
The inclusive cross sections
\be
\P_\nu^{(r)}(k_1,k_2,....,k_\nu;q^2,x)=\prod_{i=1}^\nu
\f{\d}{\d w^{r_i}(k_i)}F^{ab}(q^2,x;w)\le.\ri|_{w=1}.
\l{12}\ee
The correlation function
\be
N_\nu^{(r)}(k;q^2,x)
=\prod_{1=1}^\nu
\f{\d}{\d w^{r_i}(k_i)}\ln F^{ab}(q^2,x;w)\le.\ri|_{w=1}.
\l{13a}\ee
We can find the partial structure functions $D^{ab}(q^2,x;n)$, where
$n$ is the number of produced (time-like) gluons, using there
definitions.

It will be useful to introduce the Laplace transform over the variable
$\ln(1/x)$:
\be
F^{ab}(q^2,x;w)=\int_{\Re j<0}\f{dj}{2\pi i}\le(\f{1}{x}\ri)^j
f^{ab}(q^2,j;w)
\l{14a}\ee

The expansion parameter of our problem $\a_s\ln(-q^2/\la^2)\sim1$.
For this reason one should take into account all possible cuts of
the ladder diagrams. So, calculating $D^{ab}(q^2,x)$ in the LLA all
possible cuts of the skeleton ladder diagrams are defined by the
factor \C{lla}:
\be
\f{1}{\pi}\le\{\Ga^{ab}_{r}G_{r}\Ga_{r}^{ab}\ri\},
\l{15a}\ee
i.e. the cut line may not only get through the exact Green
function $G_{r}(k^2_i)$ but through the exact vertex functions
$\Ga^{ab}_{r}(q_i,q_{i+1},k_i)$ also ($q_i^2,q_{i+1}^2$ are
negative). We have in the LLA (see Appendix D)
$$
\la^2<<-q^2_i<<-q_{i+1}^2<<-q^2
$$
and
$$
x\leq x_{i+1}\leq x_i\leq 1.
$$
Following our approximation, see previous section, we would not
distinguish the way in which the cut line goes through the Born
amplitude
$$
a^{ab}_{r}=\le\{(\Ga^{ab}_{r})^2G_{r}\ri\}.
$$
We will simply associate $w^{r}\Im a^{ab}_{r}$ to each rung of the
ladder.

Considering the asymptotics over $n$, the time-like partons virtuality
$k_i\simeq -q^2_i/y_i$ should be maximal. Here $y_i$ is the fraction
of the longitudinal momentum of the jet. Then, slightly limiting the
jets phase space,
\be
\ln k_i^2=\ln \le|q_{i+1}\ri|^2(1+O(\ln(1/x)/|q_{i+1}|^2)).
\l{16}\ee

As a result, introducing $\tau_i=\ln(q^2_i/\La^2)$, where
$\a_s(q^2)=1/\b\tau$, $\b=(11N/3)-(2n_f/3)$ in the LLA variable, we
can find the following set of equations:
\be
\tau\f{\pa}{\pa\tau}f_{ab}(q^2,j;w)=\sum_{c,r}\vp^r_{ac}(j)w^r(\tau)
f_{ab}(q^2,x;w),
\l{17a}\ee
where
\be
\vp^r_{ac}(j)=\vp_{ac}(j)=\int^1_0\f{dx}{x}x^jP_{ac}(x)
\l{18a}\ee
and $P_{ac}(x)$ is the regular kernel of the Bethe-Salpeter equation
\C{lla}. At $w=1$ this equation is the ordinary one for
$D^{ab}(q^2,x)$.

We will search the correlation functions from eq.(\r{17a}) in terms of
the Laplace transform
$$
n_{ab}^{(r)_\nu}(k_1,k_2,....,k_\nu;q^2,j)=n_{ab}^{(r)_\nu}(k;q^2,j)
$$
Let us write:
\be
f_{ab}(q^2,j;w)=d_{ab}(q^2,j)\exp\le\{
\sum_\nu\f{1}{\nu!}\int\prod_{i=1}^\nu\le(\f{d\tau_i}{\tau_i}
(w^{r_i}(\tau_i)-1)\ri)n_{ab}^{(r)_\nu}(k;q^2,j)\ri\}
\l{19a}\ee
Inserting (\r{19a}) into (\r{17a}) and expanding over $(w-1)$ we find
the sequence of coupled equations.

Omitting the cumbersome calculations, we write in the LLA that
\ba
&\p_{ab}^{(r)_\nu}(\tau_1,\tau_2,....,\tau_\nu;q^2,j)=
\n\\
&=d_{ac_1}(j,\tau_1)\vp^{r_1}_{c_1c_2}(j)d_{c_2c_3}(j,\tau_2)\cdots
\vp^{r_\nu}_{c_\nu c_{\nu+1}}(j)d_{c_{\nu+1}b}(j,\tau_{\nu+1}).
\l{20}\ea
One should take into account the conservation laws:
\be
\tau_1\cdot\tau_2\cdots\tau_{\nu+1}=\tau,~
\tau_1<\tau_2<\dots<\tau_{\nu+1}<\tau.
\l{21}\ee
Computing the Laplace transform of this expression we find
$$\P_{ab}^{(r)_\nu}(\tau_1,\tau_2,....,\tau_\nu;q^2,x).$$

The kernel $d_{ab}(j,\tau)$ was introduced in (\r{20}). Let us write
it in the form:
\be
d_{ab}(j,\tau)=\sum_{\s=\pm}\s\f{d_{ab}(j)}{\nu_+-\nu_-}\tau^{\nu_\s(j)},
\l{22}\ee
where
\be
d^\s_{qq}=\nu_s-\vp_{gg},~d^\s_{qg}=\nu_s-\vp_{qq},~
d^\s_{qg}=\vp_{gq},~d^\s_{gq}=\vp_{qg}
\l{23}\ee
and
\be
\nu_\s=\f{1}{2}\le\{\vp_{qq}+\vp_{gg}+\s\le[(\vp_{qq}-\vp_{gg})^2-
4n_f\vp_{qg}\vp_{gq}\ri]^{1/2}\ri\}.
\l{24}\ee
If $x<<1$, then $(j-1)<<1$ are essential. In this case \C{lla},
\be
\vp_{gg}\sim\vp_{gq}>>\vp_{qg}\sim\vp_{qq}=O(1).
\l{25}\ee
This means the gluon jets dominance and
\be
n^g_{gg}=\vp_{gg}+O(1).
\l{26}\ee
One can find the following estimation of the two-jet correlation
function:
\be
n_{ab}^{r_1r_2}(\tau_1,\tau_2;\j,\tau)=
O\le({\rm max}\{(\tau_1/\tau)^{\vp_{gg}},(\tau_2/\tau)^{\vp_{gg}},
(\tau_1/\tau_2)^{\vp_{gg}}\}\ri\}.
\l{27}\ee
This correlation function is small since in the LLA
$\tau_1<\tau_2<\tau$. This means that the jet correlation becomes
high if and only if the mass of the correlated jets are comparable.
But this condition shrinks the range of integration over $\tau$ and
for this reason one may neglect the `short-range' correlations among
jets.  Therefore, as follows from (\r{19a}),
\be
f_{ab}(q^2,j;w)=d_{gg}(\tau,j)\exp\le\{
\vp_{gg}\int^\tau_{\tau_0}\f{d\tau'}{\tau'}
w^{g}(\tau')\ri\}
\l{28}\ee
We will use this expression to find the multiplicity distribution in
the DIS domain.

\vskip 0.2cm
\begin{center}
{\bf Generating function}
\end{center}
\vskip 0.2cm

To describe particle production, one should replace: $$w^{r}\Im
a^{ab}_{r}\to w^{r}_n\Im a^{ab}_{r},$$ where $w^{r}_n$ is the
$probability$ of $n$ particle production,
\be
\sum_nw^{r}_n=1.
\l{29}\ee
Having $\nu$ jets, one should take into account the conservation
condition $n_1+n_2+...+n_\nu=n$. For this reason, the generating
functions formalism is useful. In the result, one can find that if we
take (\r{28})
\be
w^g=w^g(\tau,z),~w^g(\tau,z)\le.\ri|_{z=1}=1,
\l{30}\ee
then $f_{ab}(q^2,j;w)$ defined by (\r{28}) is the generating
functional of the multiplicity distribution in the `$j$
representation'. In this expression $w^g(\tau,z)$ is the generating
function of the multiplicity distribution in the jet of mass $|k|=\la
e^{\tau/2}$.

In the result, see (\r{14a}),
\be
F^{ab}(q^2,x;w)\propto\int_{\Re j<0}\f{dj}{2\pi i}({1}/{x})^j
e^{\vp_{gg}\o(\tau,z)}
\l{31}\ee
where
\be
\o(\tau,z)=\int^\tau_{\tau_0}\f{d\tau'}{\tau'}w^{g}(\tau',z).
\l{32'}\ee
Noting the normalization condition (\r{30}),
\be
\o(\tau,z=1)=\ln\tau.
\l{33'}\ee

The integral (\r{31}) may be calculated by the steepest descent
method.  It is not hard to see that \be j\simeq
j_c=1+\le\{4N\o(\tau,z)/\ln(1/x)\ri\}^{1/2} \l{34}\ee is essential.
Notice that $j-1<<1$ should be essential but we find, instead of the
constraint (\r{2}), that
\be \o(\tau,z)<<\ln(1/x).
\l{35}\ee
In the frame of this constraint,
\be
F^{ab}(q^2,x;w)\propto
\exp\le\{4\sqrt{N\o(\tau,z)\ln(1/x)}\ri\}.
\l{36}\ee

Generally speaking, there exist such values of $z$ that
$j_c-1\sim1$.  This is possible if $\o(\tau,z)$ is a regular function
of $z$ at $z=1$.  Then $z_c$ should be an increasing function of $n$
and consequently $\o(\tau,z_c)$ would be an increasing function of
$n$.  Therefore, one may expect that in the VHM domain $j_c-1\sim1$.

Then $j\simeq 1+\o(\tau,z)/\ln(1/x)$ would be essential in the
integral (\r{31}). This leads to the following estimation:
$$
F^{ab}(q^2,x;w)\propto e^{-\o(\tau,z)}.
$$
But this is impossible since $F^{ab}(q^2,x;w)$ should be an
increasing function of $z$. This shows that the estimation (\r{36})
has a finite range of validity.

Solution of this problem with unitarity is evident. One should take
into account correlations among jets considering the expansion
(\r{19a}). Indeed, smallness of $n_{ab}^{(r)_\nu}$ may be compensated
by large values of $\prod_i^\nu w^{r_i}(\tau_i,z)$ in the VHM domain.

\renewcommand{\theequation}{I.\arabic{equation}}
\section{Appendix. Solution of the jets evolution equation}\0

One may neglect quark jets in the VHM region since the gluons mean
multiplicity $\bar{n}_g>\bar{n}_q$ - quarks multiplicity \C{140, 141}
and in the VHM region the leftmost singularities are important. Then
we can write \C{144}:
\be
\f{\pa}{\pa\tau}T_j(\tau,z)=\f{12}{11}T_j(\tau,z)
\int^\tau_{\tau_0}d\tau'(T_j(\tau',z)-1),
\l{j1}\ee
where $\tau=\ln(q^2/\la^2)$ and $T_j(\tau,z)$ is the generating
function of the distribution over the number of gluons $w_n(tau)$:
\be
T_j(\tau,z)=\sum_n z^n w_n(\tau),~T_j(\tau,z=1)=1.
\l{j2}\ee
We search a solution in the VHM region, where
\be
n>>n_j\propto\exp\{\sqrt{a\tau}\},~a=12/11.
\l{j3}\ee
Let us consider the following solution:
\be
w_n=\le(\f{n}{\bar{n}_j}\ri)^\ga e^{-\a n/\bar{n}_j}.
\l{j4}\ee
It is useful to introduce
\be
\a_k(\tau)=\f{1}{k!}\sum_{n=1}^\infty n^{k-1}w_n(\tau)
\l{j5}\ee
for this solution. Inserting (\r{j4}) in this expression,
\be
\a_k(\tau)=\bar{n}^k(\tau)\b_k,
\l{j6}\ee
where $\b_k$ (i) should be positive and (ii) $\tau$ independent.

These conditions are satisfied for the following values of $k$.
Indeed,at $k>>1$,
\be
\b_k=\f{1}{k!}\sum_{n=1}^\infty \f{1}{n}
\le(\f{n}{\bar{n}_j}\ri)^{k+\ga}e^{-\a n/\bar{n}_j}\simeq
\a^{-(k+\ga)}\f{\Ga(k+\ga)}{\Ga(k+1)}.
\l{j7}\ee

The generating function $T_j$ has the following form in terms of
$\a_k$:
\be
T_j(\tau,z)=\sum_{n=1}^\infty (\ln z)^k\a_k(\tau).
\l{j8}\ee
Inserting (\r{j8}) into (\r{j1}) and assuming that $\b_k$ is a $\tau$
independent quantity, we find the following recurrent equation for
$\b_k$:
\be
\b_k=\f{4}{k}\sum^k_{k_1=1}\f{1}{k_1}\b_{k_1}\b_{k-k_1}-
2\f{\sqrt{a\tau}}{k}\b_k.
\l{j9}\ee
Therefore, if
\be
k>> \sqrt{a\tau}
\l{j10}\ee
then we can neglect last term in the right hand side of (\r{j9})
and in this case $\b_k$ are positive and $\tau$ independent. Noting
that in (\r{j7}) $n\sim k\bar{n}_j$ are essential the inequality
(\r{j10}) means that the solution (\r{j4}) is correct if \be
n>>\bar{n}_j\ln\bar{n}_j,
\l{j11}\ee
i.e., only for this value of $n$ $w_n$ has the form (\r{j4}) and
the corresponding generating functional has singularity at
\be
z_s=1+\f{\a}{\bar{n}_j}.
\l{j12}\ee

\renewcommand{\theequation}{J.\arabic{equation}}
\section{Appendix. Condensation and type of asymptotics over
multiplicity}\0

It is important for the VHM experiment to have an upper restriction
on the asymptotics. We wish to show that $\s_n$ decreases faster than
any power of $1/n$:
\be
\s_n<O(1/n).
\l{2.1}\ee
To prove this estimation, one should know the type of singularity at
$z=1$.

One can imagine that the points, where the external particles are
created, form the system. Here we assume that this system is in
equilibrium, i.e. there is not in this system macroscopical flows
of energy, particles, charges and so on.

The lattice gas approximation is used to describe such a system. This
description is quite general and does not depend on details. Motion of
the gas particles leads to the necessity to sum over all
distributions of the particles on cells. For simplicity we will
assume that only one particle can occupy the cell.

So, we will introduce the occupation number $\s_i=\pm1$ in the $i$-th
cell: $\s_i=+1$ means that we have no particle in the cell and
$\s_i=-1$ means that a particle exists in a cell. Assuming that the
system is in equilibrium, we may use the ergodic hypothesis and sum
over all `spin' configurations of $\s_i$, with the restriction:
$\s_i^2=1$. It is evident that this restriction introduces the
interactions \C{wilson}.

The corresponding partition function in temperature representation
\C{langer}
\be
\R(\b,H)=\int D\s e^{-S\la(\s)}
\l{2.2a}\ee
where integration is performed over $|\s(x)|\leq\infty$ and,
considering the continuum limit, $D\s=\prod_x d\s(x)$. The action
\be
S_\la(\s)=\int dx\le\{\f{1}{2}(\nabla \s)^2 -\o\s^2+g\s^4-\la\s\ri\}
\l{2.3}\ee
where
\be
\o\sim\le(1-\f{\b_{cr}}{\b}\ri),~g\sim\f{\b_{cr}}{\b},~
\la\sim\le(\f{\b_{cr}}{\b}\ri)^{1/2}\b H.
\l{2.4}\ee
and $1/\b_{cr}$ is the critical temperature.

\vskip 0.2cm
\begin{center}
{\bf Unstable vacuum}
\end{center}
\vskip 0.2cm

We start this consideration from the case $\o>0$, i.e. assuming that
$\b>\b_{cr}$.  In this case the ground state is degenerate if $H=0$.
The extra term $\sim\s H$ in (\r{2.3}) can be interpreted as the
interaction with external magnetic field $H$. This term regulates the
number of `down' spins with $\s=-1$ and is related to the
activity:
\be
z^{1/2}=e^{\b H},
\l{2.5}\ee i.e. $H$ coincides with the chemical potential.

The potential
\be
v(\s)=-\o\s^2+g\s^4,~\o>0,
\l{2.6}\ee
has two minima at
$$
\s_\pm=\pm\sqrt{\o/2g}.
$$
If the dimension $d>1$, no tunnelling phenomena exist. But choosing
$H<0$ the system in the correct minimum (it corresponds to the state
without particles) becomes unstable. The system tunneling into the
state with an absolute minimum of energy.

The partition function $\R(\b,z)$ becomes singular at $H=0$ because
of this instability. The square root branch point gives
\be
\Im\R(b,z)=\f{a_1(\b)}{H^4}e^{-a_2(\b)/H^2},~a_i>0.
\l{2.7}\ee
Note, $\Im\R(b,z)=0$ at $H=0$. Deforming the contour in the Mellin
integral over $z$ on the branch line,
\be
\R_n(\b)=\f{1}{\pi}\int^\infty_1\f{dz}{z^{n+1}}\f{8a_1\b^4}{\ln^4z}
e^{-4a_2\b^2/\ln^2z}.
\l{2.8}\ee
In this integral
\be
z_c\propto \exp\le\{\f{8a_2\b^2}{n}\ri\}^{1/3}
\l{2.9}\ee
is essential. This leads to the following estimation:
\be
\R_n\propto e^{-3(a_2\b^2)^{1/3}n^{2/3}}<O(1/n).
\l{2.10}\ee

It is useful to note at the end of this section that\\
(i) The value of $\R_n$ is defined by $\Im\R(b,z)$ and the
metastable states, the decay of which gives a contribution into ${\rm
Re}\R(b,z)$, are not important.\\
(ii) It follows from (\r{2.9}) that in the VHM domain
\be
H\sim H_c\sim\ln z_c\sim (1/n)^{1/3}\to0.
\l{2.11}\ee
So, the calculations are performed for the `weak' external field case,
when the degeneracy is weakly broken. It is evident that the life
time of the unstable (without particles) state is large in this case
and for this reason the semiclassical approximation is correct.
This is an important consequence of (\r{A}).

\vskip 0.2cm
\begin{center}
{\bf Stable vacuum}
\end{center}
\vskip 0.2cm

Let us consider now $\o<0$, i.e. $\b<\b_{cr}$. The potential (\r{2.6})
has only one minimum at $\s=0$ in this case. The inclusion of
an external field shifts the minimum to the point $\s_c=\s_c(H)$. In
this case the expansion in the vicinity of $\s_c$ should be useful.
As a result,
\be
\R(\b,z)=\exp\{\int dx \la\s_c-W(\s_c)\},
\l{2.12}\ee
where $W(\s_c)$ can be expanded over $\s_c$:
\be
W(\s_c)=\sum_l\f{1}{l}\int\prod_k\{dx_k\s_c(x_k;H)\}\tilde{b}_l
(x_1,...,x_l).
\l{2.13}\ee
In this expression, $\tilde{b}_l(x_1,...,x_l)$ is the one-particle
irreducible Green function, i.e. $\tilde{b}_l$ is the
virial coefficient. Then $\s_c$ can be considered as the effective
activity of the correlated $l$-particle group.

The sum in (\r{2.13}) should be convergent and, therefore,
$|s_c|\to\infty$ if $|H|\to\infty$. But in this case the virial
decomposition is equivalent to the expansion over the inverse density
of particles \C{mayer}. In the VHM region it is high and the mean
field approximation becomes correct. In result is
\be
\s_c\simeq-\le(\f{|\la|}{4g}\ri)^{1/3}:~|s_c|\to\infty~{\rm if}~
|\la|\to\infty,
\l{2.15}\ee
and
\be
\R(\b,z)\propto e^{\f{3|\la|^{4/3}}{(4g)^{1/3}}}
\le\{12g\le(\f{|\la|}{4g}\ri)^{2/3}\ri\}^{-1/2}.
\l{2.16}\ee
We can use this expression to calculate $\R_n$. In this case
\be
z_c\propto e^{4gn^3}\to\infty~{\rm at}~n\to\infty,
\l{2.17}\ee
is essential and in the VHM domain
\be
\R_n\propto e^{-4gn^4}<O(e^{-n}).
\l{2.18}\ee
This result is an evident consequence of vacuum stability.
It should be noted once more that the conditions (\r{A}) considerably
simplify calculations.

\renewcommand{\theequation}{K.\arabic{equation}}
\section{Appendix. New multiple production formalism and integrable
systems}\0

\begin{center}
{\bf $S$-matrix unitarity constraints}
\end{center}
\vskip 0.2cm

To explain our idea, let us consider the spectral representation of
the one-particle amplitude:
\be
A_1(x_1 ,x_2 ;E)=\f{\Psi^*_{n}(x_2)\Psi_n(x_1)}{E-E_n-i\ve},
~\ve \rar +0,
\l{3a'}\ee
It describes the transition of a particle with
energy $E$ from point $x_1$ to $x_2$. According to our general idea,
see introduction to Sec.2.1, we will calculate
\be
R_1(E)=\int dx_1 dx_2 A_1(x_1 ,x_2 ;E)A^*_1 (x_1 ,x_2 ;E).
\l{4'} \ee
The integration over the end points $x_1$ and $x_2$ is performed only
for the sake of simplicity.

Inserting (\r{3a'}) into (\r{4'}) and using ortho-normalizability of
the wave functions $\Psi_{n} (x)$ we find that
\ba
&\ve R_1(E)=\ve\sum_n\le|\f{1}{E-E_n-i\ve}\ri|^2 =
\f{1}{2i}\sum_n\le\{\f{1}{E-E_n-i\ve}-\f{1}{E-E_n+i\ve}\ri\}=
\n\\
&=\Im\sum_n\frac{1}{E-E_n-i\ve}=\pi\sum_{n}\d(E-E_{n}).
\l{5'}\ea
On other hand, the closed-path amplitude, offered for calculation
in \C{dashen},
\ba
&C_1(E)=\sum_{n}\int dx\f{\Psi^*_{n}(x)\Psi_n(x)}{E-E_n-i\ve}=
\sum_{n}\f{1}{E-E_n-i\ve}=
\n\\
&=\sum_{n}\le\{{\cal P}\f{1}{E-E_n}+
i\pi\d(E-E_n)\ri\}=\sum_{n}{\cal P}\f{1}{E-E_n}+i\ve R_1(E).
\l{4''}\ea
So, we wish to calculate only the imaginary part of the closed-path
contribution:
$$
\ve R(E)=\Im C_1(E).
$$
Notice the extra factor $\ve$ in the left hand side.

The reason for this choice is evident: the real part of $C_1(E)$
is equal to zero at $E=E_n$, i.e. did not contribute to the
measurable. To calculate the bound states energy spectrum, it is
enough to know the only imaginary part of the closed-path amplitude.

This property is not accidental. It is known as the optical theorem
and is the consequence of the total probability conservation
principles.  Formal realization of it is the unitarity condition for
the ${\bf S}$-matrix: ${\bf S}{\bf S}^+ ={\bf I}$. In terms of the
amplitudes ${\bf A}$, ${\bf S}={\bf I}+i{\bf A}$, the unitarity
condition presents an infinite set of nonlinear operator equalities:
\be
i{\bf A}{\bf A}^* ={\bf A} - {\bf A}^*.
\l{2'}\ee
Notice that expressing the amplitude by the path integral one can see
that the left hand side of this equality offers the double integral
and, at the same time, the right hand side is the linear combination
of integrals.  Thus, the continuum contributions into the amplitudes
should be canceled to provide the conservation of total probability.
In this sense it is a necessary condition.

Indeed, to see the integral form of our approach, let us use the
proper-time representation:
\be
A_1(x_1 ,x_2 ;E)=\sum_{n} \Psi_{n}
(x_1)\Psi^{*}_{n} (x_2)i
\int^{\infty}_{0}dTe^{i(E-E_{n}+i\e)T}
\l{6'} \ee
and insert it into (\ref{4'}):
\be
R_1(E)=\sum_{n} \int^{\infty}_{0} dT_{+}dT_{-}
e^{-(T_{+}+T_{-})\e} e^{i(E-E_{n})(T_{+}-T_{-})}.
\l{7'} \ee

We  will introduce new time variables instead of $T_{\pm}$:
\be T_{\pm}=T\pm\tau,
\l{8'} \ee
where, as follows from the Jacobian of transformation, $|\tau|\leq
T,~0\leq T\leq \infty$. But we can put $|\tau|\leq\infty$ since
$T\sim1/\e\rar\infty$ is essential in the integral over
$T$. As a result,
\be
\R_1(E)=2\pi\sum_{n}\int^{\infty}_{0} dT
e^{-2\e T} \int^{+\infty}_{-\infty}\f{d\tau}{\pi}
e^{2i(E-E_{n})\tau}.
\l{9'} \ee
In the last integral, the continuum of contributions with $E\neq
E_{n}$ are canceled.  Note that the product of amplitudes $AA^*$ was
`linearized' after the introduction of `virtual' time \C{fok} $\tau
=(T_{+}-T_{-})/2$.

We wish to calculate the density matrix $\R(\b,z)$ including the
consequence of the unitarity condition cancelation of unnecessary
contributions. Here we demonstrate the result and the intermediate
steps we will formulate, without proof, as the statements offered in
\C{tmf, jmp}, where the formalities are described.

\vskip 0.2cm
\begin{center}
{\bf Dirac measure}
\end{center}
\vskip 0.2cm

The statement, see \C{jmp} and references cited therein,
\vskip 0.1cm
 {\it S1. The unitarity condition unambiguously determines
contributions in the  path integrals for $\R$}
\vskip 0.1cm
looks like a tautology since $e^{iS(x)}$, where $S(x)$ is the
action, is the unitary operator which shifts a system along the
trajectory\footnote{It is well known that this unitary transformation
is the analogy of the tangent transformations of classical mechanics
\C{fok}.}. So, it seems evident that the unitarity condition is
already included in the path integrals.

The rule as the path integrals should be calculated is
weel known, see e.g. \C{a.slav}. Nevertheless the general
path-integral solution contains unnecessary degrees of freedom
(unobservable states with $E\neq E_n$ in the above example).  We
would define the path integrals in such a way that the condition of
absence of unnecessary contributions in the final (measurable) result
be loaded from the very beginning. Just in this sense, the unitarity
looks like the necessary and sufficient condition unambiguously
determining the complete set of contributions.

{\it S2. The $m$- into $n$-particles transition (unnormalized)
$probability$ $R_{nm}$ would have on the Dirac measure the following
symmetrical form}:
\ba
&R_{nm}(p_1,...,p_n, q_1,...,q_m)=
<\prod^{m}_{k=1}|\Ga(q_k;u)|^2 \prod^{n}_{k=1}|\Ga(p_k;u)|^2>_u=
\n\\
&=e^{-i\K(j,e)}\int DM(u)e^{iS_O(u)-iU(u,e)}
\prod^{m}_{k=1}|\Ga(q_k;u)|^2 \prod^{n}_{k=1}|\Ga(p_k;u)|^2\equiv
\n\\
&\equiv\hat{\cal O}(u)\prod^{m}_{k=1}|\Ga(q_k;u)|^2
\prod^{n}_{k=1}|\Ga(p_k;u)|^2.
\l{1.1e}\ea
Here $p(q)$ are the in(out)-going particle momenta. It should be
underlined that this representation is strict and is valid for
arbitrary Lagrange theory of arbitrary dimensions. The eikonal
approximation for inelastic amplitudes was considered in \C{barb}

The operator $\h{\cal O}$ contains three element. The Dirac
measure $DM$, the functional $U(x,e)$ and the operator $\K(j,e)$.

The expansion over the operator
\be
\K (j,e)=\f{1}{2}\Re\int_{C_+} dx dt \f{\d}{\d j(x,t)}\f{\d}{\d
e(x,t)}
\equiv \f{1}{2}\Re\int_{C_+} dx dt \hat{j}(x,t)\hat{e}(x,t)
\l{2.2'}\ee
generates the perturbation series. We will assume that
this series exist (at least in Borel sense).

The functionals $U(u,e)$ and $S_O(u)$ are defined
by the equalities:
\be
S_O(u)=(S_0(u+e)-S_0(u-e))
+2\Re\int_{C_+}dx dte(x,t)(\pa^2+m^2)u(x,t),
\ee
\be
U(u,e)= V(u+e)-V(u-e)-2\Re\int_{C_+} dx dte(x,t)v'(u),
\l{2.3a}\ee
where $S_0(u)$ is the free part of the Lagrangian and $V(u)$
describes interactions. The quantity $S_O(u)$ is not equal to zero
if $u$ have nontrivial topological charge (see also \C{yad}).

According to S1, considering motion in the phase space $(u,p)$ the
measure $DM(u,p)$ has the Dirac form:
\be
DM(u,p)=
\prod_{x,t}
du(x,t) dp(x,t)
\d \le(\dot{u}-\frac{\d H_j (u,p)}{\d p}\ri)
\d \le(\dot{p}+\frac{\d H_j (u,p)}{\d u}\ri)
\l{2.4a}\ee
with the total Hamiltonian
\be
H_j(u,p)=\int dx \{ \f{1}{2}p^2 +\f{1}{2}(\nabla u)^2 +
v(u)-ju \}.
\l{2.5a}\ee
This last one includes the energy $ju$ of quantum fluctuations.

The measure (\r{2.4a}) contains following information:

{\bf a.} Only $strict$ solutions of equations
\be
\dot{u}-\frac{\d H_j (u,p)}{\d p}=0,~
\dot{p}+\frac{\d H_j (u,p)}{\d u}=0
\l{equ}\ee
with $j=0$ should be taken into account. This `rigidness' of the
formalism means the absence of pseudo-solutions (similar to
multi-instanton, or multi-kink) contribution.

{\bf b.} $\R_{nm}$ is described by the $sum$ of all solutions of
eq.(\r{equ}), independently from their `nearness' in the functional
space;

{\bf c.} $\R_{nm}$ did not contain the interference terms from various
topologically nonequivalent contributions. This displays the
orthogonality of corresponding Hilbert spaces;

{\bf d.} The measure (\r{2.4a}) includes $j(x)$ as the external
adiabatic source.  Its fluctuation disturbs the solutions of
eq.(\r{equ}) and {\it vice versa} since the measure (\r{2.4a}) is
strict;

{\bf e.} In the frame of the adiabaticity condition, the field
disturbed by $j(x)$ belongs to the same manifold (topology class) as
the classical field defined by (\r{equ}) \C{yad}.

{\bf f.} The Dirac measure is derived for
$real-time$ processes only, i.e. (\r{2.4a}) is not valid for
tunneling ones.  For this reason, the above conclusions should be
taken carefully.

{\bf g.} It can be shown that theory on the measure (\r{2.4a})
restores ordinary (canonical) perturbation theory.

The parameter $\Ga(q;u)$ is connected directly with $external$
particle energy, momentum, spin, polarization, charge, etc., and
is sensitive to the symmetry properties of the interacting fields
system \footnote{The following trivial analogy with ferromagnetic
may be useful.  So, the external magnetic field $\cal{H}$$\sim
\bar{\mu}$, if $\bar{\mu}$ is the magnetics order parameter, and the
phase transition means that $\bar{\mu}\neq 0$. $\Ga (q,u)$ has just
the same meaning as $\cal{H}$.}. For the sake of simplicity, $u(x)$
is the real scalar field. The generalization would be evident.

As a consequence of (\r{2.4a}), $\Ga(q;u)$ is the function of the
external particle momentum $q$ and is a $linear$ functional of
$u(x)$:
\be
\Ga(q;u)=-\int dx e^{iqx} \f{\d S_0 (u)}{\d u(x)}=
\int dx e^{iqx}(\pa^2 +m^2)u(x) ,~~q^2=m^2,
\l{1.2e}\ee
for the mass $m$ field. This parameter presents the momentum
distribution of the interacting field $u(x)$ on the remote
hypersurface $\s_\infty$ if $u(x)$ is the regular function. Notice,
the operator $(\pa^2 +m^2)$ cancels the mass-shell states of $u(x)$.

The construction (\r{1.2e}) means, because of the Klein-Gordon
operator and the external states being mass-shell by definition
\C{pei}, the solution $\R_{nm}=0$ is possible for a particular
topology (compactness and analytic properties) of $quantum$ field
$u(x)$. So, $\Ga(q;u)$ carries the following remarkable properties:

-- it directly defines the observables,

-- it is defined by the topology of $u(x)$,

-- it is the linear functional of the actions symmetry group element
$u(x)$.

Notice, the space-time topology of $u(x,t)$ becomes important in
calculating integral (\r{1.2}) by parts. This procedure is available
if and only if $u(x,t)$ is the regular function. But the $quantum$
fields are always singular. Therefore, the solution $\Ga(q;u)=0$ is
valid if and only if the quasiclassical approximation is exact. Just
this situation is realized in the soliton sector of sin-Gordon model.

Despite evident ambiguity $\Ga(q;u)$ carries the definite properties
of the order parameter since the opposite solution $\R_{nm}=0$ can
only be the dynamical display of an $unbroken$ symmetry\footnote{The
$S$-matrix was introduced `phenomenologically', see also the example
considered in \C{kadysh, slavnov}, postulating the LSZ reduction
formulae, see eq.(\r{22.5}).  So, the formal constraints, e.g. the
Haag theorem, would not be taken into account on the chosen level of
accuracy.}, i.e. of the nontrivial topology of interacting fields, as
the consequence of unbroken symmetry.

If (\r{equ}) have nontrivial solution $u_c(x,t)$, then this `extended
objects' quantization problem \C{ext.obj.} arises. We solve it by
introducing convenient dynamical variables \C{jmp}. The main formal
difficulty, see e.g.  \C{marinov}, of this program consists of
transformation of the path-integral measure which was solved in
\C{yad}\foot{Number of problems of quantum mechanics was solved using
also the `time sliced' method \C{groshe}.  This approach presents the
path integral as the finite product of well defined ordinary
integrals and, therefore, allows perform arbitrary space and
space-time transformations. But transformed `effective' Lagrangian
gains additional term $\sim\hbar^2$.  Last one crucially depends from
the way as the `slicing' was performed. This phenomena considerably
complicates calculations and the general solution of this problem is
unknown for us. It is evident that this method is especially
effective if the quantum corrections $\sim\hbar$ play no role.  Such
models are well known.  For instance, the Coulomb model in quantum
mechanics, the sine-Gordon model in field theory, where the
bound-state energies are exactly quasiclassical.}.

Then\\
{\it S3. The measure (\r{2.4a}) admits the transformation:
\be
u_c:~(u,p)\to (\x,\eta)\in W=G/G_c.
\l{o11}\ee
and the transformed measure has the form:
\be
DM(u,p)=
\prod_{x,t\it C}
d\x(t) d\eta(t)
\d \le(\dot{\x}-\frac{\d h_j (\x,\eta)}{\d\eta}\ri)
\d \le(\dot{\eta}+\frac{\d h_j (\x,\eta)}{\d\x}\ri),
\l{o12}\ee
where $h_j (\x,\eta)=H_j(u_c,p_c)$ is the transformed Hamiltonian.}
\vskip 0.1cm

It is evident that $(\x,\eta)$ are parameters of integration of
eqs.(\r{equ}) and they form the factor space $W=G/G_c$. For instance,
if one particle dynamics is considered, then one may choose $\x=x(0)$
and $\eta=p(0)$. One may consider also the following possibility:
$$
\x=\int^x\f{du}{\sqrt{2(\eta-v(u))}}
$$
and
$$
\eta=p^2/2+v(x)
$$
In this terms $h_j=\eta- j(t)u_c(\x,\eta)$ and new Hamilton equations
have the form:
\be
\dot\x=1-j\f{\pa u_c(\x,\eta)}{\pa\eta},~
\dot\eta=j\f{\pa u_c(\x,\eta)}{\pa\x}.
\l{o13}\ee
So, we have at $j=0$: $\x=t+t_0$ and $\eta=\eta_0$. By this reason\\
{\it S4. The (action, angle)-type variables are mostly useful}.

According to (\r{o11}) there exists transformation of the perturbation
generating operator:\\
{\it S5. The operator $\K$ has following transformed form:
\be
2\K=\int dt \{\h{j}_\x\cdot\h{e}_\x+\h{j}_\eta\cdot\h{e}_\eta\},
\l{o14}\ee
in the factor space, where ${j}_X,{e}_X$, $X=\x,\eta$, are new
auxiliary variables.}

As a result of mapping of the perturbation generating operator $\K$
on the manifold $W$ the equations of motion became linearized:
\be
DM=\prod_t\d\le(\dot{\x}-\frac{\d h(\eta)}{\d\eta}-j_\x\ri)
\d\le(\dot{\eta}-j_\eta\ri).
\l{o21}\ee
Then\\
{\it S6. If the Feynman's $i\e$-prescription is adopted, then the
Green function of eq.(\r{o21})}
\be
g(t-t')=\Th(t-t')
\l{o22}\ee

Later on we will consider the soliton sector of sin-Gordon model. In
this case $\x_i$ is the coordinate and $\eta_i$ is the momentum of
$i$-th soliton and $N$ is the number of solitons.

Expansion of $\exp\{\K(je)\}$ gives the `strong coupling'
perturbation series. Its analysis shows that\\
{\it S7. Action of the integro-differential operator $\h{\cal O}$
leads to the following representation}:
\ba
&R_{nm}(p,q)=\int_{W}\le\{
d\x(0)\cdot\f{\pa}{\pa\x(0)}R^\x_{nm}(p,q)\ri.+
\n\\
&\le. d\eta(0)\cdot\f{\pa}{\pa\x(0)}R^\eta_{nm}(p,q)\ri\}.
\l{o20}\ea
This means that the contributions into $R_{nm}(p,q)$ are accumulated
strictly on the boundary `bifurcation manifold' $\pa W$ \C{<<<},
i.e. depends directly on the topology of $W$.

\vskip 0.2cm
\begin{center}
{\bf Multiple production in sin-Gordon model}
\end{center}
\vskip 0.2cm

Let us consider now the completely integrable sin-Gordon
model. For the sake of simplicity the integral:
\be
R_2 (q) = e^{-i \hat K (j,e)}\int DM (u,p)
|\Ga (q;u)|^2 e^{is_0 (u)-i U(u,e)},
\l{2.1a}\ee
where $\Ga (q;u)$ was defined in (\r{1.2}), will be calculated.

The effective potential of the sin-Gordon model
\be
U(u_N;e_c)=-\f{2m^2}{\la^2}\int dx dt \sin\la u_N~
(\sin \la e -\la e)
\l{4.15}\ee
with
\be
e_c=e_\x\cdot\f{\pa u_c}{\pa\eta}-e_\eta\cdot\f{\pa u_c}{\pa\x}.
\l{o17}\ee

Performing the shifts in (\r{o21}):
\ba
\x_i (t) \rar \x_i (t) + \int dt' g(t-t') j_{\x,i}(t') \equiv
\x_i (t) +\x'_i (t),
\n \\
\eta_i (t) \rar \eta_i (t) + \int dt' g(t-t') j_{\eta ,i}(t') \equiv
\eta_i (t) +\eta'_i (t),
\l{4.16}\ea
we can get the Green function $g(t-t')$ into the operator
exponent:
\be
\K(ej) =\f{1}{2}\int dt dt' \Th (t-t')\{\hat{\x}'(t)\cdot
\hat{e}_{\x}(t') +\hat{\eta}'(t)\cdot \hat{e}_{\eta}(t')\}.
\ee
Note the Lorentz
noncovariantness of our perturbation theory with Green function
(\r{o22}).

As a result:
\be
D^N M(\x,\eta)=\prod^{N}_{i=1}\prod_{t}d\x_i(t) d\eta_i(t)
\d (\dot{\x}_i- \o (\eta+\eta'))\d (\dot{\eta}_i),~
\o(\eta)=\f{\pa h}{\pa\eta}
\l{c}\ee
with
\be
u_N=u_N (x;\x+\x',\eta+\eta').
\l{4.19}\ee

Using the definition:
$$
\int Dx \d (\dot{x})=\int dx(0)=\int dx_0,
$$
the functional integrals on the measure (\r{c}) are reduced to the
ordinary integrals over initial data $(\x,\eta)_{0}$. These integrals
define zero modes volume. Notice that the zero-modes measure was
defined without the Faddeev-Popov $anzats$.

We would divide the calculations into two parts. First of all, we
would consider the quasiclassical approximation and then we will show
that this approximation is exact.

This strategy is necessary since it seems to be important to show the
role of quantum corrections noting that for all physically
acceptable field theories $R_{nm}=0$ in the quasiclassical
approximation.

The $N$-soliton solution $u_N$ depends upon $2N$ parameters. Half
of them, $N$, can be considered as the position of solitons and the
other $N$ as the solitons momentum. Generally at $|t|\to\infty$ the
$u_N$ solution decomposed on the single solitons $u_s$ and on the
double soliton bound states $u_b$ \C{takh}:
$$
u_N(x,t)=\sum^{n_1}_{j=1}u_{s,j}(x,t)+\sum^{n_2}_{k=1}u_{b,k}(x,t)+
O(e^{-|t|})
$$
Note that this asymptotic is achieved if $\x_i\rar\infty$ or/and
$\eta_i\rar\infty$. This last one defines the bifurcation line of our
model. So, the one soliton $u_s$ and two-soliton bound state $u_b$
would be the main elements of our formalism.  Its $(\x,\eta)$
parametrizations have the form:
\be
u_s(x;\x,\eta)=-\f{4}{\la}\arctan\{\exp(mx\cosh\b\eta
-\x)\},~~~ \b =\f{\la^2}{8}
\l{so} \ee
and
\be
u_b(x;\x,\eta)=
-\f{4}{\la}\arctan\{\tan\f{\b\eta_2}{2}
\f{mx\sinh \f{\b\eta_1}{2}\cos \f{\b\eta_2}{2}-\x_2}
{mx\cosh \f{\b\eta_1}{2}\sin \f{\b\eta_2}{2}-\x_1}\}.
\l{bo}\ee

Performing last integration we find:
\be
R_2(q)=
\sum_N \int \prod^{N}_{i=1} \{d\x_0 d\eta_0\}_i
e^{-i \hat{K}}e^{iS_O(u_N)}
e^{-iU(u_N;e_{\x},e_{\eta})}
|\Ga (q;u_N)|^2
\l{5.1}\ee
where
\be
u_N=u_N (\eta_0 +\eta',\x_0 + \o (t) +\x').
\l{5.2}\ee
and
\be
\o (t)=\int dt'  \th (t-t') \o (\eta_0 +\eta')(t')
\l{5.3}\ee

In the quasiclassical approximation $\x'=\eta'=0$ we have:
\be
u_N=u_N (x;\eta_0 ,\x_0 + \o (\eta_0)t).
\l{5.7}\ee
Notice that the surface term
\be
\int dx^\mu\pa_{\mu}(e^{iqx}u_N)=0.
\l{5.8}\ee
Then
\be
\int d^2x e^{iqx}(\pa^2 +m^2)u_N (x,t)
=-(q^2-m^2)\int d^2x e^{iqx}u_N (x,t) =0
\l{5.9}\ee
since $q^2$  belongs to the mass shell by definition. The condition
(\r{5.8}) is satisfied for all $q_{\mu}\neq0$ since $u_N$ belongs
to Schwarz space (the periodic boundary condition for $u(x,t)$ do not
alter this conclusion). Therefore, in the quasiclassical approximation
$R_2=0$.

Expanding the operator exponent in (\ref{5.1}) we find that action of
operators $\hat{\x}'$, $\hat{\eta}'$ creates terms
\be
\sim\int d^2x e^{iqx} \th (t-t') (\pa^2 +m^2)u_N (x,t) \neq 0.
\l{5.12}\ee
So, generally, if the quantum corrections are included, $R_2\neq0$.

Now we will show that the quasiclassical approximation is exact in
the soliton sector of sin-Gordon model. The structure of the
perturbation theory is readily seen in the `normal-product' form:
\be
R_2(q)=\sum_N \int \prod^{N}_{i=1} \{d\x_0 d\eta_0\}_i
:e^{-iU(u_N;\hat{j}/2i)}e^{is_0(u_N)}|\Ga (q;u_N)|^2:,
\l{6.1}\ee
where
\be
\hat{j}=\hat{j}_{\x}\cdot\f{\pa u_N}{\pa \eta}- \hat{j}_{\eta}\cdot
\f{\pa u_N}{\pa \x}=\O \hat{j}_{X}\f{\pa u_N}{\pa X}
\l{6.2}\ee
and
\be
\hat{j}_{X}=\int dt' \Th(t-t')\hat{X}(t')
\l{6.3}\ee
with the $2N$-dimensional vector $X=(\x ,\eta)$. In eq.(\r{6.2}) $\O$
is the ordinary symplectic matrix.

The colons in (\r{6.1}) mean that the operator $\hat{j}$ should stay
to the left of all functions in the perturbation theory expansion
over it. The structure (\r{6.2}) shows that each order over
$\hat{j}_{X_i}$ is proportional at least to the first order
derivative of $u_N$ over the variable conjugate to $X_i$.

The expansion of (\r{6.1}) over $\hat{j}_{X}$ can be written
\C{yad} in the form (omitting the quasiclassical approximation):
\be
R_2(q)=\sum_N \int \prod^{N}_{i=1} \{d\x_0 d\eta_0\}_i
\{\sum^{2N}_{i=1}\f{\pa}{\pa X_{0i}}P_{X_i}(u_N)\},
\l{6.4}\ee
where $P_{X_i}(u_N)$ is the infinite sum of `time-ordered' polynomials
(see \C{yad}) over $u_N$ and its derivatives. The explicit form of
$P_{X_i}(u_N)$ is complicated since the interaction potential is
nonpolynomial. But it is enough to know, see (\r{6.2}), that
\be
P_{X_i}(u_N)\sim \O_{ij} \f{\pa u_N}{\pa X_{0j}}.
\l{6.5}\ee

Therefore,
\be
R_2(q)=0
\l{6.6}\ee
since (i) each term in (\r{6.4}) is the total derivative, (ii) we have
(\r{6.5}) and (iii) $u_N$ belongs to Schwarz space.

\newpage
\vspace{0.3in}
{\large \bf Acknowledgement}
\vspace{0.2in}

Authors would like to take the opportunity to thank J. Allaby,
A. M. Baldin, J. A. Budagov, R. Cashmore, G. A. Chelkov, M. Della
Negra, D.  Denegri, A.  T.  Filippov, D.  Froidevaux, F. Gianotti,
I. A.  Golutvin, P.  Jenni, V.  G.  Kadyshevski, O. V.  Kancheli, Z.
Krumstein, E.  A.  Kuraev, E.  M.  Levin, L. N.  Lipatov, L. Maiani,
V. A.  Matveev, V.  A.  Nikitin, A.  Nikitenko, A. G. Olchevski, G.
S.  Pogosyan, N.  A.  Russakovich, V.  I.  Savrin, M.  V.  Saveliev,
Ju.  Schukraft, N.  B. Skachkov, A.  N.  Tavkhelidze, V.  M.
Ter-Antonyan, H.  T.  Torosyan, N.E.  Tyurin, D.  V.  Shirkov, E.
Sarkisian, S.  Tapprogge, O.  I.  Zavialov, G.  Zinovyev for valuable
discussions and important comments. The scientific communications
with them on the various stages were important. We acknowledged also
to members of the seminar `Symmetries and integrable systems' of the
N.N.Bogolyubov Laboratory of theoretical physics (JINR), to the
scientific seminars of the V.  P.  Dzelepov Laboratory of nuclear
problems, D. V.  Scobeltsyn INP of the Moscow state University, of
the ATLAS experiment community for a number of important discussions.
We would like also to thank M.  Gostkin and N.  Shubitidze for the
help in the formulation of the Monte Carlo simulation programs and
for the discussions. One of us (J.M.) was supported in part by
Georgian Academy of Sciences.

\newpage
\begin{figure}
\begin{center}
\epsfig{file=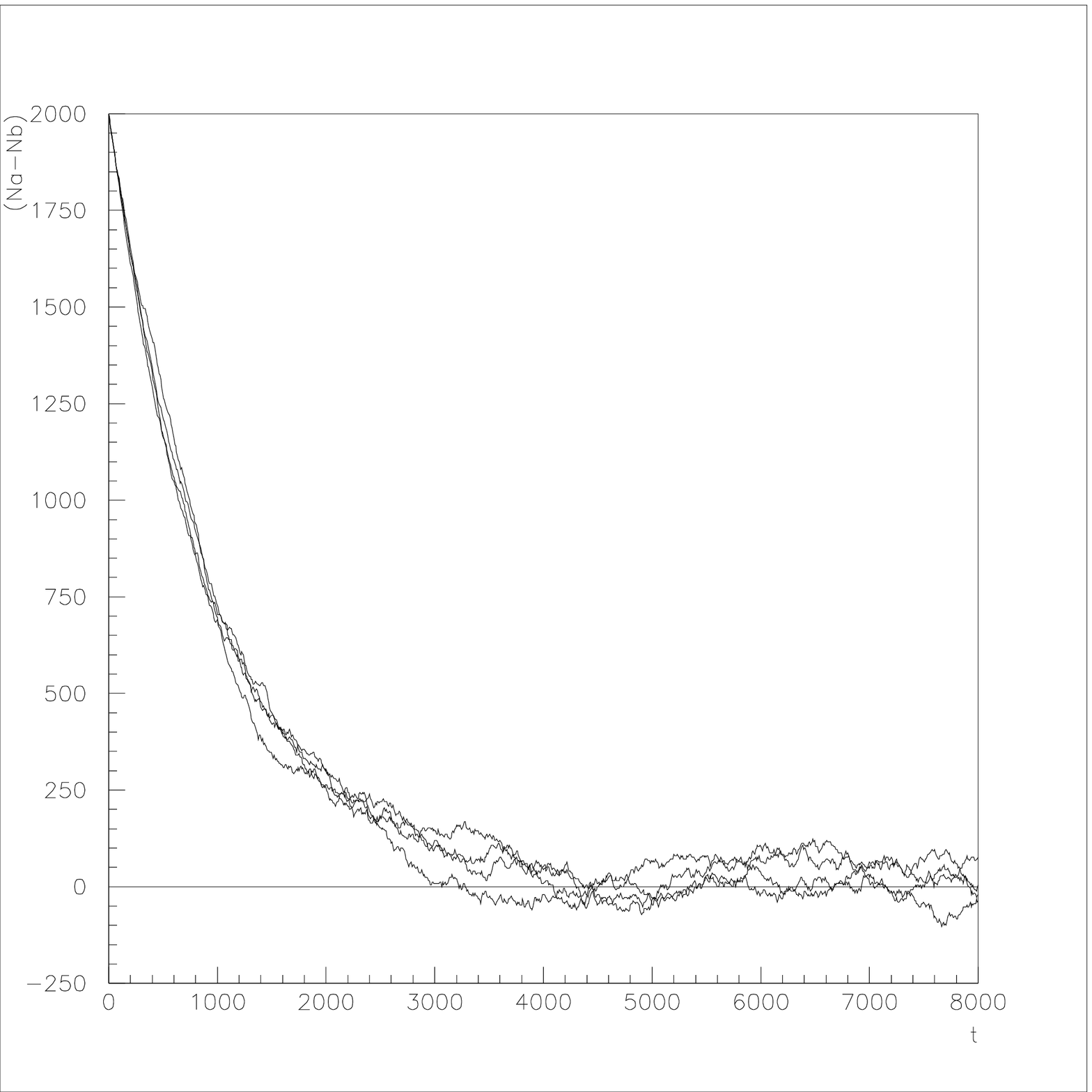,width=14cm}
\caption{Predictions of Euhrenfest model. Four simulations are
displayed.}
\end{center}
\end{figure}
\begin{figure}
\begin{center}
\epsfig{file=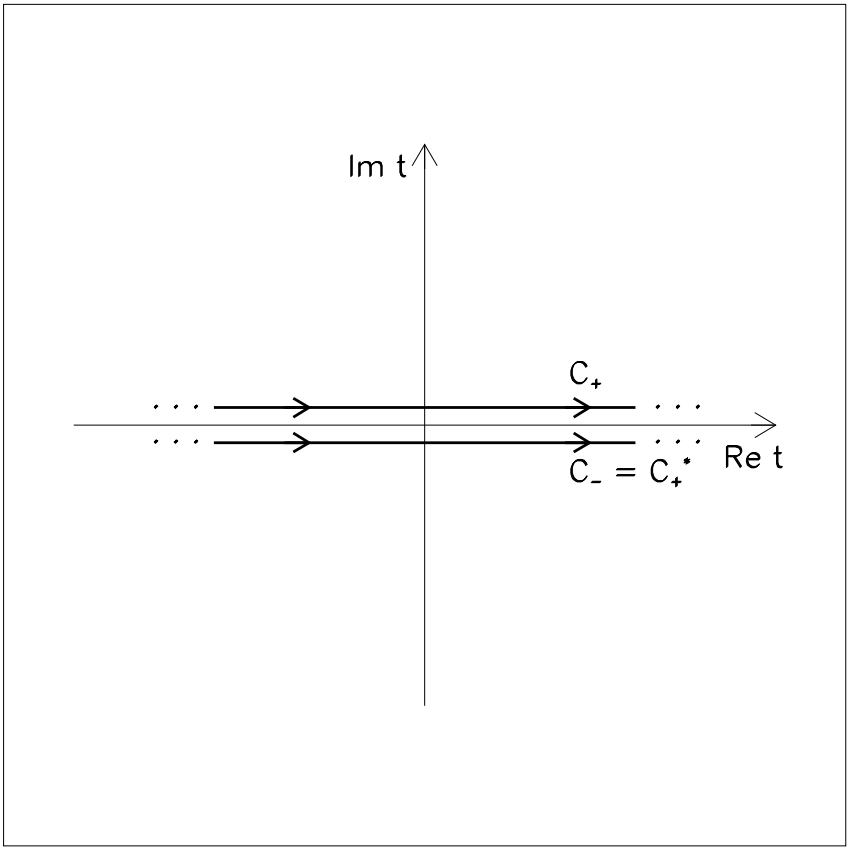,width=14cm}
\caption{Keldysh time contour.}
\end{center}
\end{figure}
\begin{figure}
\begin{center}
\epsfig{file=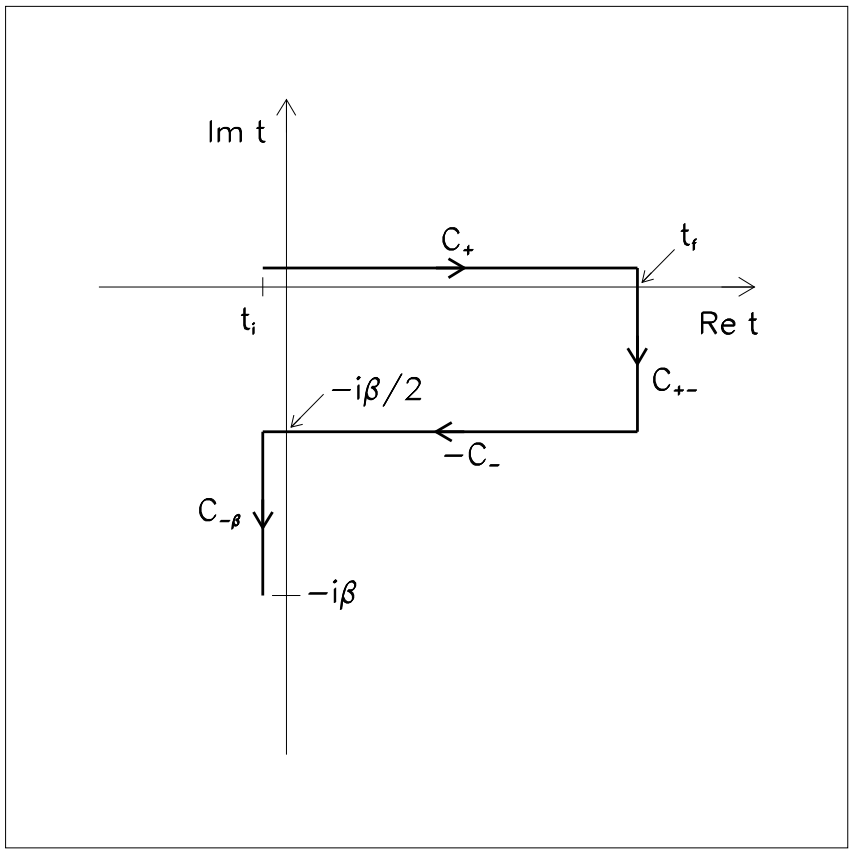,width=14cm}
\caption{Niemi-Semenoff time contour.}
\end{center}
\end{figure}
\begin{figure}
\begin{center}
\epsfig{file=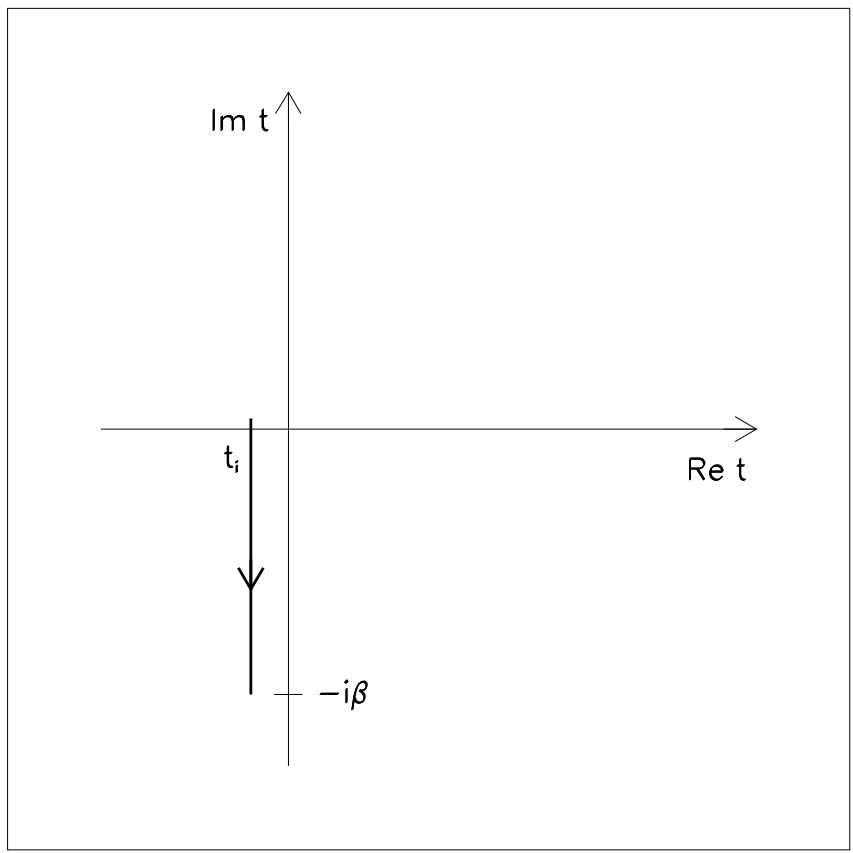,width=14cm}
\caption{Matsubara time contour.}
\end{center}
\end{figure}
\end{document}